%% file: Assessing_DPDL_by_MI.tex
\documentclass{article}

\usepackage{arxiv}

\usepackage[utf8]{inputenc} 
\usepackage[T1]{fontenc}    
\usepackage{hyperref}       
\usepackage{url}            
\usepackage{booktabs}       
\usepackage{amsfonts}       
\usepackage{nicefrac}       
\usepackage{microtype}      
\usepackage{lipsum}
\input{defi}

\usepackage{amsmath,amssymb,amsfonts}
\usepackage{graphicx}
\usepackage{textcomp}
\usepackage{xcolor}
\usepackage[font=footnotesize]{subfig}
\usepackage{todonotes}
\usepackage[normalem]{ulem}

\title{Assessing differentially private deep learning with Membership Inference}

\author{
  Daniel Bernau, Philip-William Grassal\thanks{Both authors contributed equally.}, Jonas Robl\footnotemark[1] \\
  SAP SE\\
  Karlsruhe, Germany \\
  \texttt{firstname.lastname@sap.com} \\
   \And
 Florian Kerschbaum \\
  University of Waterloo \\
  Waterloo, Canada \\
  \texttt{florian.kerschbaum@uwaterloo.ca} \\
}

\begin{document}
\maketitle

\begin{abstract}
{\input{src/abstract.tex}}
\end{abstract}

\input{src/intro.tex}
\input{src/prel-DP.tex}
\input{src/prel-MI.tex}
\input{src/datasets.tex}

\input{src/experiments.tex}
\input{src/discussion.tex}
\input{src/related_work.tex}
\input{src/conclusion}
\input{src/acknowledgements}

\bibliographystyle{abbrv}
\bibliography{Assessing_DPDL_by_MI}

\end{document}

%% file: src/abstract.tex
Attacks that aim to identify the training data of public neural networks represent a severe threat to the privacy of individuals participating in the training data set. 
A possible protection is offered by anonymization of the training data or training function with differential privacy. 
However, data scientists can choose between local and central differential privacy and need to select meaningful privacy parameters \eps~which is challenging for non-privacy experts. 
We empirically compare local and central differential privacy mechanisms under white- and black-box membership inference to evaluate their relative privacy-accuracy trade-offs. 
We experiment with several datasets and show that this trade-off is similar for both types of mechanisms. 
This suggests that local differential privacy is a sound alternative to central differential privacy for differentially private deep learning, since small \eps~in central differential privacy and large \eps~in local differential privacy result in similar membership inference attack risk.

%% file: src/intro.tex
\section{Introduction}
Deep neural networks have successfully been applied to a wide range of learning tasks, each requiring its own specific set of training data, architecture and hyperparameters to achieve meaningful classification accuracy and foster generalization. 
In some learning tasks data scientists have to deal with personally identifiable or sensitive information, which results in two challenges. 
First, legal restrictions might not permit collecting, processing and publishing certain data, such as National Health Service data~\cite{bbc2017}. 
Second, membership inference (MI) ~\cite{HMD+19, shokri2017, NSH18} and model inversion attacks~\cite{fredrikson2015, tramer2016} are capable of identifying and reconstructing training data based on information leakage from a trained, published neural network model. 
A mitigation to both challenges is offered by anonymized deep neural network training with differential privacy (DP). 
However, a data scientist has the choice between two types of DP mechanisms: local DP (LDP)~\cite{wang2017} and central DP (CDP)~\cite{dwork2006b}. 
LDP perturbs the training data before any processing takes place, whereas CDP perturbs the gradient update steps during training. 
The degree of perturbation, which affects the accuracy of the trained neural network on test data, is calibrated for both DP categories by adjusting their respective privacy parameter \eps. 
Choosing \eps~too large will unlikely mitigate privacy attacks such as MI, and choosing \eps~too small will significantly reduce model accuracy. 
Thus, balancing the privacy-accuracy trade-off is a challenging problem especially for data scientists who are not experts in DP.

Furthermore, data scientists might rule out LDP when designing differentially private neural networks due to accuracy concerns raised by the comparatively higher privacy parameter \eps~in LDP. 
In this work, we compare the empirical privacy protection by applying the black-box MI attack of Shokri et al.~\cite{shokri2017} as well as the white-box MI attack of Nasr et al.~\cite{NSH18} using LDP and CDP mechanisms for learning problems from diverse domains: consumer matrices, social graphs, face recognition and health data. These MI attacks indicate a lower bound on the inference risk whereas DP formulates an upper bound, and thus even high privacy parameters \eps~such as experienced in LDP may already offer protection.

In summary, this work makes the following contributions:
\begin{itemize}
	\item We compare LDP and CDP based on MI receiver operating characteristic Area under Curve (AUC) as privacy measure and show that under this measure they have similar privacy-accuracy trade-offs despite vastly different privacy parameters \eps. We provide extensive evaluation on various datasets and reference models, such as face and graph data.
	\item We show that CDP mechanisms are not achieving a consistently better privacy-accuracy trade-off. The trade-off rather depends on the specific dataset and MI assumptions.
	\item We analyze the relative privacy-accuracy trade-off and show that it is not linear over \eps, but that for each data set there are ranges where the relative trade-off is greater for protection against MI than accuracy.
\end{itemize}

The remainder of this paper is structured as follows: 
Section~\ref{sec:prel:dp} outlines local and central differential privacy in the data science process for deep learning and Section~\ref{sec:prel:mi} membership inference. We introduce and analyze experiments and datasets in Section~\ref{sec:data} and~\ref{sec:experiments}. We discuss our findings in Section~\ref{sec:phi}. Related work and conclusions are provided in Section~\ref{sec:rel} and~\ref{sec:conc}.

%% file: src/prel-DP.tex
\section{Differential Privacy}
\label{sec:prel:dp}
DP~\cite{dwork2006a} anonymizes a dataset $\cali{D} = \{d_1,\ldots,d_n\}$ by perturbation and can be either enforced locally to each entry $d\in\cali{D}$, or centrally to an aggregation function $f(\cali{D})$.
\subsection{Central DP}
\label{sec:prel:cdp}

In the central model the aggregation function $f(\cdot)$ is evaluated and perturbed by a trusted server. Due to perturbation, it is no longer possible for an adversary to confidently determine whether $f(\cdot)$ was evaluated on \cali{D}, or some neighboring dataset \cali{D'} differing in one element.
Assuming that every participant is represented by one element, privacy is provided to participants in \cali{D} as their impact of presence (absence) on $f(\cdot)$ is limited. 
\textit{Mechanisms} \cali{M} fulfilling Definition~\ref{def:differential-privacy} are used for perturbation of $f(\cdot)$~\cite{dwork2006b}. We refer to the application of a mechanism \cali{M} to a function $f(\cdot)$ as \textit{central differential privacy}. CDP holds for all possible differences $\|f(\cali{D}) - f(\cali{D'})\|_2$ by adapting to the global sensitivity of $f(\cdot)$ per Definition~\ref{def:gs}.
\begin{definition}[\epsdlt-central differential privacy] A~mechanism~\cali{M} gives \epsdlt-central differential privacy if $\cali{D},\cali{D'}\subseteq\cali{DOM}$ differing in at most one element, and all outputs $\cali{S}\subseteq\cali{R}$
	 \begin{equation*}
	 \Pr[\cali{M}(\cali{D}) \in \cali{S}] \leq e^{\eps} \cdot \Pr[\cali{M}(\cali{D'}) \in \cali{S}] + \dlt
	 \end{equation*}
	 \label{def:differential-privacy}
\end{definition}
\begin{definition}[Global $\ell_2$ Sensitivity]
Let \cali{D} and \cali{D'} be neighboring. The global $\ell_2$ sensitivity of a function $f$, denoted by $\Delta f$, is defined as
	\begin{equation*}
		\Delta f = max_{\cali{D},\cali{D'}}\|f(\cali{D}) - f(\cali{D'})\|_2.
	\end{equation*}
	\label{def:gs}
\end{definition}
For CDP in deep learning we use differentially private versions of two standard gradient optimizers: SGD and Adam\footnote{The Tensorflow privacy package was used throughout this work: \url{https://github.com/tensorflow/privacy}.}. We refer to these CDP optimizers as DP-SGD and DP-Adam. A CDP optimizer represents a differentially private training mechanism $\cali{M}_{nn}$ that updates the weight coefficients $\theta_t$ of a neural network per training step $t \in T$ with $\theta_t \leftarrow \theta_{t-1}-\alpha(\tilde g)$, where $\tilde g~=~\cali{M}_{nn}(\partial loss / \partial \theta_{t-1})$ denotes a Gaussian perturbed gradient and $\alpha$ is some scaling function on $\tilde g$ to compute an update, i.e., learning rate or running moment estimations. Differentially private noise is added by the Gaussian mechanism of Definition~\ref{def:dp:gauss}~\cite{abadi2016}.
\begin{definition}[Gaussian Mechanism~\cite{dwork2014}]
	~\\Let $\eps\in(0,1)$ be arbitrary. For $c^2>2ln(\frac{1.25}{\dlt})$, the Gaussian mechanism with parameter $\sgm\ge c\frac{\Delta f}{\eps}$ gives \epsdlt-CDP, adding noise scaled to $\mathcal{N}(0,\sgm^2)$.
	\label{def:dp:gauss}
\end{definition}
After $T$ update steps, $\cali{M}_{nn}$ outputs a differentially private weight matrix $\theta$ which is used by the prediction function $h(\cdot)$ of a neural network. A CDP gradient optimizer bounds the sensitivity of the computed gradients by a clipping norm \cali{C} based on which the gradients get clipped before perturbation.
Since weight updates are performed iteratively during training a composition of $\cali{M}_{nn}$ is required until the the training step $T$ is reached and the final private weights $\theta$ are obtained. For CDP we measure privacy decay under composition by tracking the noise levels $\sigma$ we used to invoke the Gaussian mechanism. After training we transform and compose $\sigma$ under Renyi DP~\cite{Mironov17}, and transform the aggregate again to CDP. We choose this accumulation method over other composition schemes~\cite{KOV17, abadi2016}, as it provides tighter bounds for heterogeneous mechanism invocations.

\subsection{Local DP}
\label{sec:prel:ldp}
We refer to the perturbation of entries $d\in\cali{D}$ as local differential privacy~\cite{wang2017}. LDP is the standard choice when the server which evaluates a function $f(D)$ is untrusted. We adapt the definitions of Kasiviswanathan et al.~\cite{KLN+08} to achieve LDP by using local randomizers \cali{LR}. In the experiments within this work we use a local randomizer to perturb each record $d\in\cali{D}$ independently. Since a record may contain multiple correlated features (e.g., items in a preference vector) a local randomizer must be applied sequentially which results in a linearly increasing privacy loss. A series of local randomizer executions per record composes a local algorithm according to Definition~\ref{def:la}. \eps-local algorithms are \eps-local differentially private~\cite{KLN+08}, where \eps~is a summation of all composed local randomizer guarantees. We perturb low domain data with randomized response~\cite{warner1965}, a (composed) local randomizer. By Equation~\eqref{eq:rr} randomized response yields $\eps=\ln(3)$ LDP for a one-time collection of values from binary domains (e.g., $\{\var{yes},\var{no}\}$) with two fair coins~\cite{erlingsson2014}. That is, retention of the original value with probability $\rho=0.5$ and uniform sampling with probability $(1-\rho)\cdot0.5$.

\begin{definition}[Local differential privacy]
A local randomizer (mechanism) $\cali{LR}: \cali{DOM}\to\cali{S}$ is \eps-local differentially private, if $\eps\geq 0$  and for all possible inputs $v,v'\in \cali{DOM}$ and all possible outcomes $s\in \cali{S}$ of \cali{LR}
	\begin{equation*}
		\Pr[\cali{LR}(v)=s] \le e^{\eps} \cdot \Pr[\cali{LR}(v')=s]
	\end{equation*}
\label{def:localrand}
\end{definition}
\begin{definition}[Local Algorithm] An algorithm is \eps-local if it accesses the database $\cali{D}$ via $\cali{LR}$ with the following restriction: for all $i\in\{1,\ldots,|\cali{D}|\}$, if $\cali{LR}_1(i),\ldots,\cali{LR}_k(i)$ are the algorithms invocations of $\cali{LR}$ on index $i$, where each $\cali{LR}_j$ is an $\eps_j$-local randomizer, then $\eps_1+\ldots+\eps_k\leq\eps$. \label{def:la}
\end{definition}
\begin{equation}
\label{eq:rr}
	 \eps = \ln{\left(\frac{\rho + (1-\rho)\cdot 0.5}{(1-\rho)\cdot0.5}\right)} = \ln{\left(\frac{\Pr[\var{yes}|\var{yes}]}{\Pr[\var{yes}|\var{no}]}\right)}.
 \end{equation}
In our evaluation we also look at image data for which we rely on the local randomizer by Fan~\cite{Fan18} for LDP image pixelation. The randomizer applies the Laplace mechanism of Definition~\ref{def:lap-mech} with scale $\lambda=\frac{255 \cdot m}{b^2 \cdot \eps}$ to each pixel. Parameter $m$ represents the neighborhood in which LDP is provided. Full neighborhood for an image dataset would require that any picture can become any other picture. In general, providing DP within a large neighborhood will require high \eps~values to retain meaningful image structure. High privacy will result in random black and white images.

\begin{table}[t]
	\caption{Notations and context}
	\small
	\begin{tabularx}{\columnwidth}{lX}
		Symbol & Description \\ 
		\hline
		\addlinespace
		$X$ & Set of vectors $\vec{x_1},\ldots,\vec{x_j}$
		~where $x^1_j,\ldots,x^i_j$ denote attribute values (\textit{features}) of $\vec{x_j}$.\\ 
		\addlinespace
		$Y$ & Set of $k$ target variables $y_1,\ldots,y_k$ (\textit{labels}).\\
		\addlinespace
		$\mathbb{C}$ & $|Y|$.\\
		\addlinespace
		$\vec{y}$ & Vector of target variables (\textit{labels}) where variable $y_j \in \vec{y}$ represents the label for $\vec{x_j}\in X$.\\
		\addlinespace
		$\hat{y}$ & Predicted target variable, i.e., $\hat{y}=h(\vec{x})$.\\
		\addlinespace
		$p(\vec{x})$ & Softmax confidence for $\vec{x}$.\\
		\addlinespace
		$\cali{D}$ & $\cali{D}:=(X,\vec{y})$.\\
		\addlinespace
		$d$ & A record $d\in\cali{D}$, where $d := (\vec{x},y)$.\\
		\addlinespace
		$n$ & $|\subsup{\cali{D}}{\var{target}}{\var{train}}|$.\\
		$h(\cdot)$ & Model function (e.g., classifier). \\
		$L(h(x;W),y)$ & Losses of the model function with learned weights $W$ on record $x$ with true label $y$. \\
		$\frac{\delta L}{\delta W}$ & Gradients of the losses and weights. \\
		\label{tab:notation}
	\end{tabularx}
\end{table}

\begin{definition}[Laplace Mechanism~\cite{dwork2014}]
\label{def:lap-mech}
Given a numerical query function $f: DOM \rightarrow \mathbb{R}^k$, the Laplace mechanism with parameter $\lambda=\frac{\Delta_f}{\eps}$ is an $\eps$-differentially private mechanism, adding noise scaled to $Lap(\lambda,\mu=0)$.  
\end{definition}

Within this work we consider the use of LDP and CDP for deep learning along a generic data science process (e.g., 
CRISP-DM~\cite{WH00}). 
In such a processes the dataset \cali{D} of a data owner \cali{DO} is 
\begin{enumerate*}[label=(\roman*)]
 \item transformed, and
 \item used to learn a model function $h(\cdot)$ (e.g., classification), which
 \item afterwards is deployed for evaluation by third parties. 
\end{enumerate*} 
In the following $h(\cdot)$ will represent a neural network. 
DP is applicable at every stage in the data science process. In the form of LDP by perturbing each record $d\in\cali{D}$, 
while learning $h(\cdot)$ centrally with a CDP gradient optimizer,
or to the evaluation of $h(\cdot)$ by federated learning with CDP voting~\cite{PAE+17}.
We focus on the data science process without collaboration and leave federated learning as future work.

When applying DP in the data science process the privacy-accuracy trade-off is of particular interest. Similar to the evaluation of regularization techniques that apply noise to the training data to foster generalization (e.g.,~\cite{GBC16,GC95,Mat92}), we judge utility by the test accuracy of $h(\cdot)$. 

%% file: src/prel-MI.tex
\\
\section{Membership Inference}
\label{sec:prel:mi}
MI attacks aim at identifying the presence or absence of individual records $d$ in the training data of a machine learning model of data owner \cali{DO}. 
This paper considers two membership inference attacks against machine learning models: black-box MI by Shokri et al.~\cite{shokri2017} and white-box MI by Nasr et al.~\cite{NSH18}. Both attacks assume an honest-but-curious adversary \cali{A} with access to a trained prediction function $h(\cdot)$ and knowledge about the hyperparameters and DP mechanisms that were used for training. Black-box MI limits the access to just predictions from $h(\cdot)$ (e.g., softmax confidence values) whereas white-box MI also assumes access to the internal features of $h(\cdot)$ (e.g., gradients, losses, softmax confidence values). We refer to the trained prediction function as \textit{target model} and the training data of \cali{DO} as \subsup{\cali{D}}{\var{target}}{\var{train}}. 
Given this accessible information \cali{A} wants to learn a binary classifier, the \textit{attack model}, that allows to classify data into members and non-members w.r.t.~the target model training dataset with high accuracy. The accuracy of an MI attack model is evaluated on a balanced dataset including all members (target model training data) and an equal number of non-members (target model test data), which simulates the \cali{DO} worst case where \cali{A} tests membership for all training records. Throughout the next sections we will first outline the black- and white-box MI attacks and, second, illustrate how these attacks are used to assess LDP and CDP privacy parameters. Our notations are summarized in Table~\ref{tab:notation}.

\subsection{Black-Box MI attack}
\label{sec:prel:mi:bb}

The black-box (BB) MI attack is limited to external features of a trained machine learning model. This is for example the case when a model is exposed through an API. 
Black-box MI exploits that an ML classifier such as a neural network (NN) tends to classify a record $d$ from its training dataset \subsup{\cali{D}}{\var{target}}{\var{train}} with different high softmax confidence $p(\vec{x})$ given $h(\vec{x})$ at its true label $y$ in than a record $d\not\in\subsup{\cali{D}}{\var{target}}{\var{train}}$. Therefore, \cali{A} follows two steps. 
First, \cali{A} trains copies of the target model w.r.t.~structure and hyper-parameters, so called \textit{shadow models}, on data statistically similar to \subsup{\cali{D}}{\var{target}}{\var{train}} and \subsup{\cali{D}}{\var{target}}{\var{test}}. 
It applies that $\abs{\subsup{\cali{D}}{\var{\subscr{shadow}{i}}}{\var{train}}}=\abs{\subsup{\cali{D}}{\var{\subscr{shadow}{i}}}{\var{test}}} \land \subsup{\cali{D}}{\var{\subscr{shadow}{i}}}{\var{train}}\cap\subsup{\cali{D}}{\var{\subscr{shadow}{i}}}{\var{test}} = \varnothing \land \abs{\subsup{\cali{D}}{\var{\subscr{shadow}{i}}}{}\cap\subsup{\cali{D}}{\var{\subscr{shadow}{j}}}{}}\ge 0$ 
for any $i\neq j$. 
After training, each shadow model is invoked by \cali{A} to classify all respective training data (member records) and test data (non-member records), i.e., $p(\vec{x})$, $\forall$ $d\in\subsup{\cali{D}}{\var{\subscr{shadow}{i}}}{\var{train}}\cup \subsup{\cali{D}}{\var{\subscr{shadow}{i}}}{\var{test}}$. Since \cali{A} has full control over $\subsup{\cali{D}}{\var{\subscr{shadow}{i}}}{\var{train}}$ and $\subsup{\cali{D}}{\var{\subscr{shadow}{i}}}{\var{test}}$, each shadow model's output $(p(\vec{x}),y)$ is appended with a label \enquote{$\var{in}$} if the corresponding record $d \in \subsup{\cali{D}}{\var{\subscr{shadow}{i}}}{\var{train}}$. Otherwise, its label is \enquote{$\var{out}$}. 
Second, \cali{A} trains a binary classification attack model per target variable $y \in Y$ to map $p(\vec{x})$ to the indicator \enquote{\var{in}} or \enquote{\var{out}}. The triples $(p(\vec{x}),y,\var{in}/\var{out})$ serve as attack model training data, i.e., \subsup{\cali{D}}{\var{attack}}{\var{train}}. Thus, the attack model exploits the imbalance between predictions on $d\in \subsup{\cali{D}}{\var{target}}{\var{train}}$ and $d\not\in\subsup{\cali{D}}{\var{target}}{\var{train}}$. 

\begin{figure}
	\centering
	\begin{tikzpicture}[thick,scale=0.8, every node/.style={transform shape}]
	\footnotesize
	\node[] (aa) {};
	\node[draw,cylinder,shape border rotate=90,minimum height=10ex,minimum width=4.5em, dotted, shape aspect=.20, above=0.5ex of aa, double color fill={violet!10}{white!10}] (a) {\subsup{\cali{D}}{}{train}};
	\node[draw,cylinder,shape border rotate=90,minimum height=10ex,minimum width=4.5em,shape aspect=.20,below=0.5ex of aa, double color fill={green!10}{white!10}] (b) {\subsup{\cali{D}}{}{test}};
	
	\node[right=3.5em of a] (c) {};
	\node[draw,cylinder,shape border rotate=90, minimum width=4.5em, above= 1ex of c, shape aspect=.10, fill=violet!10] (d) {\subsup{\cali{D}}{target}{train}};
	\node[minimum width=4em, above=0ex of d] (uu) {\subscr{\cali{D}}{target}};
	\node[draw,cylinder,shape border rotate=90, minimum width=4.5em, below= 1ex of d, shape aspect=.10, fill=green!10] (e) {\subsup{\cali{D}}{target}{test}};
	\node[draw,chamfered rectangle, align=center, color=black, dashed, right=5em of c] (u) {target \\ model};
	\node[right=3.5em of b] (f) {};
	\node[above= 4ex of f] (fa) {};
	\node[below= 0.5ex of f] (fb) {};
	\node[draw,cylinder,shape border rotate=90, minimum width=4.5em, above= 1ex of fb, shape aspect=.10] (i) {\subsup{\cali{D}}{shadow_i}{train}};
	\node[minimum width=4em, above=0ex of i] (ii) {\subscr{\cali{D}}{shadows}};
	
	\node[draw,cylinder,shape border rotate=90, minimum width=4.5em, below= 1ex of i, shape aspect=.10] (j) {\subsup{\cali{D}}{shadow_i}{test}};
	\node[draw,chamfered rectangle, align=center, color=black, dashed, right=5em of fb] (t) {shadow \\ model$_i$};
	\node[draw,cylinder,shape border rotate=90, minimum width=4.5em, right= 11.5em of i, shape aspect=.10] (n) {\subsup{\cali{D}}{shadow_i}{in}};
	\node[minimum width=4em, above=0ex of n] (nn) {\subsup{\cali{D}}{attack}{train}};
	
	\node[draw,cylinder,shape border rotate=90, minimum width=4.5em, right= 11.5em of j, shape aspect=.10] (o) {\subsup{\cali{D}}{shadow_i}{out}};
	\node[draw,cylinder,shape border rotate=90, minimum width=4.5em, right= 11.5em of d, shape aspect=.10, fill=violet!10] (q) {\subsup{\cali{D}}{target}{in}};
	\node[draw,cylinder,shape border rotate=90, minimum width=4.5em, right= 11.5em of e, shape aspect=.10, fill=green!10] (r) {\subsup{\cali{D}}{target}{out}};
	\node[minimum width=4em, right=11.5em of uu] (vv) {\subsup{\cali{D}}{attack}{test}};
	\node[draw,chamfered rectangle, align=center, below right=0.5ex and 3em of r] (v) {attack \\ model};
	
	\draw [->] (a.east) -- (d.west);
	\draw [->] (a.east) -- (i.west);
	
	\draw [->] (b.east) -- (e.west);
	\draw [->] (b.east) -- (j.west);
	
	\draw [->] (d.east) -- node [near start, above=1.5ex,sloped]{\small training}(u.west);
	\draw [->] (e.east) -- (u.west);
	
	\draw [->] (u.east) -- node [midway, above=1.5ex,sloped]{\small inference}(q.west);
	\draw [->] (u.east) -- (r.west);
	
	\draw [->] (i.east) -- (t.west);
	\draw [->] (j.east) -- node [near start, below=1.5ex,sloped]{\small training}(t.west);
	
	\draw [->] (t.east) -- (n.west);
	\draw [->] (t.east) -- node [midway, below=1.5ex,sloped]{\small inference}(o.west);
	
	\draw [->] (q.east) -- (v.north west);
	\draw [->] (r.east) -- (v.north west);
	
	\draw [->] (n.east) -- (v.south west);
	\draw [->] (o.east) -- (v.south west);
	
	\end{tikzpicture}
	\caption{Black-box MI with attack features $(y^*, p(\vec{x}))$. LDP perturbation on \subscr{\cali{D}}{train} (dotted) and CDP on target model and shadow model training (dashed). Data that was used by \cali{DO} during target model training is colored: training (violet) and validation (green).} 
	\label{fig:bb_mia}
\end{figure}
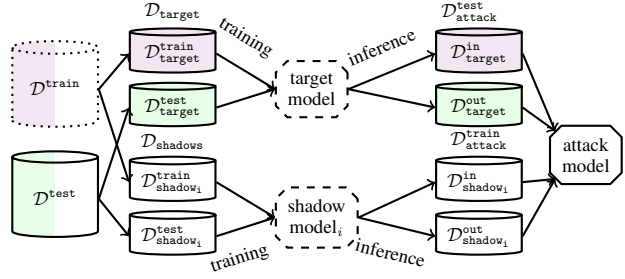

An illustration of the BB MI attack is given in Figure~\ref{fig:bb_mia}. 
The features for attack model training are generated by passing shadow model training and test data again through the trained shadow model. The attack model AUC is computed on features extracted from the target model in the same manner.

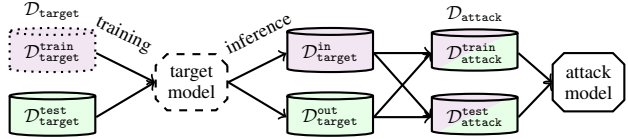
\begin{figure}
	\centering
	\begin{tikzpicture}[thick,scale=0.8, every node/.style={transform shape}]
	\footnotesize
	\node[] (aa) {};
	\node[draw,cylinder,shape border rotate=90, minimum width=4.5em, dotted, above= 0.25ex of aa, shape aspect=.10, fill=violet!10] (a) {\subsup{\cali{D}}{target}{train}};
	\node[minimum width=4em, above=0ex of a] (uu) {\subscr{\cali{D}}{target}};
	\node[draw,cylinder,shape border rotate=90, minimum width=4.5em, below= 0.25ex of aa, shape aspect=.10, fill=green!10] (b) {\subsup{\cali{D}}{target}{test}};
	\node[draw,cylinder,shape border rotate=90, minimum width=4.5em, right= 10em of a, shape aspect=.10, fill=violet!10] (c) {\subsup{\cali{D}}{target}{in}};
	\node[draw,cylinder,shape border rotate=90, minimum width=4.5em, right= 10em of b, shape aspect=.10, fill=green!10] (d) {\subsup{\cali{D}}{target}{out}};
	\node[draw,cylinder,shape border rotate=90, minimum width=4.5em, right= 3em of c, shape aspect=.10, double color fill={violet!10}{green!10}, shading angle=45] (e) {\subsup{\cali{D}}{attack}{train}};
	\node[draw,cylinder,shape border rotate=90, minimum width=4.5em, right= 3em of d, shape aspect=.10, double color fill={violet!10}{green!10}, shading angle=45] (f) {\subsup{\cali{D}}{attack}{test}};
	\node[minimum width=4em, above=0ex of e] (uu) {\subscr{\cali{D}}{attack}};
	\node[draw,chamfered rectangle, dashed, align=center, right=5em of aa] (z) {target \\ model};
	\node[draw,chamfered rectangle, align=center, right=17em of z] (y) {attack \\ model};

	\draw [->] (a.east) --  node [near start, above=1.5ex,sloped]{\small training}(z.west);
	\draw [->] (b.east) --  (z.west);
	
	\draw [->] (z.east) -- node [near end, above=2ex,sloped]{\small inference}(c.west);
	\draw [->] (z.east) -- (d.west);
	
	\draw [->] (c.east) -- (e.west);
	\draw [->] (c.east) -- (f.west);
	
	\draw [->] (d.east) -- (e.west);
	\draw [->] (d.east) -- (f.west);
	
	\draw [->] (e.east) -- (y.west);
	\draw [->] (f.east) -- (y.west);
	
	\end{tikzpicture}
	\caption{White-box MI with attack features $(y^*, p(\vec{x}),L(h(x;W),y), \frac{\delta L}{\delta W})$. LDP perturbation on \subsup{\cali{D}}{target}{train} (dotted) and CDP on target model training (dashed). Data that was used by \cali{DO} during target model training is colored: training (violet) and validation (green).} 
	\label{fig:wb_mia}
\end{figure}
\subsection{White-box MI attack}
\label{sec:prel:mi:wb}

The introduced notations for BB MI also hold for white-box (WB) MI. However, WB MI makes two additional assumptions about \cali{A}. First, \cali{A} is able to observe internal features of the ML model in addition to the external features (i.e., model outputs). The internal features comprise observed losses $L(h(x;W))$, gradients $\frac{\delta L}{\delta W}$ and the learned weights $W$ of $h(\cdot)$. Second, \cali{A} is aware of a portion of $\cali{D}_{target}^{train}$ and $\cali{D}_{target}^{test}$. These portions were set to $50\%$ by Nasr et al.~\cite{NSH18} and thus will be the same within this work.
The second assumption simplifies the process for training an attack model in comparison to BB MI since no shadow models are required to learn the external features of members and non-members. Instead, \cali{A} extracts internal and external features of a balanced set of confirmed members and non-members. An illustration of the WB MI attack is given in Figure~\ref{fig:wb_mia}. \cali{A} is assumed to know a portion of $\cali{D}_{target}^{train}$ and $\cali{D}_{target}^{test}$ and generates attack features by passing these records through the trained target model. The attack model accuracy is computed on features extracted from the target model in the same manner.

\subsection{Evaluating CDP and LDP under MI}
\label{sec:prel:mi:dp}
\cali{A} is pursuing a binary classification task: classifying a balanced set of records into \var{in} and \var{out} with high accuracy and thus identifying members and non-members of the training data. DP in contrast hides the presence or absence of any record in a dataset by perturbation. Thus, the use of DP should impact the classification accuracy of the MI accuracy of \cali{A}, since differences between members and non-members become obfuscated. To illustrate effect of changes in privacy parameter \eps~on the MI attack model we use the receiver operating characteristics (ROC). The ROC allows us to compare \cali{A}'s ability of correctly identifying members (i.e., sensitivity) and non-members (i.e., specificity) of the training data. Concretely, the ROC states the connection of the true positive rate (TPR; sensitivity) and the false positive rate (FPR; 1-specificity) for various classification threshold settings, and thus reflects the possible MI attack models of \cali{A}. We calculate and compare the ROC's area under curve (AUC) to asses the overall effect of DP on MI. Good MI attack models will realize an AUC of close to $1$ while poor MI attack models will be close to the baseline of uniform random guessing, i.e., $AUC = 0.5$.
The data owner \cali{DO} has two options to apply DP against MI within the data science process introduced in Section~\ref{sec:prel:dp}. Either in the form of LDP by applying a local randomizer on the training data and using the resulting $\cali{LR}(\subsup{\cali{D}}{target}{train})$ for training, or CDP with a differentially private optimizer on $\subsup{\cali{D}}{target}{train}$. A discussion and comparison of LDP and CDP based on the privacy parameter \eps~likely falls short and potentially leads data scientists to incorrect conclusions. 
Thus, data scientists give up flexibility w.r.t.~applicable learning algorithms if ruling out the use of LDP due to comparatively greater \eps~and instead solely investigating CDP (e.g., DP-SGD). This might even lead to a favorable privacy-accuracy trade-off. We suggest to compare LDP and CDP by their concrete effect on the MI AUC. While we consider two specific MI attack instances our methodology is applicable to other MI attacks as well. The use of LDP on the training data of the target model or CDP on the optimization step of the target model slightly changes the MI attack scheme described earlier in this section. For CDP the BB MI attack achieves the highest MI AUC when shadow models are trained similarly to the target model. This means that shadow models are assumed to use the same CDP optimizer and the same privacy parameters \epsdlt. We assume that shadow models use different seeds to initialize the randomness in comparison to the target model (cf.~Kerckhoff's principle). Models that use CDP are represented by dashed lines in Figure~\ref{fig:bb_mia} and~\ref{fig:wb_mia}. In the LDP setup, the target model is trained with perturbed records from a local randomizer, i.e., $\cali{LR}(\subsup{\cali{D}}{target}{train})$. However, in order to increase his attack accuracy \cali{A} needs to learn attack models with high accuracy on the original data from which the perturbed records stem, i.e., $\subsup{\cali{D}}{target}{train}$. Thus, \cali{A} will again train shadow models similar to the target model and use perturbed training data, but generates attack features for the attack model solely from the shadow models unperturbed member and non-member data. This avoids just learning the difference between two distributions (i.e., perturbed and unperturbed data). Instead the confidence difference of the trained shadow models on unperturbed member and non-member data is learned. Perturbation with LDP is represented by dotted lines in Figure~\ref{fig:bb_mia} and ~\ref{fig:wb_mia}. 
\subsection{Relative privacy-accuracy trade-off}
\label{sec:trade_off}
We calculate the relative privacy-accuracy trade-off for LDP and CDP as the relative difference between \cali{A}'s change in MI AUC to \cali{DO}'s change in test accuracy. Let $AUC_{orig}$, $AUC_{\eps}$ be the MI AUCs and $ACC_{orig}$, $ACC_{\eps}$ be the test accuracies for original and DP. Furthermore, let $ACC_{base}$ be the baseline test accuracy of uniform random guessing $1/\mathbb{C}$ and $AUC_{base}$ be the baseline MI AUC at $0.5$.
We fix $ACC_{base}$, $AUC_{base}$, since \cali{A} or \cali{DO} would perform worse than uniform random guessing at lower values:
\begin{equation*}
\varphi=\frac{(AUC_{orig}-AUC_{\epsilon})/(AUC_{orig} - AUC_{base})}{(ACC_{orig}-ACC_{\eps})/(ACC_{orig}-ACC_{base})}
\end{equation*}
Rearranging and bounding the cases where AUC and ACC increases over $\epsilon$ yields:
\begin{equation*}
\varphi=\frac{\max(0, AUC_{orig} - AUC_{\eps})\cdot(ACC_{orig}-ACC_{base})}{\max(0, ACC_{orig}-ACC_{\epsilon})\cdot(AUC_{orig}-AUC_{base})}
\end{equation*}
However, in the above definition $\varphi$ might approach infinitely large values when the accuracy remains stable while $AUC$ decreases significantly, and $\varphi$ might also be undefined for $ACC_{orig} \le ACC_{\eps} $. We bound $\varphi$ at 2 for these cases. In consequence, when the relative gain in privacy (lower MI AUC) exceeds the relative loss in accuracy, it applies that $1 < \varphi < 2$, and $0 < \varphi < 1$ when the loss in test accuracy exceeds the gain in privacy. 
\begin{equation*}
\min\left(2,\frac{\max(0, (AUC_{orig} - AUC_{\eps})\cdot(ACC_{orig}-ACC_{base}))}{\max(0, (ACC_{orig}-ACC_{\epsilon})\cdot(AUC_{orig}-AUC_{base}))}\right)
\end{equation*}

$\varphi$ quantifies the relative loss in accuracy and the relative gains in privacy for a given privacy parameter \eps. Hence, $\varphi$ captures the relative privacy-accuracy trade-off as a ratio which we seek to maximize.

%% file: src/datasets.tex
\section{Datasets and learning tasks}
\label{sec:data}
We use five datasets in our experiments. Each dataset is summarized in Table~\ref{tab:datasets}. The datasets have been used in state-of-the-art literature on MI, for graph classification and face recognition tasks. We found both MI attacks to be especially effective when some classes within the training data comprise a comparatively low number of records and the overall training data distribution is imbalanced. We use reference architectures involving convolutional neural networks (CNN) that exploit locality of features for the COLLAB and LFW datasets. 

\begin{table}[t]
\caption{Overview of datasets considered in evaluation.}
\label{tab:datasets}
\tiny
\centering
\begin{tabular}{|l|l|l|}
\hline
\multicolumn{1}{|c|}{\textbf{Dataset}} & \multicolumn{1}{c|}{\textbf{Model}} & \multicolumn{1}{c|}{\textbf{LDP}} \\ \hline
\begin{tabular}[c]{@{}l@{}}Texas Hospital \\Stays~\cite{shokri2017}\end{tabular} & \begin{tabular}[c]{@{}l@{}}Fully connected NN\\ with three layers \\ ($512\times128\times\mathbb{C}$)~\cite{shokri2017}.\end{tabular} &\begin{tabular}[c]{@{}l@{}} $19,125$ -- $638$ \\ ($6382 \times \eps_i$) \end{tabular} \\ \hline
\begin{tabular}[c]{@{}l@{}}Purchases Shopping \\ Carts~\cite{shokri2017}\end{tabular} & \begin{tabular}[c]{@{}l@{}}Fully connected NN\\ with two layers ($128\times\mathbb{C}$)~\cite{shokri2017}\\ (i.e., logistic regression).\end{tabular} & \begin{tabular}[c]{@{}l@{}} $1800$ -- $60$ \\ ($600 \times \eps_i$) \end{tabular} \\ \hline
COLLAB~\cite{Yanardag15} & \begin{tabular}[c]{@{}l@{}} Graph Convolutional \\ Network~\cite{Zhang18} \end{tabular} & \begin{tabular}[c]{@{}l@{}}$6,000$ -- $200$ \\ ($2000 \times \eps_i$)\end{tabular} \\ \hline
\begin{tabular}[c]{@{}l@{}}Labeled Faces \\ in the Wild~\cite{Huang07}\end{tabular}& \begin{tabular}[c]{@{}l@{}}VGG-Very-Deep-16 \\ CNN~\cite{Parkhi15}\end{tabular} & \begin{tabular}[c]{@{}l@{}}$62.5\times 10^6$ -- $6,250$ \\ ($250 \times 250 \times \eps_i$) \end{tabular} \\ \hline
\begin{tabular}[c]{@{}l@{}} Skewed \\ Purchases \end{tabular}& \begin{tabular}[c]{@{}l@{}}Fully connected NN\\ with two layers ($128\times\mathbb{C}$)~\cite{shokri2017}\\ (i.e., logistic regression).\end{tabular} & \begin{tabular}[c]{@{}l@{}} $1,800$ -- $60$ \\ ($600 \times eps_i$)\end{tabular} \\ \hline
\end{tabular}
\end{table}

\subsection{Texas Hospital Stays} The Texas Hospital Stays dataset~\cite{shokri2017} is an unbalanced dataset (i.e., varying amounts of records per label) and consists of high dimensional binary vectors representing patient health features. Each record within the dataset is labeled with a procedure. The learning task is to train a fully connected neural network for classification of patient features to a procedure and we do not try to re-identify a known individual, and fully comply with the data use agreement for the original public use data file. We train and evaluate models for a set of most common procedures $\mathbb{C}\in\{100,150,200,300\}$. Depending on the number of procedures the dataset comprises $67,330$ -- $89,815$ records and $6,170$ -- $6,382$ features. To allow comparison between BB MI, WB MI and related work, we train and test the target model on $n=10,000$ records respectively. The remaining set is used for shadow model training and test under BB MI.

\subsection{Purchases Shopping Carts} This dataset~\cite{shokri2017} is also unbalanced and consists of binary vectors which represent customer shopping carts. However, a significant difference to the Texas Hospital Stays dataset is that the dimensionality (i.e., number of features) is almost $90\%$ lower. Each vector is labeled with a customer group. The learning task is to classify shopping carts to customer groups. The dataset is provided in four variations with varying numbers of labels: $\mathbb{C}\in\{10,20,50,100\}$. The dataset comprises $38,551$ -- $197,324$ records. We sample $n=8,000$ records each for training and testing the target model and leave the rest for shadow model training under BB MI. This methodology ensures comparability between BB MI, WB MI and related work.

\subsection{COLLAB} The COLLAB dataset~\cite{Yanardag15} consists of undirected social graphs stored as adjacency matrix. Each graph represents the collaboration network of a researcher from one of three different fields, namely \textit{high energy}, \textit{condensed matter} or \textit{astrophysics}. The learning problem is to predict the correct research field per researcher. We implement the graph classification architecture suggested by Zhang et al.~\cite{Zhang18} which consists of a Graph Convolution Network (GCN) for feature extraction and a final fully connected classifier. In between these two components, a sort-pooling layer is added to support classification of arbitrary graph sizes which we require to classify collaboration networks of different sizes from the COLLAB dataset. However, we omit the final dropout layer. Their model requires an adjacency matrix, a degree vector and a node feature matrix as inputs per graph. 
The complete dataset contains $5,000$ entries with no separation into train and test splits. We choose to split all records into sets of $2,500$ training instances and $2,500$ testing examples. The maximum edge degree is $2,000$. For LDP we randomize all entries in the adjacency matrices with randomized response. This LDP approach achieves edge-local differential privacy, the plausible deniability of any connection within a graph. Randomized adjacency matrices are post-processed by modifying it such that every node has at least one symmetric edge.

\subsection{Labeled Faces in the Wild} The Labeled Faces in the Wild (LFW)~\cite{Huang07} dataset contains labeled images each depicting a specific person with a resolution of $250\times250$ pixels. The dataset has a long distribution tail w.r.t.~to the number of images per label. We thus focus on learning the topmost classes $\mathbb{C}\in\{20, 50, 100\}$ with $1906$, $2773$ and $3651$ overall records respectively and train the target model on $n \in \{368, 544, 720\}$ due to the high amount of data required for shadow model training. Additionally, we transform all images to grayscale. We start our comparison of LDP and CDP from a pre-trained VGG-Very-Deep-16 CNN faces model~\cite{Parkhi15} by keeping the convolutional core, exchanging the dense layer at the end of the model and training for LFW grayscale faces. For LDP, we apply differentially private image pixelation within neighborhood $m=\sqrt{250\times250}$ and avoid coarsening by setting $b=1$. The pre-trained VGG is then trained on LDP images from the training dataset. For CDP we use the DP-Adam optimizer.

\subsection{Skewed Purchases} We specifically crafted this dataset
to mimic a situation for transfer learning, i.e., the application of a trained model to novel data which is similar to the training data w.r.t.~format but following a different distribution. This situation arises for Purchases Shopping Carts, if for example not enough high-quality shopping cart data for a specific retailer are available yet. 
Thus, only few high-quality data (e.g., manually crafted examples) can be used for testing and large amounts of low quality data from potentially different distributions for training (e.g., from other retailers). In effect the distribution between train and test data varies for this dataset. Similar to Purchases Shopping Carts the dataset consists of $200,000$ records with $600$ features and is available in four versions with $\mathbb{C}\in\{10,20,50,100\}$ labels. However, each vector $x$ in the training dataset $X$ is generated by using two independent random coins to sample a value from $\{0, 1\}$ per position $i=1,\ldots,600$. The first coin steers the probability $\Pr[x_i=1]$ for a fraction of $600$ positions per $x$. We refer to these positions as indicator bits (\textit{ind}) which indicate products frequently purchased together. The second coin steers the probability $\Pr[x_i=1]$ for a fraction of $600-(\frac{600}{|\mathbb{C}|})$ positions per $x$. We refer to these positions as noise bits (\textit{noise}) that introduce scatter in addition to \textit{ind}. We let $\Pr_{ind}[x_i = 1] = 0.8 \land \Pr_{noise}[x_i=1]= 0.2$, $\forall x \in X_{train}$ and $\Pr_{ind}[x_i = 1]=0.8 \land Pr_{noise}[x_i = 1] = 0.5, \land x \in X_{test}, 1 \le i \le 600$. This dataset has a difference in information entropy between test between test and train data of $\approx0.3$. The difference is $\approx0$ if there is no skew. 

%% file: src/experiments.tex
\section{Experiments}
\label{sec:experiments}
We perform an experiment per dataset which measures MI AUC for LDP and CDP for privacy comparison instead of privacy parameter \eps. The results of each experiment are visualized by three sets of figures. 
First, by plotting test accuracy and MI AUC over \eps~for BB and WB MI and $\varphi$ over \eps~for WB MI.
We present this information for CDP per dataset in Figures \ref{fig:eval:texas:figures}--\ref{fig:eval:purch_skew:figures} a -- d and for LDP in Figures \ref{fig:eval:texas:figures}--\ref{fig:eval:purch_skew:figures} e -- h. 
This information serves as basis to identify privacy parameters at which the MI AUC is converging towards the MI AUC baseline.
We provide an overview of the train accuracy of all trained models in Table~\ref{tab:comp:train_acc}. Second, we state the ROC curves from which MI AUC was calculated to illustrate the slope with which TPR and FPR are diverging from the baseline for LDP and CDP under WB and BB MI in Figures \ref{fig:eval:texas:figures}--\ref{fig:eval:purch_skew:figures} i,j. Third, we depict the test accuracy over MI AUC and \eps~for both LDP and CDP in a scatterplot to compare the absolute privacy-accuracy trade-offs per dataset. We present this information for WB and BB MI in Figures \ref{fig:eval:texas:figures}--\ref{fig:eval:purch_skew:figures}k. For each experiment model training stops once the test data loss is stagnating (i.e., early stopping) or a maximum number of epochs is reached. 
For all executions of the experiment CDP noise is sampled from a Gaussian distribution (cf.~Definition~\ref{def:dp:gauss}) with $\sgm=\textit{noise multiplier}~z \times \textit{clipping norm}~\cali{C}$. We evaluate increasing noise regimes per dataset by evaluating noise multipliers $z\in\{2,4,6,8,16\}$ and calculate the resulting $\eps$~at a fixed $\dlt=\frac{1}{n}$. However, since batch size, dataset size and number of epochs are also influencing the Renyi differential privacy accounting a fixed $z$ will result in different \eps~for different datasets. We provide the resulting \eps~for all datasets and $z$ in Table~\ref{tab:comp:train_acc}. For LDP we use the same hyperparameters as in the original training and evaluate two local randomizers, namely randomized response and LDP image pixelation with the Laplace mechanism. 
For each randomizer we state the individual $\eps_i$~per invocation (i.e., per anonymized value) and \eps~per record (i.e., collection of dependent values). 
We apply randomized response to all datasets except LFW with a range of privacy parameter values $\eps_i\in\{0.1,0.5,1,2,3\}$ that reflect retention probabilities $\rho$ from 5\% -- 90\% (cf.~Equation~\eqref{eq:rr}). 
For LFW we perturb each pixel with Laplace noise, and also investigate a wide range of resulting noise regimes by varying $\eps_i$.
For ease of reading we provide the composed \eps~for LDP per dataset in Table~\ref{tab:datasets}. We provide hyper-parameters for all experiments in Table~\ref{tab:comp:train_acc}. The experiment was repeated five times per dataset and we report mean values with error bars unless otherwise stated. ROC curves depict the mean linear interpolation of the TPR over the FPR.

\subsection{Texas Hospital Stays}\label{sec:experiments:texas}\input{src/experiments-th.tex}
\subsection{Purchases Shopping Carts}\label{sec:experiments:purch}\input{src/experiments-purch.tex}
\subsection{COLLAB}\label{sec:experiments:gcn}\input{src/experiments-gcn.tex}
\subsection{LFW}\label{sec:experiments:lfw}\input{src/experiments-lfw.tex}
\subsection{Skewed Purchases}\label{sec:experiments:purch_skew}\input{src/experiments-purch_skew.tex}

\begin{table*}[t]
\centering
\caption{Target Model training accuracy (from orig. to smallest \eps), CDP \eps~values (from $z = 0.5$ to $z=16$) and hyperparameters}
\label{tab:comp:train_acc}
\tiny
\begin{tabular}{|c|l|c|c|c|c|c|c|c|c|c|c|c|c|c|c|c|c|}
\hline
\multicolumn{2}{|c|}{} & \multicolumn{4}{c|}{Texas Hospital Stays} & \multicolumn{4}{c|}{Purchases Shopping Carts} & GCN & \multicolumn{3}{c|}{LFW} & \multicolumn{4}{c|}{Skewed Purchases} \\ \hline
\multicolumn{2}{|c|}{$\mathbb{C}$} & 100 & 150 & 200 & 300 & 10 & 20 & 50 & 100 & 3 & 20 & 50 & 100 & 10 & 20 & 50 & 100 \\ \hline
\multicolumn{2}{|c|}{LDP} & \begin{tabular}[c]{@{}c@{}}$0.86$\\ $1.0$\\ $1.0$\\ $1.0$\\ $0.99$\\ $0.82$\end{tabular} & \begin{tabular}[c]{@{}c@{}}$0.92$\\ $1.0$\\ $1.0$\\ $1.0$\\ $0.95$\\ $0.71$\end{tabular} & \begin{tabular}[c]{@{}c@{}}$0.83$\\ $1.0$\\ $1.0$\\ $0.98$\\ $0.86$\\ $0.59$\end{tabular} & \begin{tabular}[c]{@{}c@{}}$0.81$\\ $1.0$\\ $1.0$\\ $0.92$\\ $0.72$\\ $0.53$\end{tabular} & \begin{tabular}[c]{@{}c@{}}$0.99$\\ $0.97$\\ $0.88$\\ $0.64$\\ $0.58$\\ $0.44$\end{tabular} & \begin{tabular}[c]{@{}c@{}}$1.0$\\ $0.97$\\ $0.85$\\ $0.58$\\ $0.47$\\ $0.38$\end{tabular} & \begin{tabular}[c]{@{}c@{}}$1.0$\\ $0.95$\\ $0.86$\\ $0.69$\\ $0.62$\\ $0.49$\end{tabular} & \begin{tabular}[c]{@{}c@{}}$0.99$\\ $0.94$\\ $0.90$\\ $0.79$\\ $0.75$\\ $0.51$\end{tabular} & \begin{tabular}[c]{@{}c@{}}$0.74$\\ $0.72$\\ $0.72$\\ $0.61$\\ $0.59$\\ $0.59$\end{tabular} & \begin{tabular}[c]{@{}c@{}}$1.0$\\ $1.0$\\ $1.0$\\ $0.22$\\ $0.24$\\ $0.25$\end{tabular} & \begin{tabular}[c]{@{}c@{}}$1.0$\\ $1.0$\\ $0.96$\\ $0.18$\\ $0.17$\\ $0.17$\end{tabular} & \begin{tabular}[c]{@{}c@{}}$1.0$\\ $1.0$\\ $1.0$\\ $0.13$\\ $0.13$\\ $0.13$\end{tabular} & \begin{tabular}[c]{@{}c@{}}$1.0$\\ $1.0$\\ $1.0$\\ $1.0$\\ $0.93$\\ $0.52$\end{tabular} & \begin{tabular}[c]{@{}c@{}}$1.0$\\ $1.0$\\ $1.0$\\ $0.99$\\ $0.98$\\ $0.55$\end{tabular} & \begin{tabular}[c]{@{}c@{}}$1.0$\\ $1.0$\\ $1.0$\\ $0.97$\\ $0.9$\\ $0.71$\end{tabular} & \begin{tabular}[c]{@{}c@{}}$1.0$\\ $0.99$\\ $0.97$\\ $0.89$\\ $0.80$\\ $0.45$\end{tabular} \\ \hline
\multicolumn{2}{|c|}{CDP} & \begin{tabular}[c]{@{}c@{}}$0.86$\\ $0.74$\\ $0.57$\\ $0.35$\\ $0.22$\\ $0.05$\end{tabular} & \begin{tabular}[c]{@{}c@{}}$0.92$\\ $0.75$\\ $0.54$\\ $0.31$\\ $0.19$\\ $0.04$\end{tabular} & \begin{tabular}[c]{@{}c@{}}$0.83$\\ $0.69$\\ $0.48$\\ $0.26$\\ $0.16$\\ $0.03$\end{tabular} & \begin{tabular}[c]{@{}c@{}}$0.81$\\ $0.62$\\ $0.42$\\ $0.22$\\ $0.13$\\ $0.02$\end{tabular} & \begin{tabular}[c]{@{}c@{}}$1.0$\\ $0.95$\\ $0.91$\\ $0.80$\\ $0.69$\\ $0.28$\end{tabular} & \begin{tabular}[c]{@{}c@{}}$1.0$\\ $0.91$\\ $0.84$\\ $0.69$\\ $0.51$\\ $0.14$\end{tabular} & \begin{tabular}[c]{@{}c@{}}$1.0$\\ $0.82$\\ $0.71$\\ $0.46$\\ $0.28$\\ $0.05$\end{tabular} & \begin{tabular}[c]{@{}c@{}}$0.99$\\ $0.63$\\ $0.51$\\ $0.27$\\ $0.14$\\ $0.02$\end{tabular} & \begin{tabular}[c]{@{}c@{}} $0.74$\\ $0.67$\\ $0.59$\\ $0.56$\\ $0.54$\\ $0.52$\end{tabular} & \begin{tabular}[c]{@{}c@{}}$1.0$\\ $0.99$\\ $0.76$\\ $0.44$\\ $0.36$\\ $0.32$\end{tabular} & \begin{tabular}[c]{@{}c@{}}$1.0$\\ $0.87$\\ $0.5$\\ $0.28$\\ $0.23$\\ $0.19$\end{tabular} & \begin{tabular}[c]{@{}c@{}}$1.0$\\ $0.79$\\ $0.35$\\ $0.25$\\ $0.18$\\ $0.13$\end{tabular} & \begin{tabular}[c]{@{}c@{}}$1.0$\\ $1.0$\\ $1.0$\\ $0.92$\\ $0.89$\\ $0.66$\end{tabular} & \begin{tabular}[c]{@{}c@{}}$1.0$\\ $1.0$\\ $0.96$\\ $0.8$\\ $0.64$\\ $0.24$\end{tabular} & \begin{tabular}[c]{@{}c@{}}$1.0$\\ $0.97$\\ $0.6$\\ $0.25$\\ $0.12$\\ $0.03$\end{tabular} & \begin{tabular}[c]{@{}c@{}}$1.0$\\ $0.58$\\ $0.1$\\ $0.02$\\ $0.02$\\ $0.01$\end{tabular} \\ \hline
\multicolumn{2}{|c|}{$\epsilon$}& 
    \begin{tabular}[c]{@{}c@{}}$222.6$\\ $6.3$\\ $2.3$\\ $0.9$\\ $0.3$\\\end{tabular} & 
    \begin{tabular}[c]{@{}c@{}}$259.8$\\ $6.6$\\ $2.0$\\ $1.1$\\ $0.2$\\\end{tabular} & 
    \begin{tabular}[c]{@{}c@{}}$251.5$\\ $7.3$\\ $2.2$\\ $1.0$\\ $0.3$\\\end{tabular} & 
    \begin{tabular}[c]{@{}c@{}}$259.8$\\ $7.4$\\ $2.1$\\ $1.1$\\ $0.3$\\
\end{tabular} & 
    
    \begin{tabular}[c]{@{}c@{}}$88.1$\\ $4.6$\\ $2.0$\\ $1.3$\\ $0.4$\\\end{tabular} & 
    \begin{tabular}[c]{@{}c@{}}$88.1$\\ $4.1$\\ $1.8$\\ $1.2$\\ $0.4$\\\end{tabular} & 
    \begin{tabular}[c]{@{}c@{}}$88.1$\\ $4.1$\\ $1.8$\\ $1.2$\\ $0.4$\\\end{tabular} & 
    \begin{tabular}[c]{@{}c@{}}$88.1$\\ $4.1$\\ $1.8$\\ $1.2$\\ $0.3$\\\end{tabular} & 
    
    \begin{tabular}[c]{@{}c@{}}$98.4$\\ $6.1$\\ $2.6$\\ $1.7$\\ $0.6$\\\end{tabular} & 
    
    \begin{tabular}[c]{@{}c@{}}$84.3$\\ $4.8$\\ $2.1$\\ $1.3$\\ $0.5$\\\end{tabular} & 
    \begin{tabular}[c]{@{}l@{}}$70.4$\\ $3.9$\\ $1.7$\\ $0.8$\\ $0.4$\\\end{tabular}
    & \begin{tabular}[c]{@{}c@{}}$62.4$\\ $3.4$\\ $1.5$\\ $1.0$\\ $0.3$\\\end{tabular} 
    
    & \begin{tabular}[c]{@{}c@{}}$28.9$\\ $1.6$\\ $0.7$\\ $0.9$\\ $0.4$\\\end{tabular} 
    & \begin{tabular}[c]{@{}c@{}}$29.8$\\ $1.7$\\ $1.6$\\ $0.9$\\ $0.4$\\\end{tabular} 
    & \begin{tabular}[c]{@{}c@{}}$42.2$\\ $3.5$\\ $1.3$\\ $0.7$\\ $0.3$\\\end{tabular} 
    & \begin{tabular}[c]{@{}c@{}}$73.5$\\ $2.1$\\ $1.3$\\ $0.6$\\ $0.3$\\\end{tabular} \\ \hline
    
	\multirow{3}{*}{\begin{tabular}[c]{@{}c@{}}Learning\\rate \end{tabular}} & Orig. & 0.01& 0.01& 0.01& 0.01& 0.001 & 0.001& 0.001& 0.001& 0.0001 & 0.001 &  0.001 & 0.001 &  0.001 &  0.001 &  0.001 & 0.001\\ \cline{2-18} 
	& CDP & 0.01& 0.01& 0.01& 0.01& 0.01& 0.01& 0.01& 0.01& 0.001 & \begin{tabular}[c]{@{}c@{}} 0.001 \end{tabular} & \begin{tabular}[c]{@{}c@{}}0.0008 \end{tabular} & \begin{tabular}[c]{@{}c@{}} 0.0008 \end{tabular} & 0.001 & 0.001 & 0.001 & 0.001\\ \cline{2-18} 
	& LDP & 0.01& 0.01 & 0.01 & 0.01 & 0.001 & 0.001 & 0.001 & 0.001 & 0.0001 & 0.001 & 0.001 & 0.001 & 0.001 & 0.001 & 0.001 & 0.001\\ \hline
	\multirow{3}{*}{\begin{tabular}[c]{@{}c@{}}Batch \\ size \end{tabular}} & Orig. & 128 & 128 & 128 & 128 & 128 & 128 & 128 & 128 & 50 & 32 & 32 & 32 & 100 & 100 & 100 & 100\\ \cline{2-17} 
	& CDP & 128& 128& 128& 128& 128& 128& 128& 128& 50& 16 & 16 & 16 & 100& 100& 100& 100\\ \cline{2-18} 
	& LDP & 128& 128& 128& 128& 128& 128& 128& 128& 50& 32 & 32 & 32 & 100& 100& 100& 100\\ \hline
	\multirow{3}{*}{Epochs} & Orig. & 200& 200& 200& 200& 200& 200& 200& 200& 300& 30 & 30 & 30 & 200& 200& 200& 200\\ \cline{2-18} 
	& CDP & 1000& 1000& 1000& 1000& 200& 200& 200& 200& 100& 110 & 110 & 110 & 200& 200& 200& 200\\ \cline{2-18} 
	& LDP & 200& 200& 200& 200& 200& 200& 200& 200& 300& 30 & 30 & 30 & 200 & 200& 200& 200\\ \hline
	\begin{tabular}[c]{@{}c@{}}Clipping \\ Norm\end{tabular} & CDP & 4 & 4 & 4 & 4 & 4 & 4 & 4 & 4 & 3 & 3 & 3 & 3 & 4 & 4 & 4 & 4 \\ \hline
\end{tabular}
\end{table*}

%% file: src/experiments-th.tex
\begin{figure*}[t!]
	\centering
	\begin{subfigure}{0.25\linewidth}
		\includegraphics[width=1\linewidth]{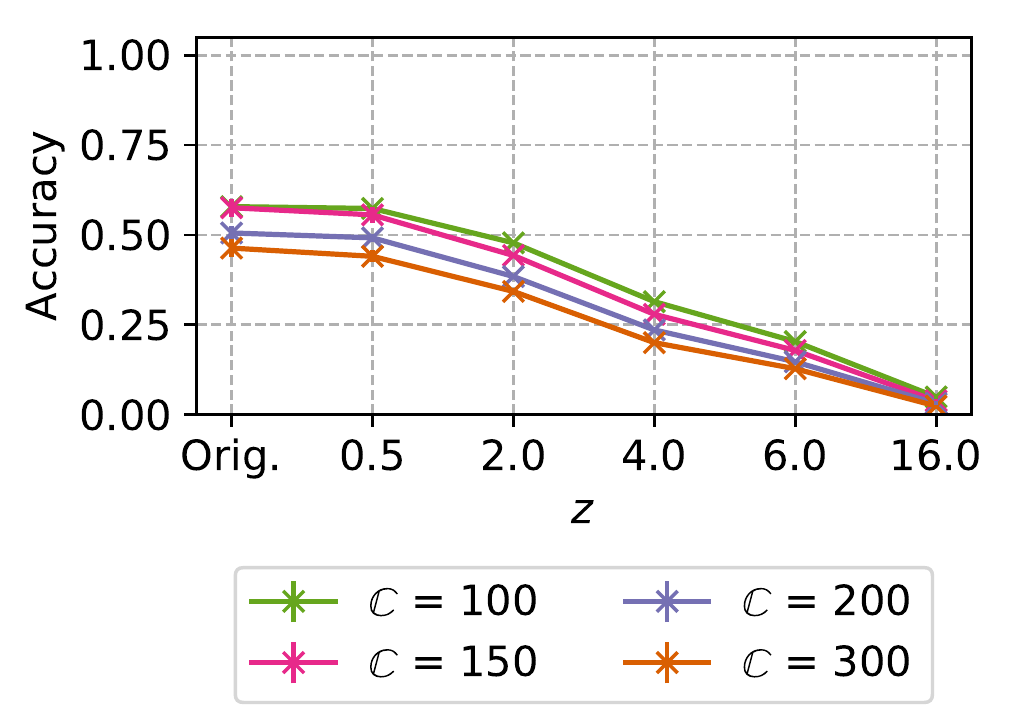}
		\caption{Target model accuracy (CDP)}
		\label{fig:eval:texas_cdp_acc}
	\end{subfigure}%
	\begin{subfigure}{0.25\linewidth}
		\includegraphics[width=1\linewidth]{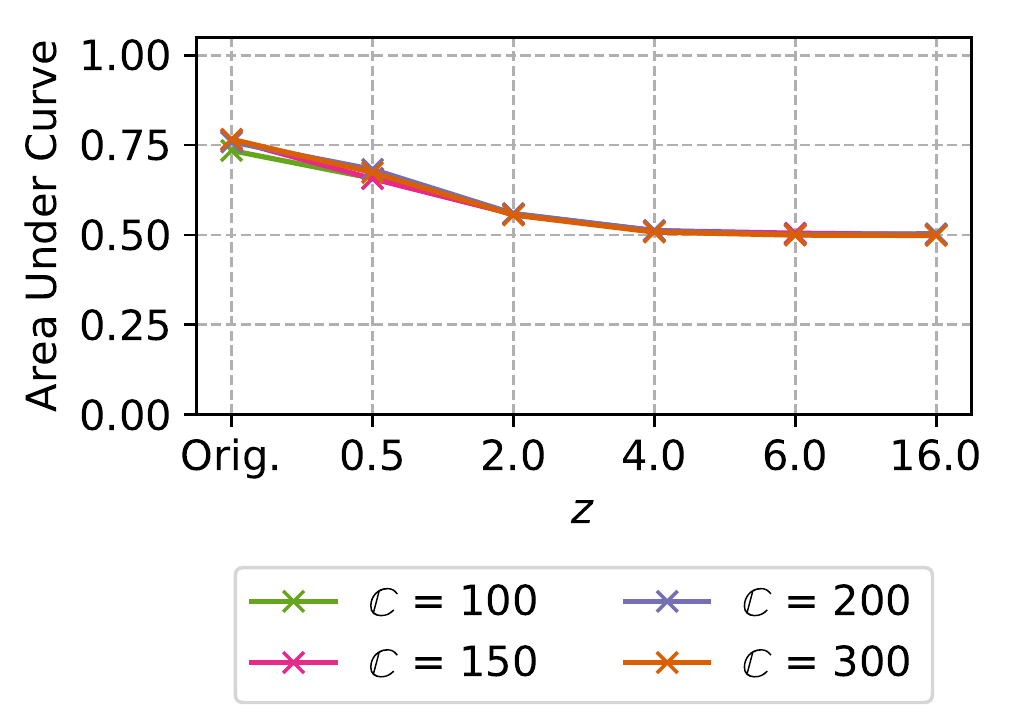}
		\caption{Black-box AUC (CDP)}
		\label{fig:eval:texas_cdp_auc_bb}
	\end{subfigure}%
	\begin{subfigure}{0.25\linewidth}
		\includegraphics[width=1\linewidth]{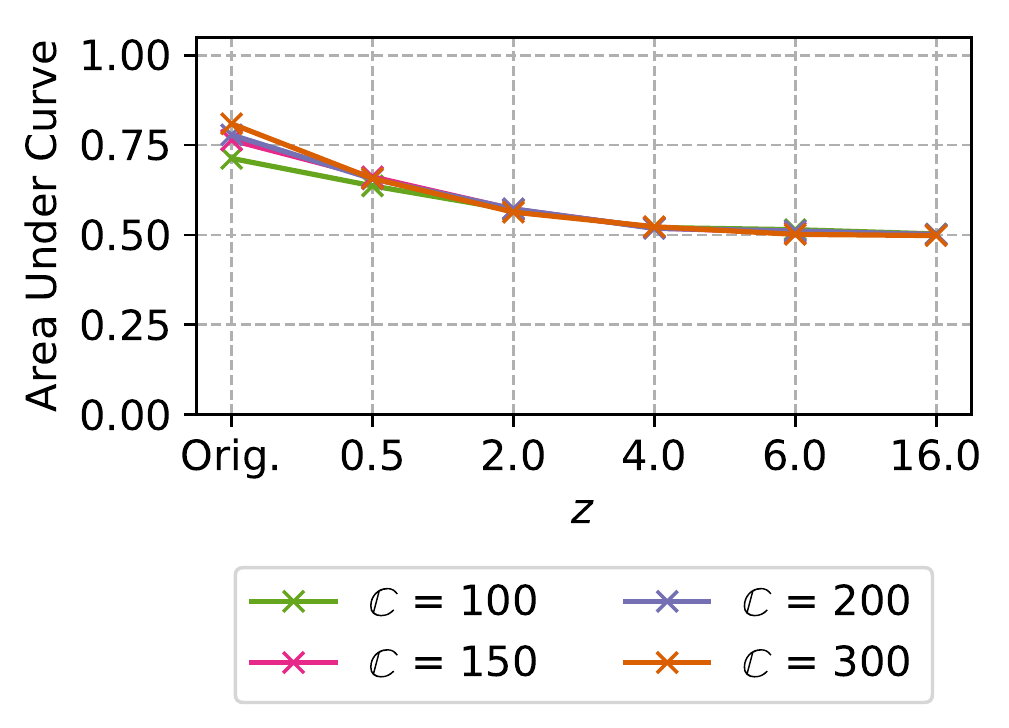}
		\caption{White-box AUC (CDP)}
		\label{fig:eval:texas_cdp_auc_wb}
	\end{subfigure}%
	\begin{subfigure}{0.25\linewidth}
		\includegraphics[width=1\linewidth]{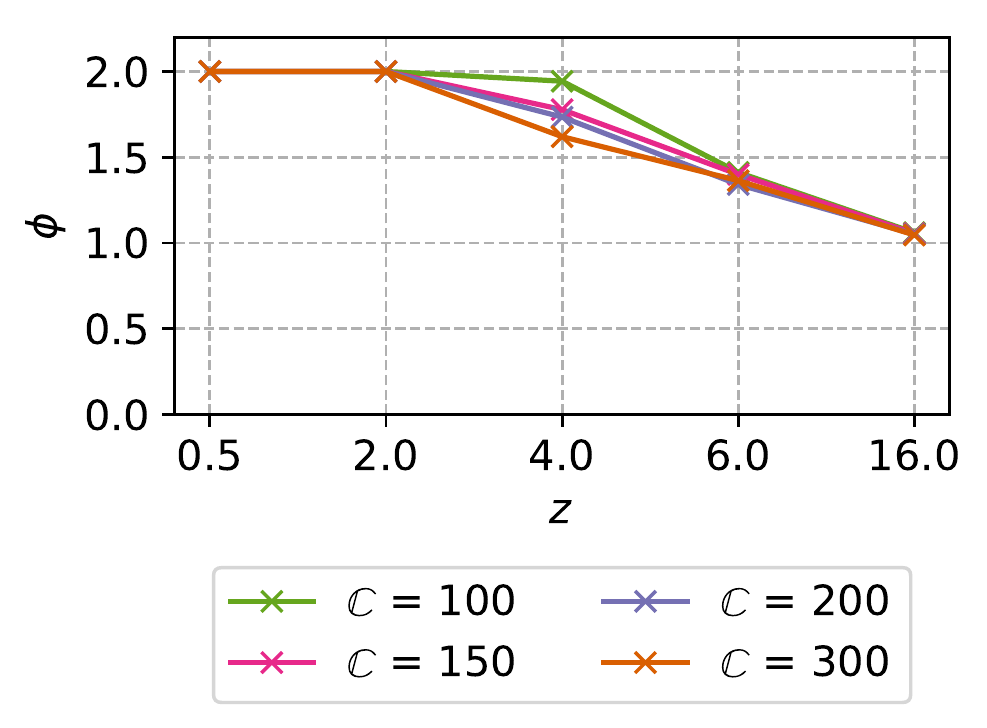}
		\caption{White-box $\varphi$ (CDP) }
		\label{fig:eval:texas_cdp_phi_wb}
	\end{subfigure}\\
	\begin{subfigure}{0.25\linewidth}
		\includegraphics[width=1\linewidth]{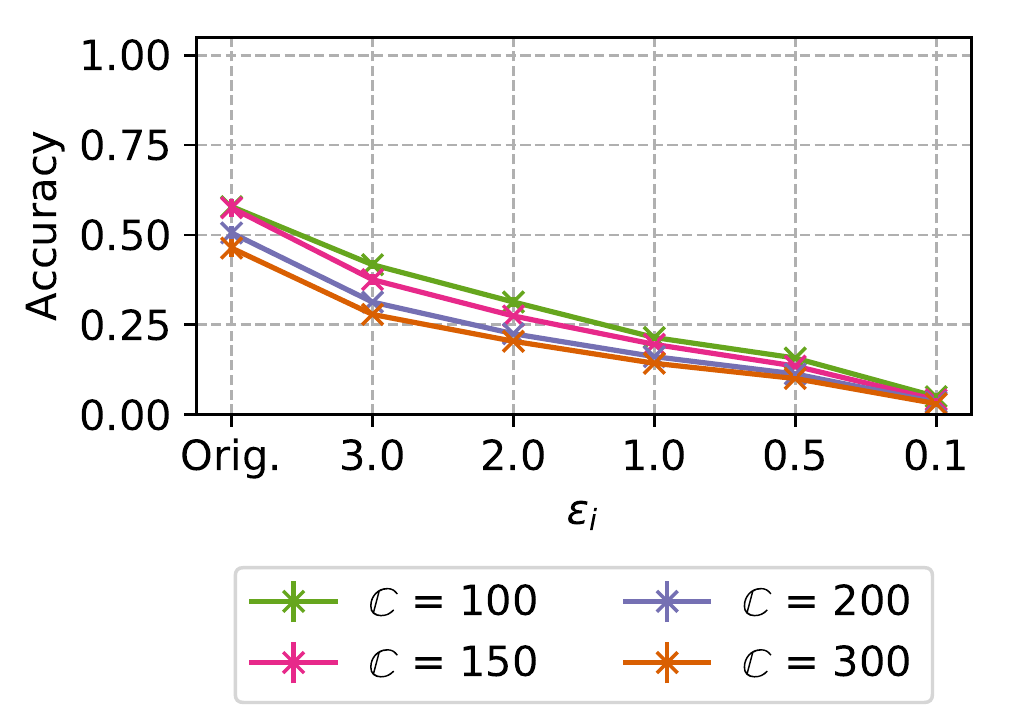}
		\caption{Target model accuracy (LDP)}
		\label{fig:eval:texas_ldp_acc}
	\end{subfigure}%
	\begin{subfigure}{0.25\linewidth}
		\includegraphics[width=1\linewidth]{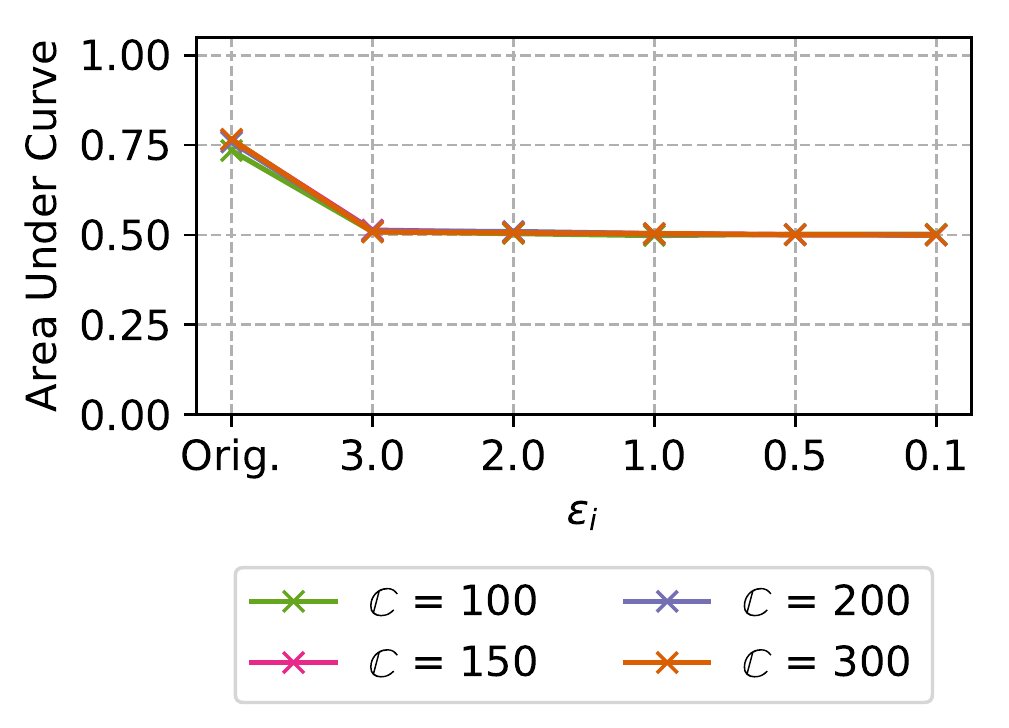}
		\caption{Black-box AUC (LDP)}
		\label{fig:eval:texas_ldp_auc_bb}
	\end{subfigure}%
	\begin{subfigure}{0.25\linewidth}
		\includegraphics[width=1\linewidth]{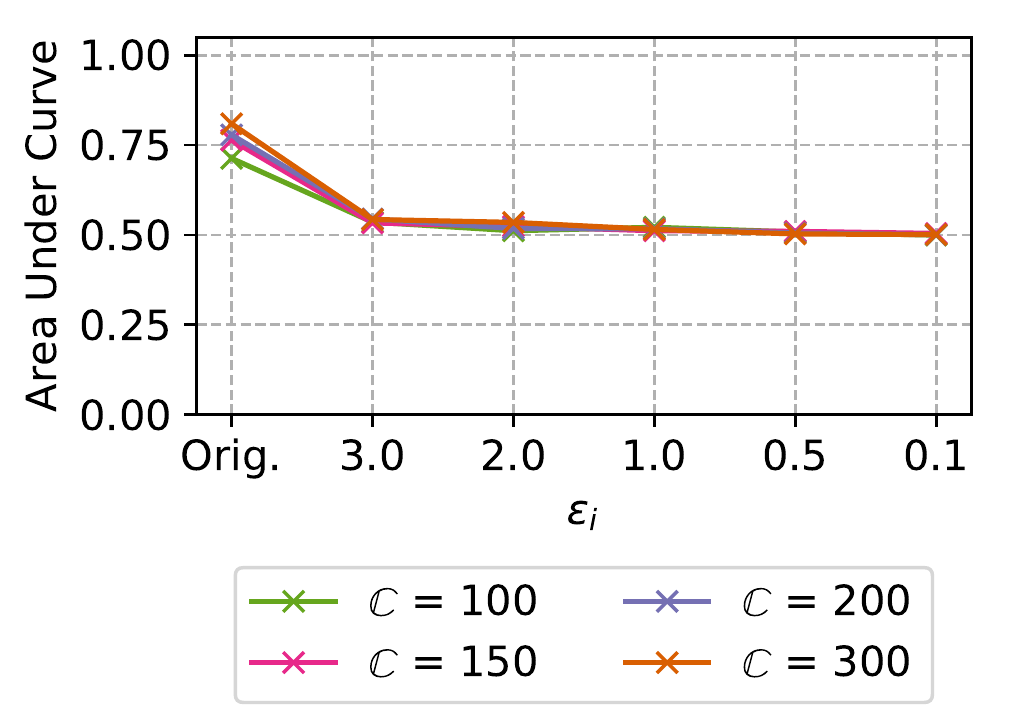}
		\caption{White-box AUC (LDP)}
		\label{fig:eval:texas_ldp_auc_wb}
	\end{subfigure}%
	\begin{subfigure}{0.25\linewidth}
		\includegraphics[width=1\linewidth]{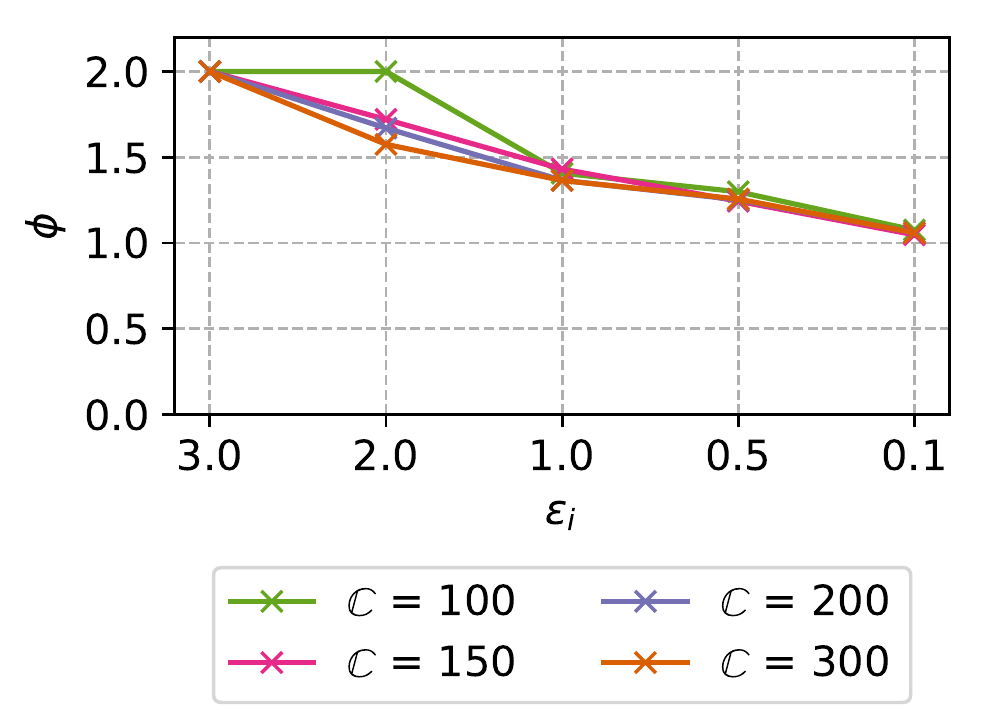}
		\caption{White-box $\varphi$ (LDP)}
		\label{fig:eval:texas_ldp_phi_wb}
	\end{subfigure}\\
	\begin{subfigure}{0.45\linewidth}
		\includegraphics[width=1\linewidth]{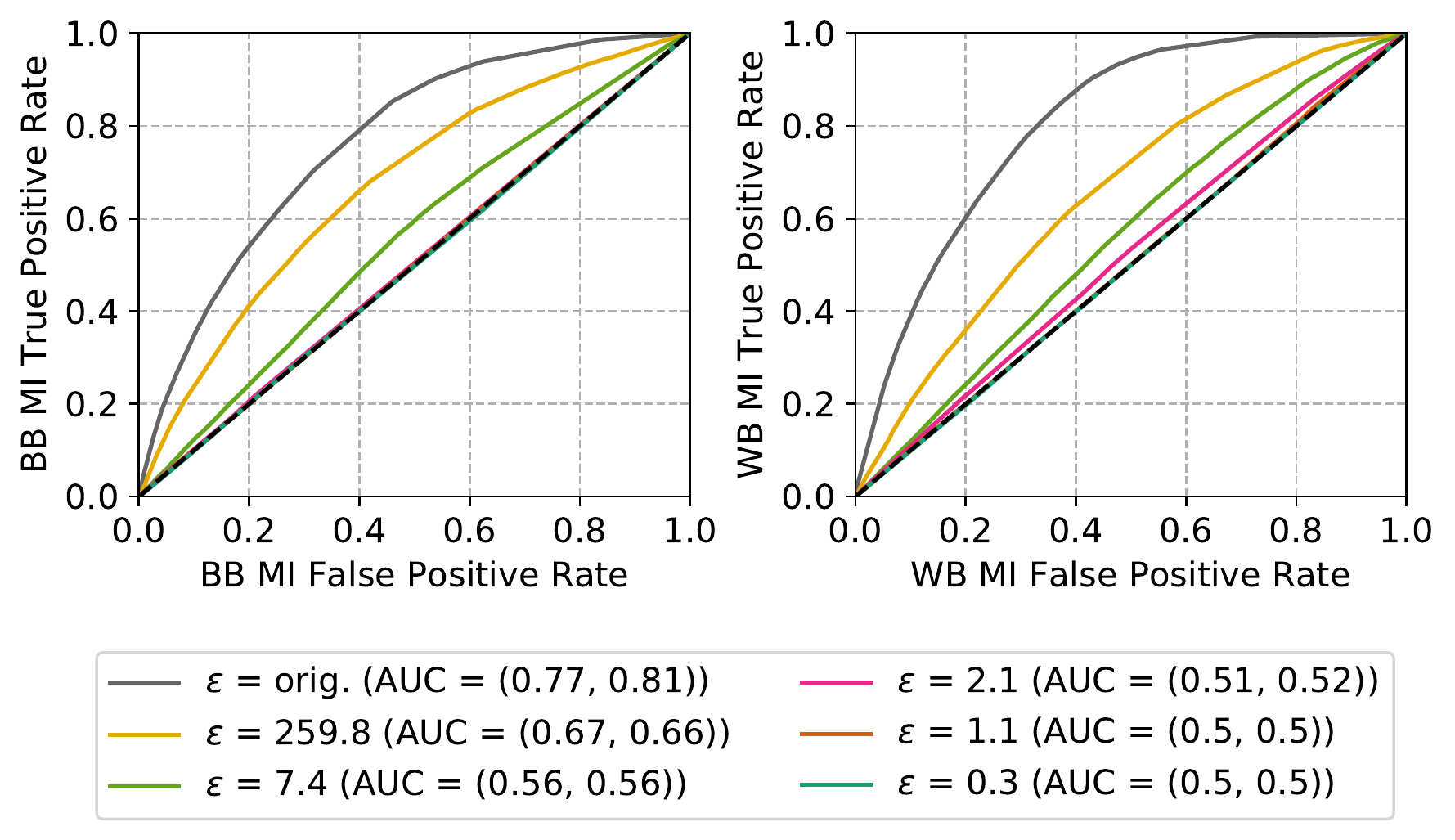}
		\caption{Receiver operating curve for $\mathbb{C}=300$ (CDP)}
		\label{fig:eval:texas_unified_roc_cdp}	
	\end{subfigure}%
		\begin{subfigure}{0.45\linewidth}
		\includegraphics[width=1\linewidth]{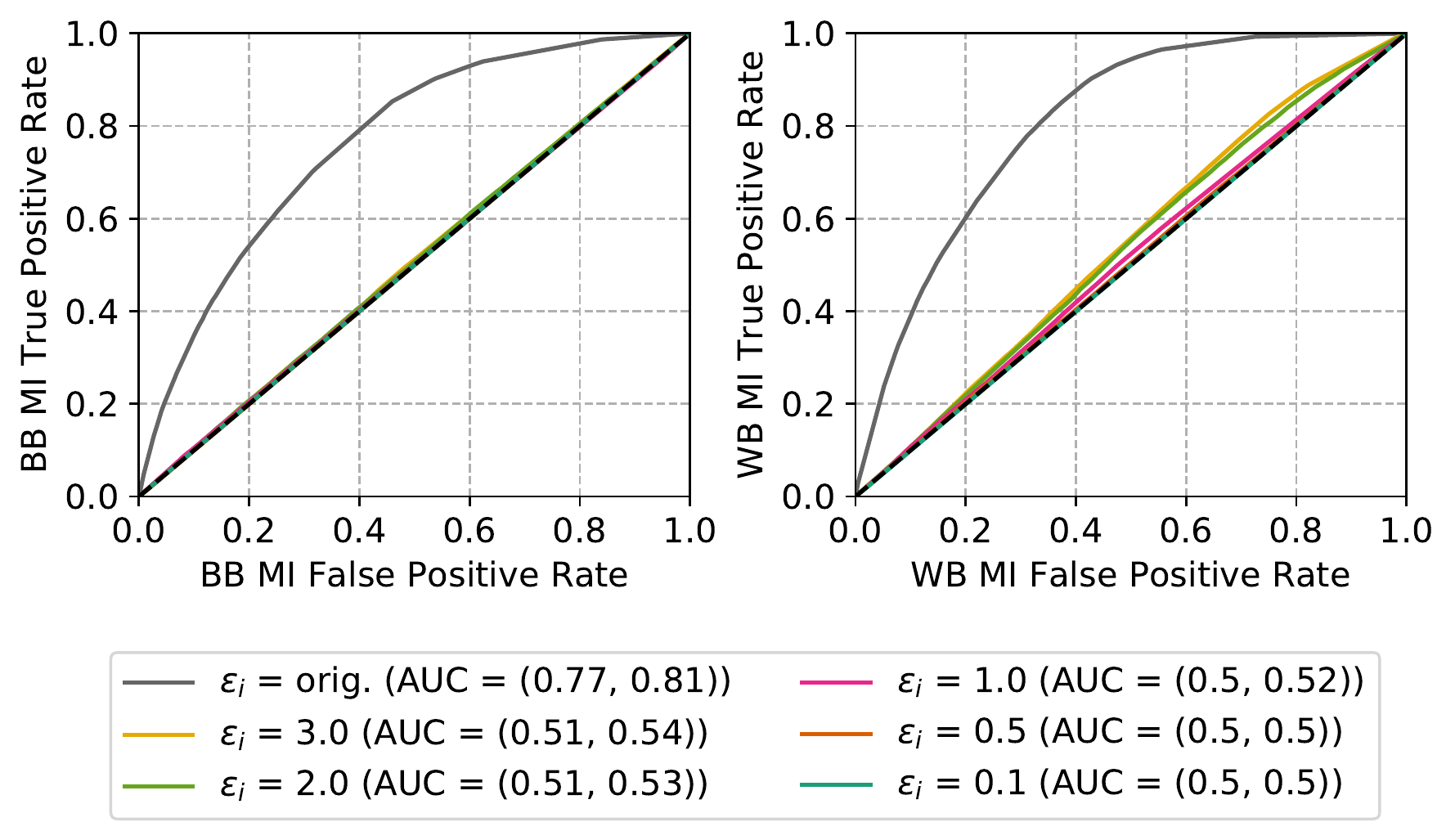}
		\caption{Receiver operating curve for $\mathbb{C}=300$ (LDP)}
		\label{fig:eval:texas_unified_roc_ldp}	
	\end{subfigure}\\
	\begin{subfigure}{0.6\linewidth}
		\includegraphics[width=1\linewidth]{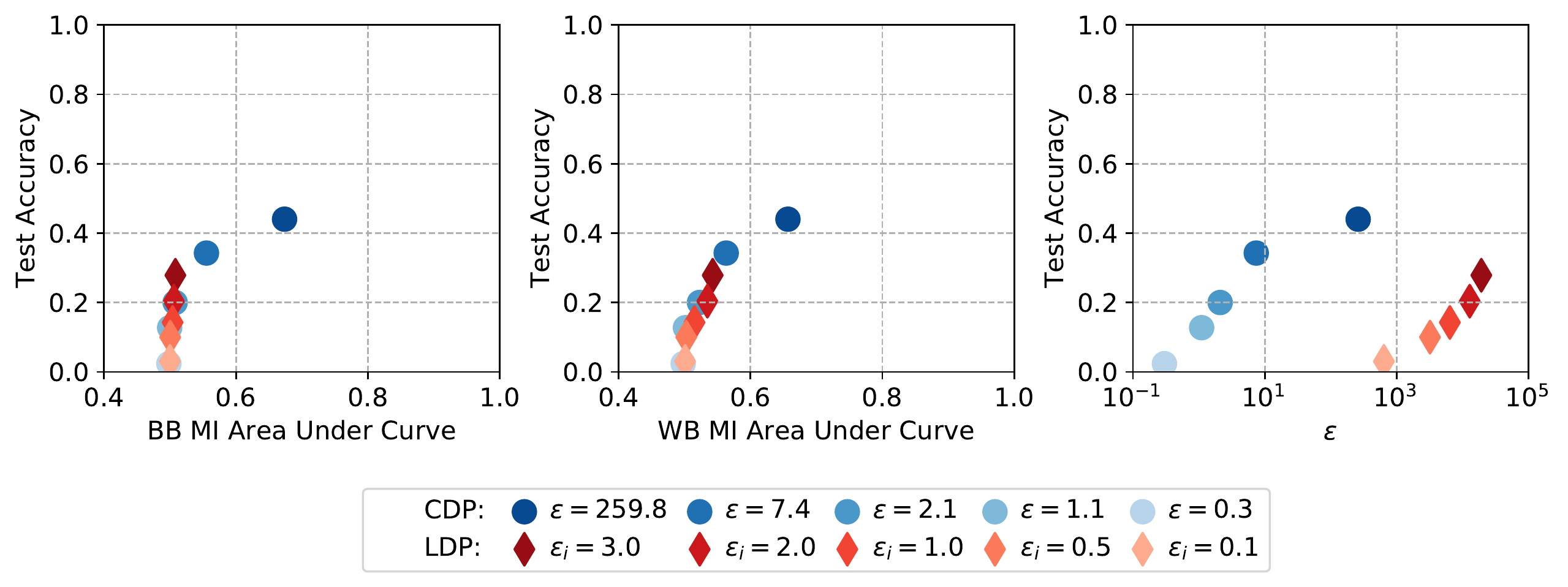}
		\caption{Privacy-accuracy trade-off comparison over black-box AUC, white-box AUC and $\eps$ for $\mathbb{C}=300$}
		\label{fig:eval:texas_unified_scatter}	
	\end{subfigure}
	\caption{\cali{DO} accuracy and privacy analysis on Texas Hospital Stays (error bars lie within most points)}
	\label{fig:eval:texas:figures}
\end{figure*}

For Texas Hospital Stays we can observe that LDP and CDP are achieving very similar privacy-accuracy trade-offs under both MI attacks. The main difference in LDP and CDP is observable in smoother decrease of target model accuracy for CDP in contrast to LDP, which are depicted in Figures~\ref{fig:eval:texas_cdp_acc} and~\ref{fig:eval:texas_ldp_acc}. The smoother decay also manifests in a slower drop of MI AUC for CDP in comparison to LDP as stated in Figures~\ref{fig:eval:texas_cdp_auc_bb},~\ref{fig:eval:texas_cdp_auc_wb} and ~\ref{fig:eval:texas_ldp_auc_bb},~\ref{fig:eval:texas_ldp_auc_wb}. Texas Hospital Stays represents an unbalanced high dimensional dataset and both factors foster MI. However, the increase in dataset imbalance by increasing $\mathbb{C}$ is negligible w.r.t.~MI AUC under both considered MI attacks and the additional WB assumptions do not lead to a significant gain in MI AUC. The relative privacy-accuracy trade-off of LDP and CDP is also close and for example the baseline MI AUC of $0.5$ is reached at $\varphi\approx1.5$. \cali{DO} might however prefer to use CDP since the space of achievable MI AUCs in LDP is narrow while CDP also yields AUCs in between original and baseline as illustrated for $\mathbb{C}=300$ in the ROC curves in Figures~\ref{fig:eval:texas_unified_roc_cdp} and ~\ref{fig:eval:texas_unified_roc_ldp}, and the scatterplots in Figure~\ref{fig:eval:texas_unified_scatter}.

%% file: src/experiments-purch.tex
\begin{figure*}[t!]
	\centering
	\begin{subfigure}{0.25\linewidth}
		\includegraphics[width=1\linewidth]{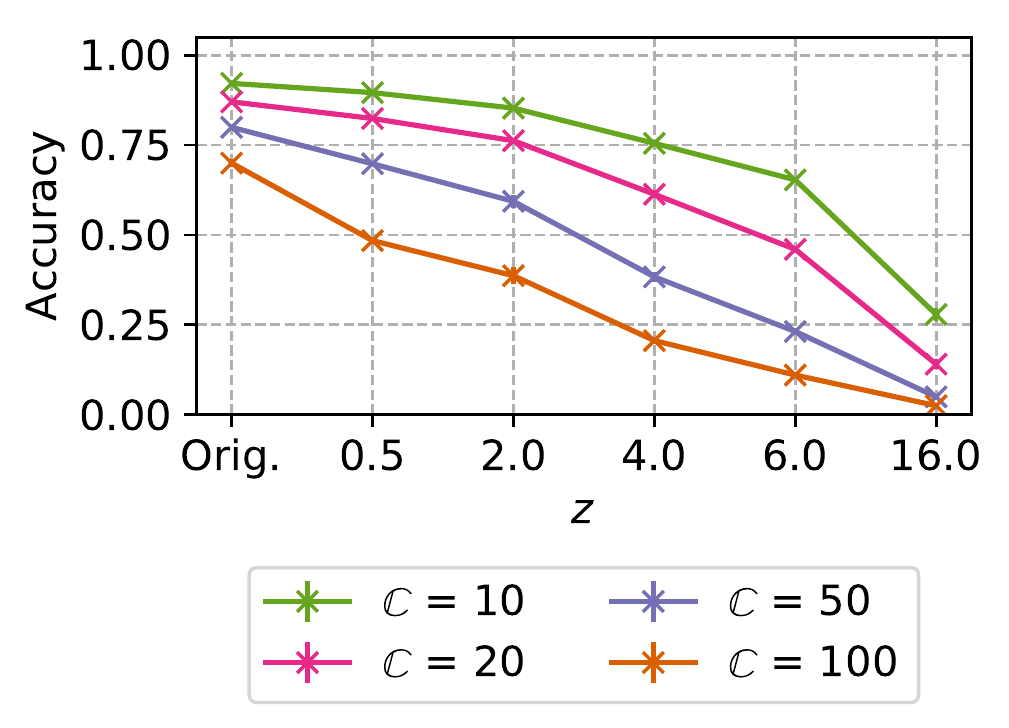}
		\caption{Target model accuracy (CDP)}
		\label{fig:eval:purch_cdp_acc}
	\end{subfigure}%
	\begin{subfigure}{0.25\linewidth}
		\includegraphics[width=1\linewidth]{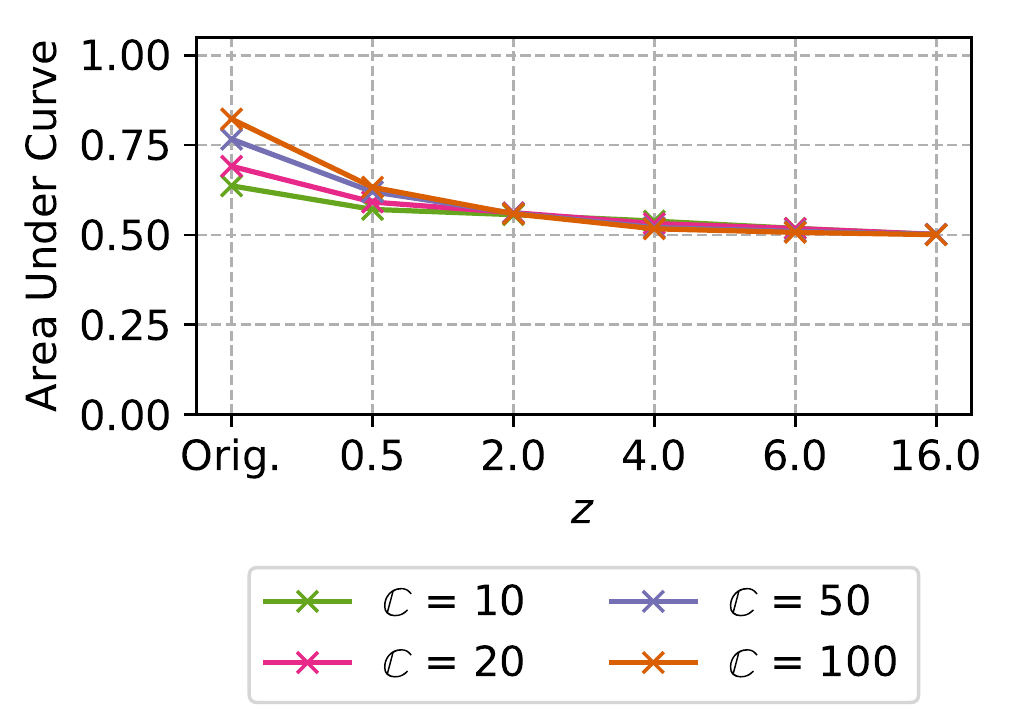}
		\caption{Black-box AUC (CDP)}
		\label{fig:eval:purch_cdp_auc_bb}
	\end{subfigure}%
	\begin{subfigure}{0.25\linewidth}
		\includegraphics[width=1\linewidth]{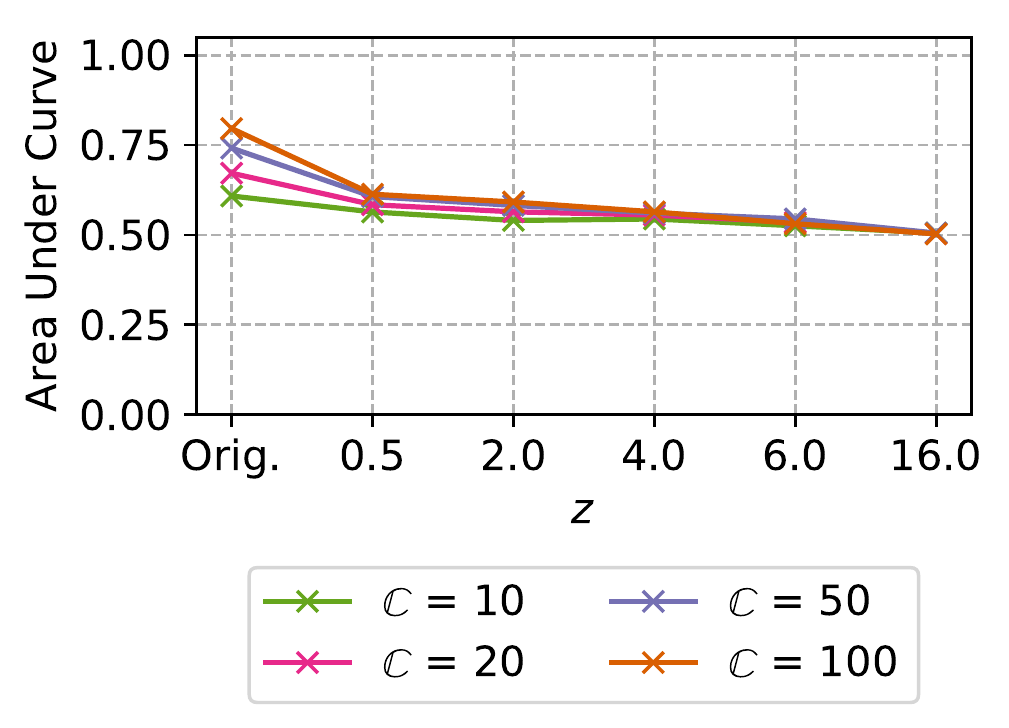}
		\caption{White-box AUC (CDP)}
		\label{fig:eval:purch_cdp_auc_wb}
	\end{subfigure}%
	\begin{subfigure}{0.25\linewidth}
		\includegraphics[width=1\linewidth]{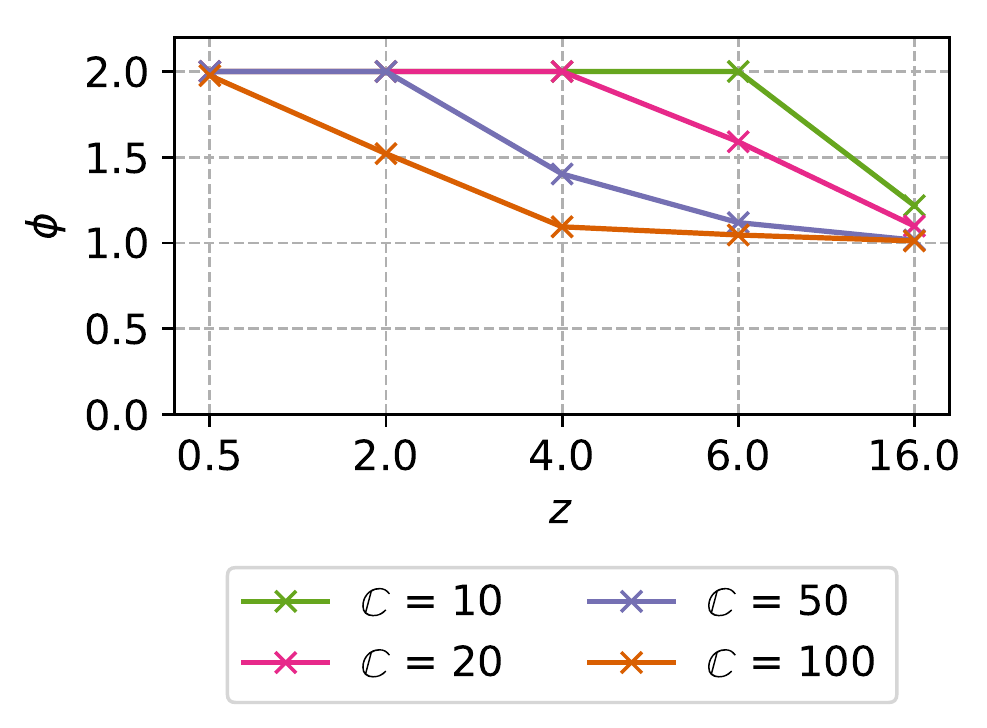}
		\caption{White-box $\varphi$ (CDP)}
		\label{fig:eval:purch_cdp_phi_wb}
	\end{subfigure}\\
	\begin{subfigure}{0.25\linewidth}
		\includegraphics[width=1\linewidth]{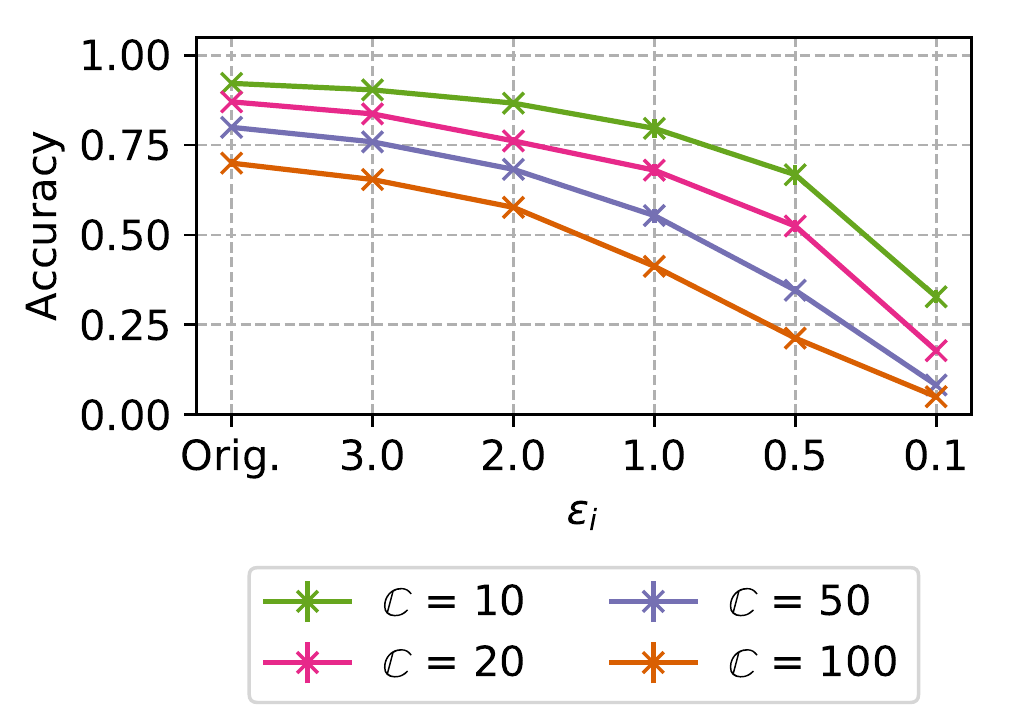}
		\caption{Target model accuracy (LDP)}
		\label{fig:eval:purch_ldp_acc}
	\end{subfigure}%
	\begin{subfigure}{0.25\linewidth}
		\includegraphics[width=1\linewidth]{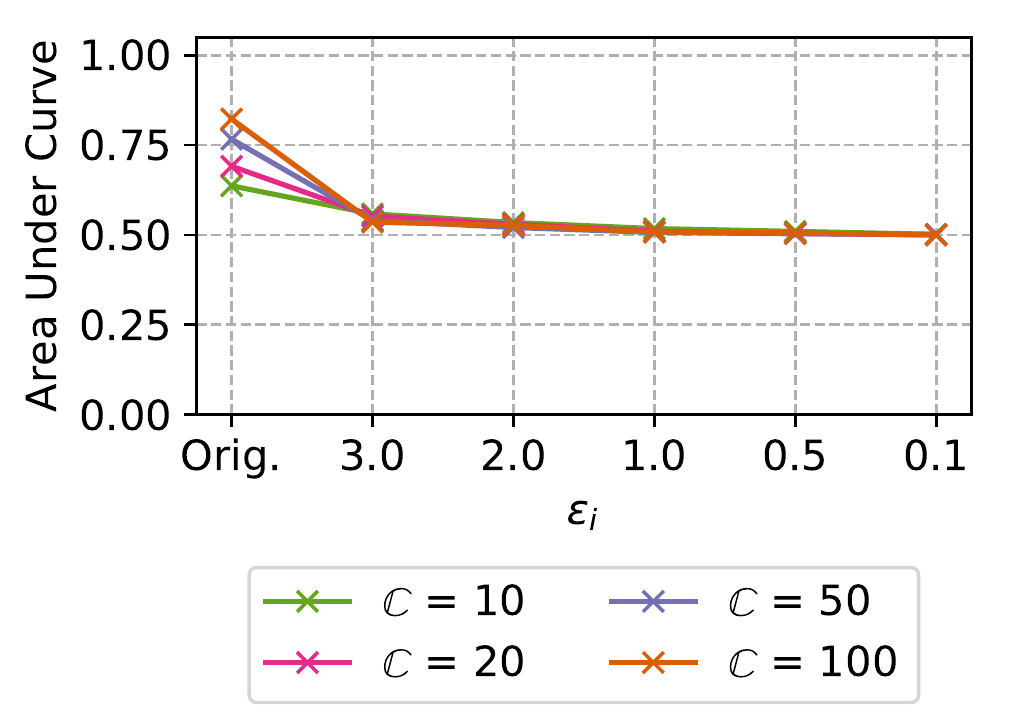}
		\caption{Black-box AUC (LDP)}
		\label{fig:eval:purch_ldp_auc_bb}
	\end{subfigure}%
	\begin{subfigure}{0.25\linewidth}
		\includegraphics[width=1\linewidth]{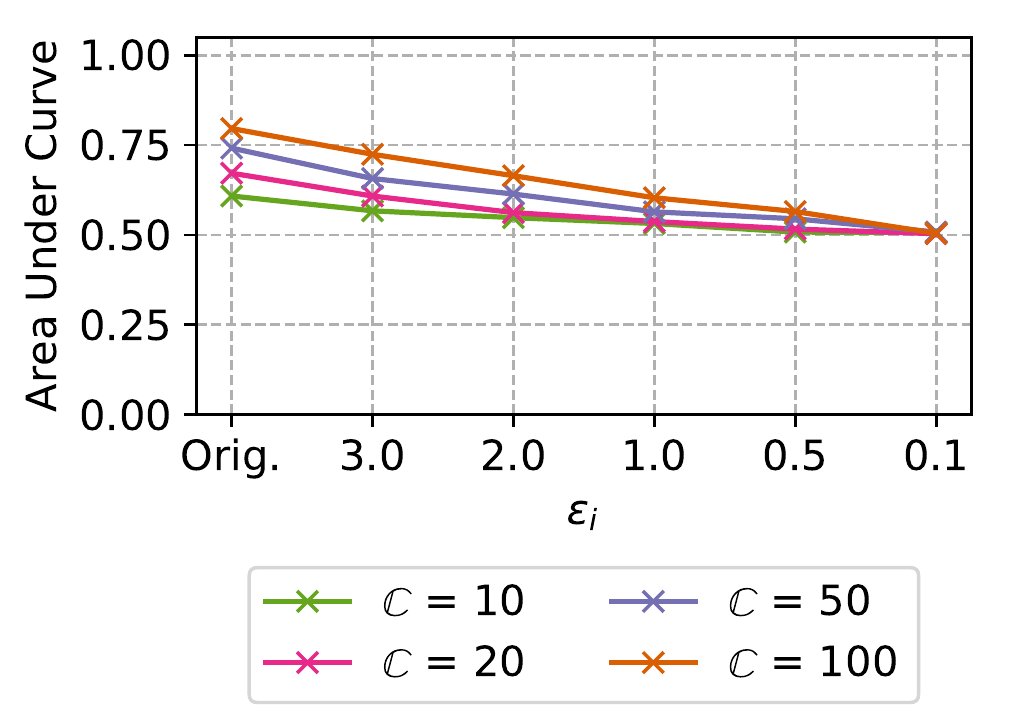}
		\caption{White-box AUC (LDP)}
		\label{fig:eval:purch_ldp_auc_wb}
	\end{subfigure}%
	\begin{subfigure}{0.25\linewidth}
		\includegraphics[width=1\linewidth]{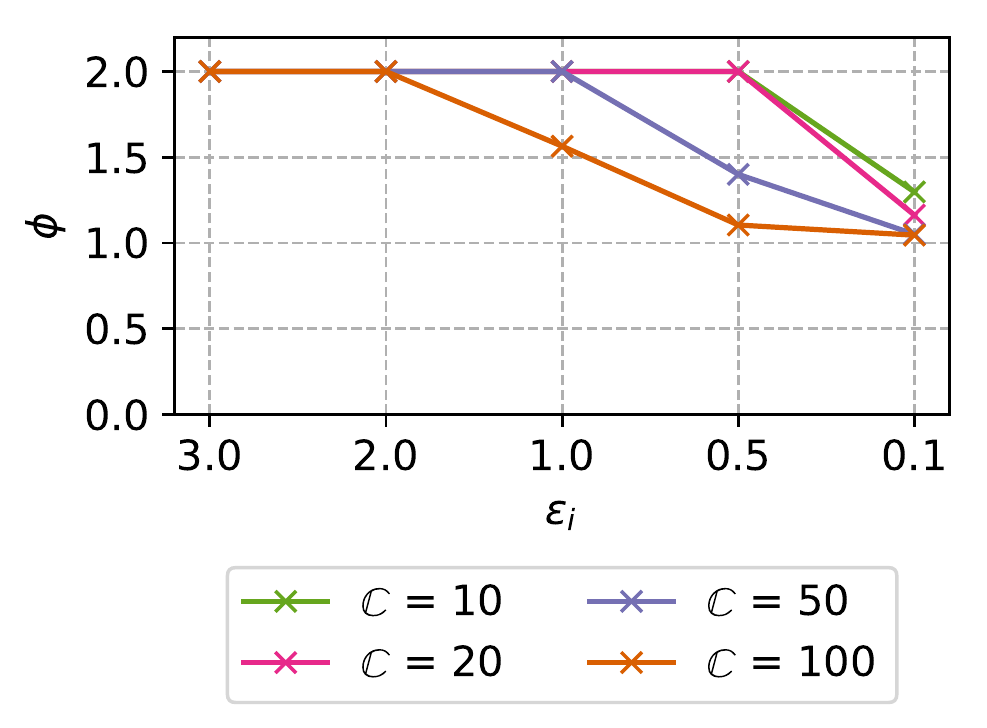}
		\caption{White-box $\varphi$ (LDP)}
		\label{fig:eval:purch_ldp_phi_wb}
	\end{subfigure}\\
	\begin{subfigure}{0.45\linewidth}
		\includegraphics[width=1\linewidth]{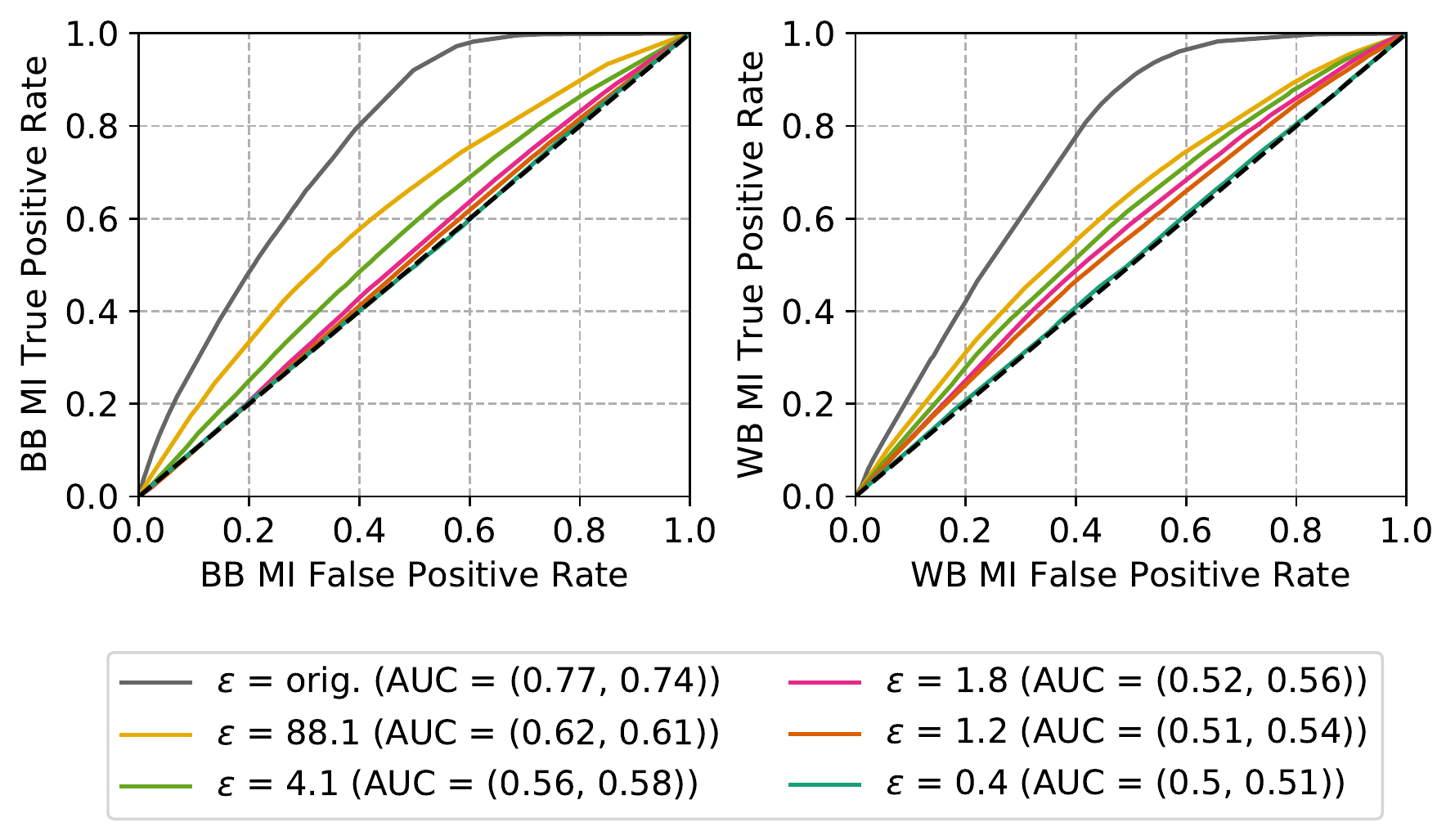}
		\caption{Receiver operating curve for $\mathbb{C}=50$ (CDP)}
		\label{fig:eval:purch_unified_roc_cdp}	
	\end{subfigure}%
	\begin{subfigure}{0.45\linewidth}
		\includegraphics[width=1\linewidth]{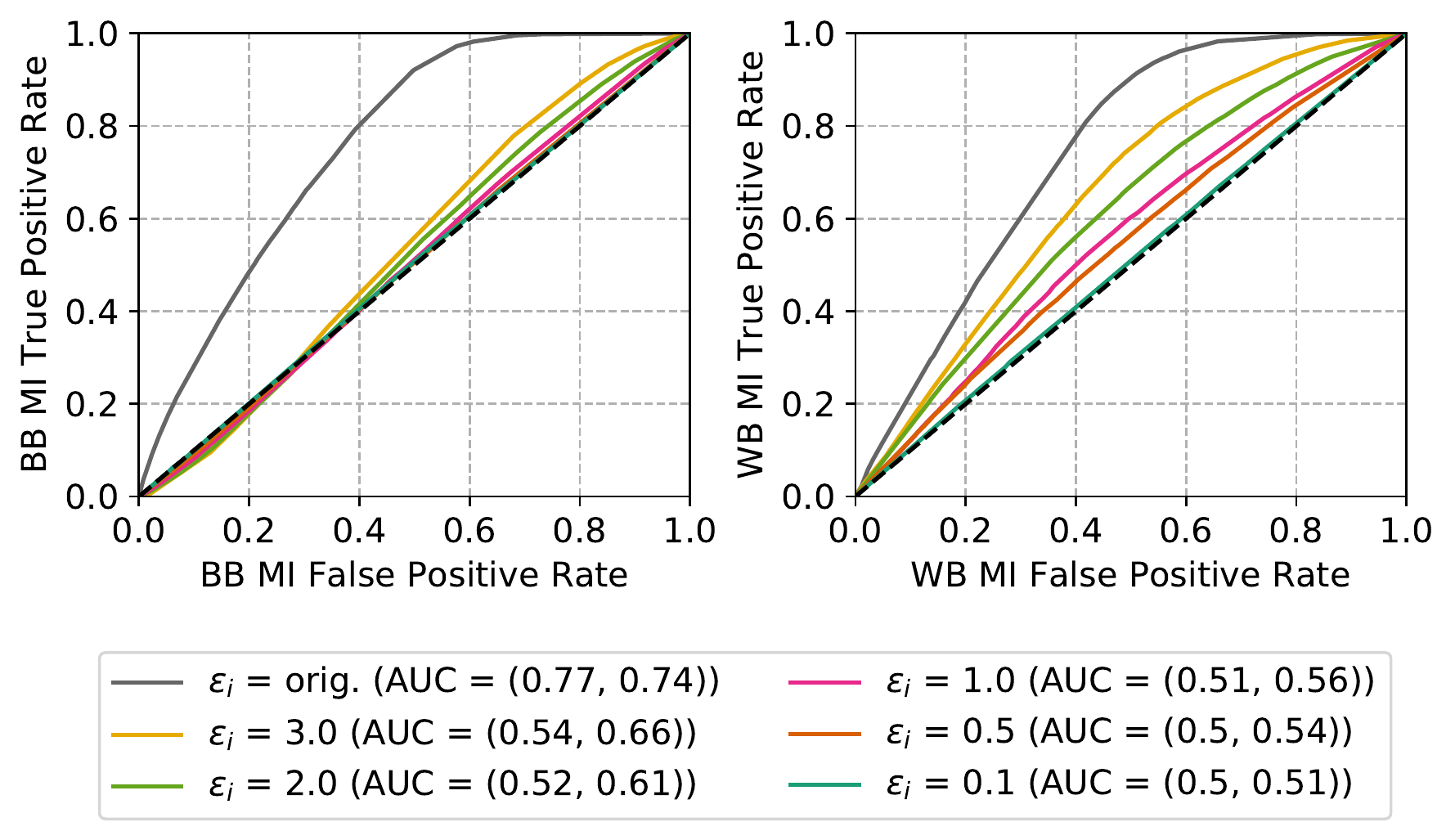}
		\caption{Receiver operating curve for $\mathbb{C}=50$ (LDP)}
		\label{fig:eval:purch_unified_roc_ldp}	
	\end{subfigure}\\
	\begin{subfigure}{0.6\linewidth}
		\includegraphics[width=1\linewidth]{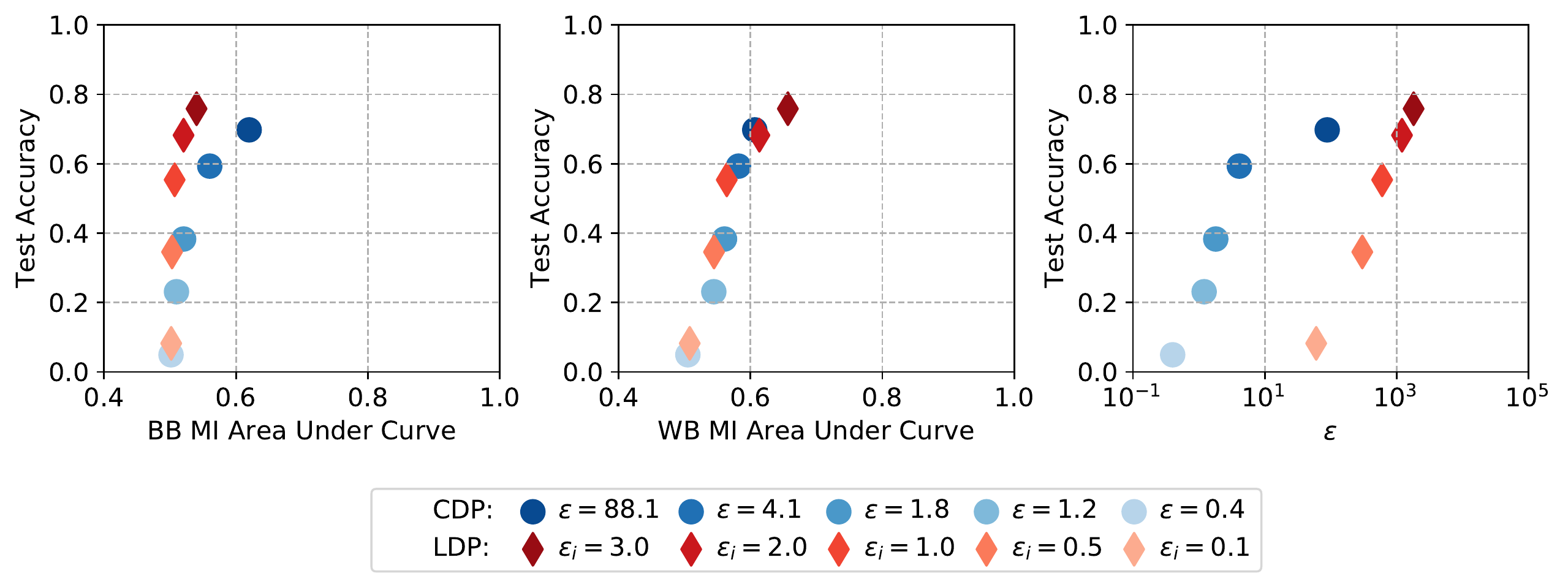}
		\caption{Privacy-accuracy trade-off comparison over black-box AUC, white-box AUC and $\eps$~for $\mathbb{C}=50$}
		\label{fig:eval:purch_unified_scatter}	
	\end{subfigure}
	\caption{\cali{DO} accuracy and privacy analysis on Purchases Shopping Carts}
	\label{fig:eval:purch:figures}
\end{figure*}

CDP and LDP are achieving similar target model test accuracy on the Purchases dataset as depicted in ~\ref{fig:eval:purch_cdp_acc} and~\ref{fig:eval:purch_ldp_acc}. However, LDP is allowing a slightly smoother decrease in test accuracy over \eps. The MI AUC for CDP Figures~\ref{fig:eval:purch_cdp_auc_bb} and~\ref{fig:eval:purch_cdp_auc_wb} illustrate that WB MI is slightly more resistant to noise and than BB MI and maintains an MI AUC above $0.5$ until $z=6.0$ while for BB MI the AUC stays above $0.5$ until $z=16.0$. The differences for MI AUC for LDP are stronger especially for large $\mathbb{C}$ where WB MI realizes significantly higher AUCs and decreases slower to the MI AUC baseline as depicted by Figures~\ref{fig:eval:purch_ldp_auc_bb} and~\ref{fig:eval:purch_ldp_auc_wb}. A comparison of the relative privacy-accuracy trade-offs $\varphi$ in Figures~\ref{fig:eval:purch_cdp_phi_wb} and ~\ref{fig:eval:purch_ldp_phi_wb} underline that CDP and LDP achieve similar trade-offs and CDP again allows for smoother drops in the MI AUC in contrast to LDP. Thus, LDP is the preferred choice if \cali{DO} desires to lower the MI AUC to a level between original and baseline as illustrated for $\mathbb{C}=50$ in the ROC curves in Figures~\ref{fig:eval:purch_unified_roc_cdp},~\ref{fig:eval:purch_unified_roc_ldp} and the scatterplots in Figure~\ref{fig:eval:purch_unified_scatter}. It is remarkable that while the overall \eps~for LDP and CDP differs by a magnitude of up to $10$ times the privacy-accuracy trade-offs are very close to each other.

%% file: src/experiments-gcn.tex
\begin{figure}[t!]
	\centering
	\begin{subfigure}{0.5\linewidth}
		\includegraphics[width=1\linewidth]{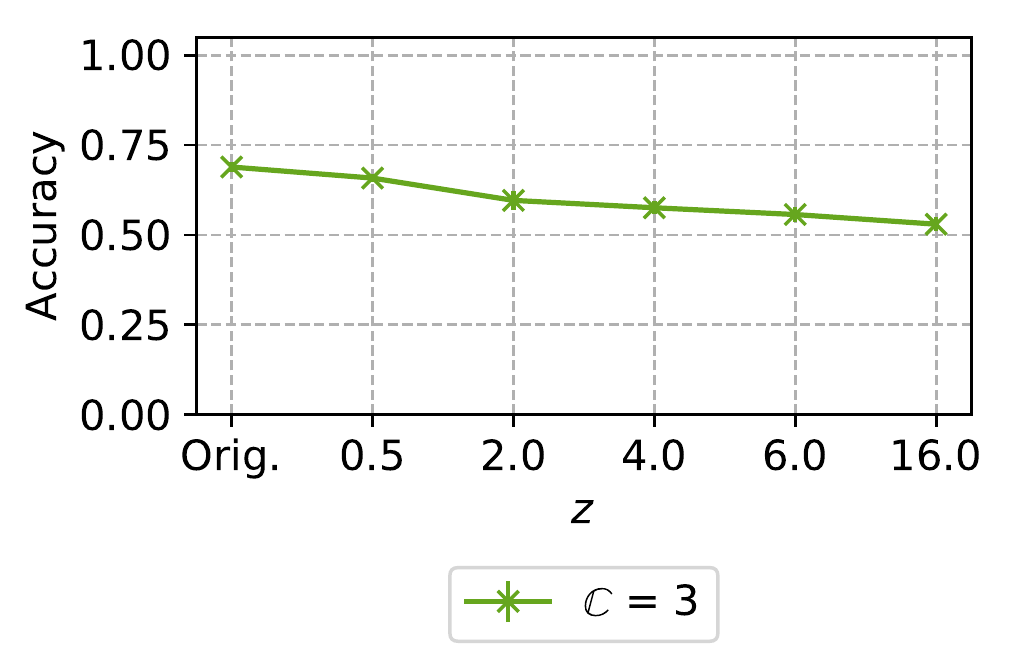}
		\caption{Accuracy (CDP)}
		\label{fig:eval:gcn_cdp_acc}
	\end{subfigure}%
	\begin{subfigure}{0.5\linewidth}
		\includegraphics[width=1\linewidth]{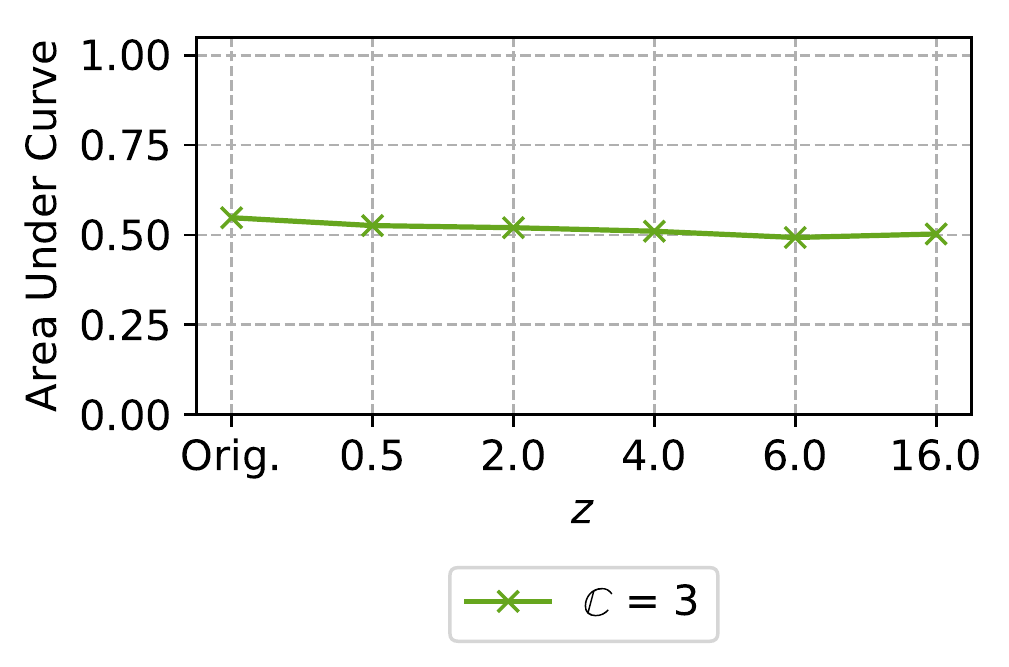}
		\caption{White-box AUC (CDP)}
		\label{fig:eval:gcn_cdp_auc_wb}
	\end{subfigure}\\
	\begin{subfigure}{0.5\linewidth}
		\includegraphics[width=1\linewidth]{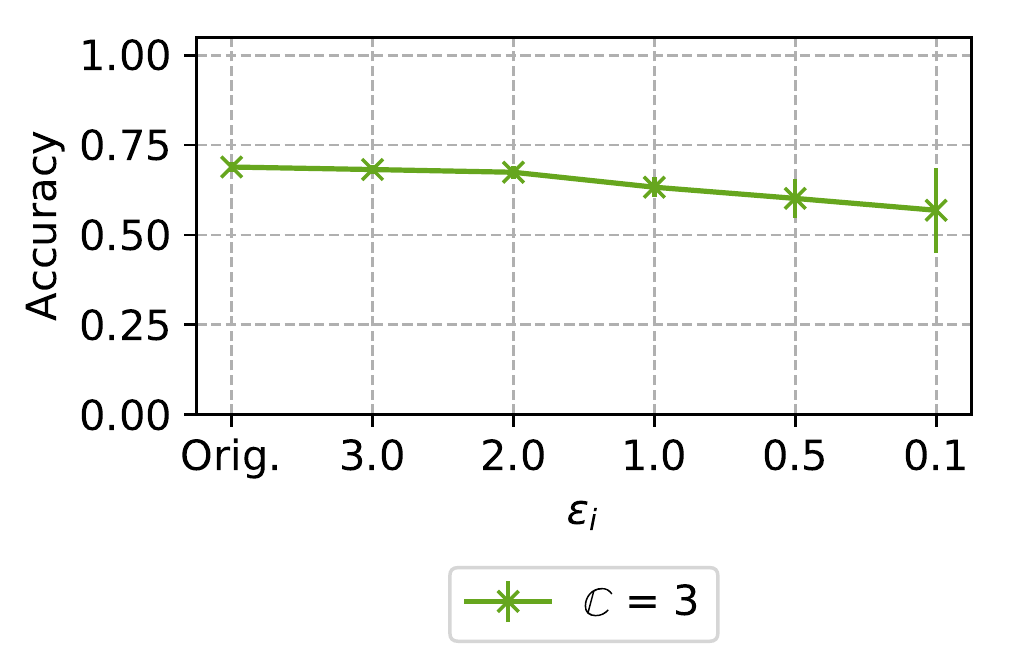}
		\caption{Accuracy (LDP)}
		\label{fig:eval:gcn_ldp_acc}
	\end{subfigure}%
	\begin{subfigure}{0.5\linewidth}
		\includegraphics[width=1\linewidth]{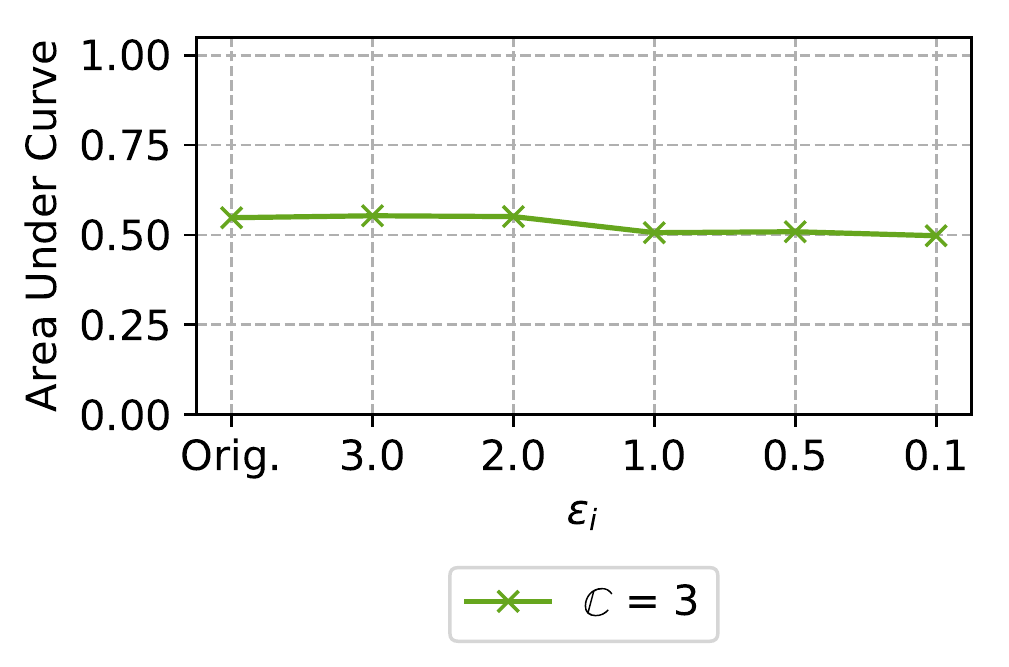}
		\caption{White-box AUC (LDP)}
		\label{fig:eval:gcn_ldp_auc_wb}
	\end{subfigure}%
	\caption{\cali{DO} accuracy and privacy analysis on COLLAB}
	\label{fig:eval:gcn:figures}
\end{figure}

For COLLAB we can observe that the test accuracy neither diminishes to the uniform random guessing baseline for CDP nor LDP as stated Figures~\ref{fig:eval:gcn_cdp_acc} and~\ref{fig:eval:gcn_ldp_acc}. The effect is caused by the fact that the target model's reference architecture resorts to classifying all records as the majority class in the dataset which results in a test accuracy of $0.52$ when high noise is added. Due to the diminishing target accuracy, the  BB MI AUC does not leave baseline and thus we omit the plots. The WB MI AUCs for CDP and LDP in Figures~\ref{fig:eval:gcn_cdp_auc_wb} and~\ref{fig:eval:gcn_ldp_auc_wb} are  only slightly above the baseline MI AUC but robust to noise and remain over the baseline until perturbation is high. We notice that LDP has a slight advantage over CDP for the COLLAB graph dataset which we account to the comparatively weak privacy notion of edge local differential privacy that was achieved under LDP. We omit showing ROC curves and scatterplots, since they are not providing additional insights due to to the small range of the MI AUCs and baseline test accuracy.

%% file: src/experiments-lfw.tex
\begin{figure*}[t!]
	\centering
	\begin{subfigure}{0.25\linewidth}
		\includegraphics[width=1\linewidth]{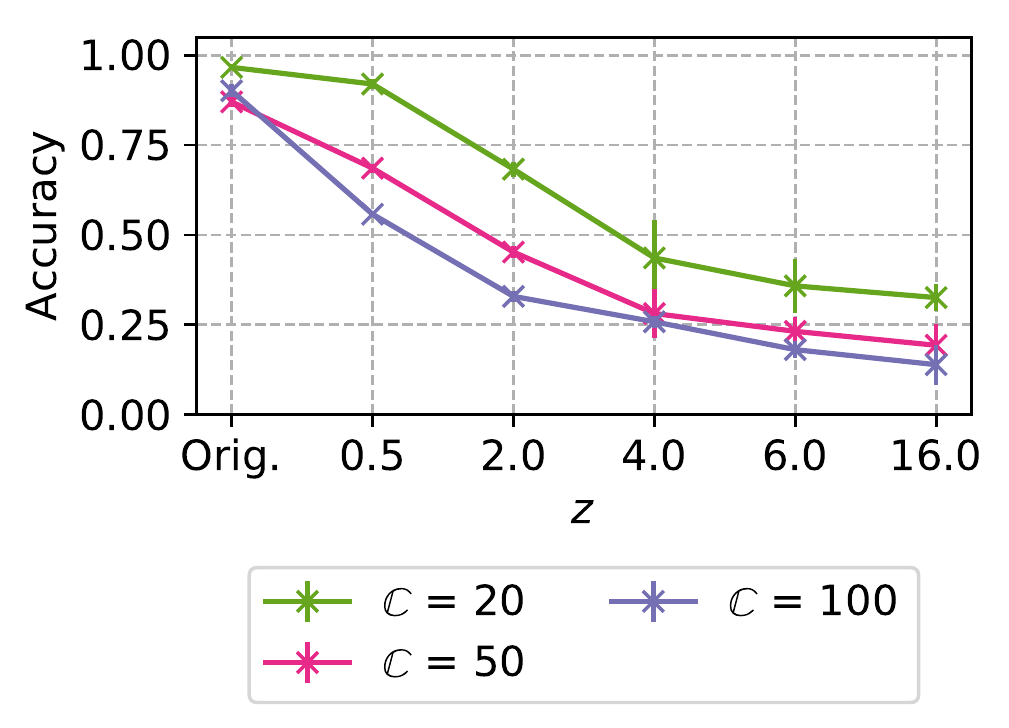}
		\caption{Target model accuracy (CDP)}
		\label{fig:eval:lfw_cdp_acc}
	\end{subfigure}%
	\begin{subfigure}{0.25\linewidth}
		\includegraphics[width=1\linewidth]{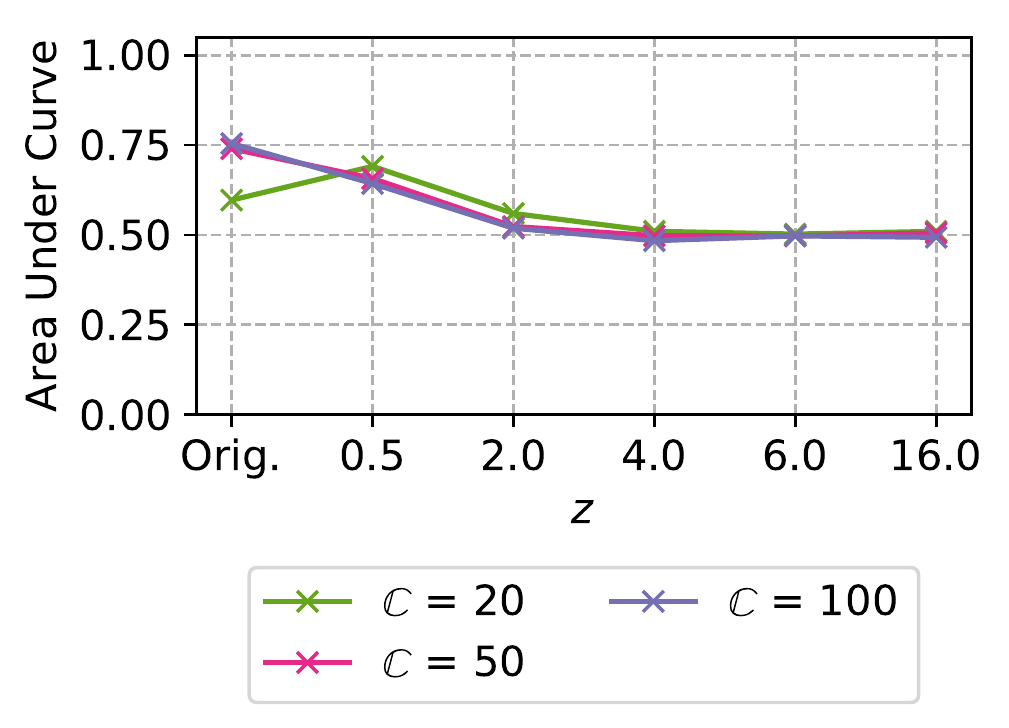}
		\caption{Black-box AUC (CDP)}
		\label{fig:eval:lfw_cdp_auc_bb}
	\end{subfigure}%
	\begin{subfigure}{0.25\linewidth}
		\includegraphics[width=1\linewidth]{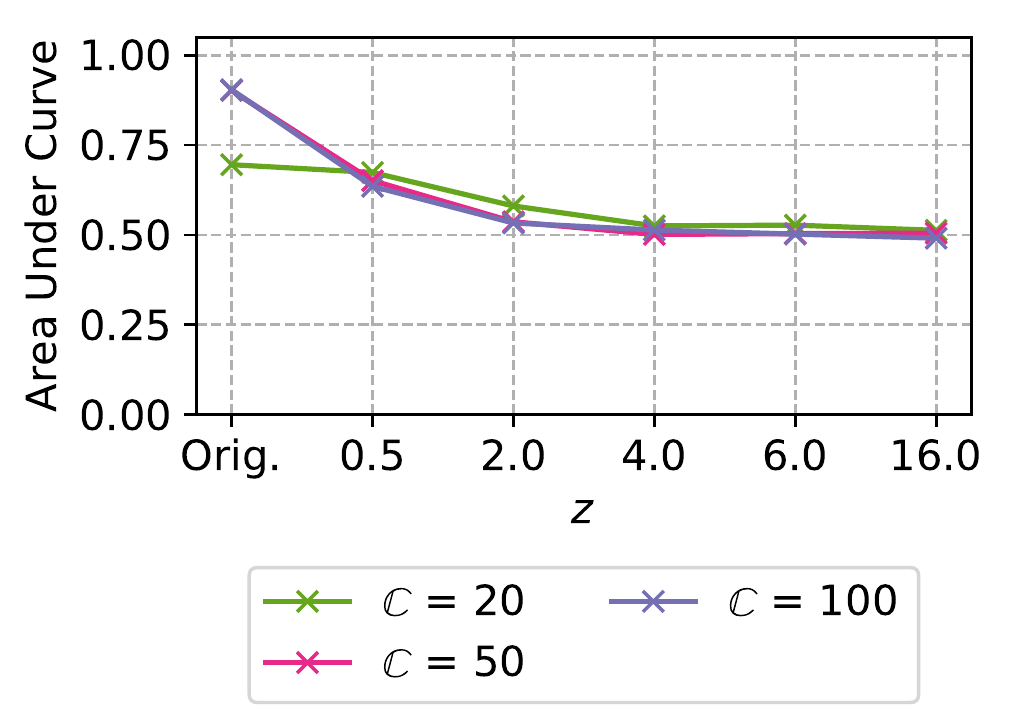}
		\caption{White-box AUC (CDP)}
		\label{fig:eval:lfw_cdp_auc_wb}
	\end{subfigure}%
	\begin{subfigure}{0.25\linewidth}
		\includegraphics[width=1\linewidth]{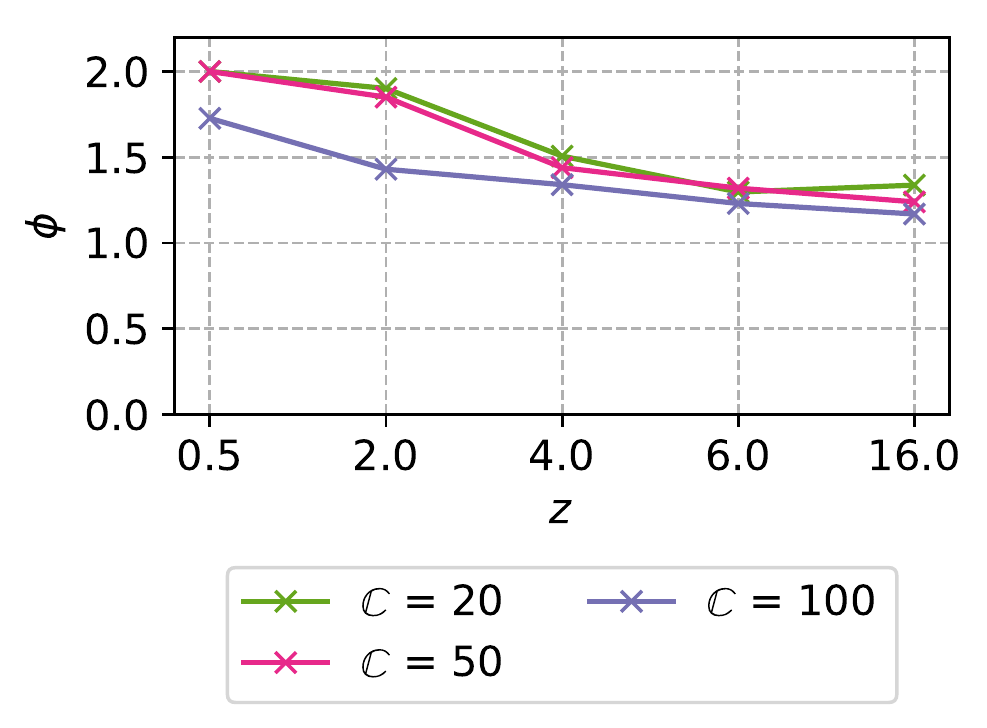}
		\caption{White-box $\varphi$ (CDP)}
		\label{fig:eval:lfw_cdp_phi_wb_new}
	\end{subfigure}\\
	\begin{subfigure}{0.25\linewidth}
		\includegraphics[width=1\linewidth]{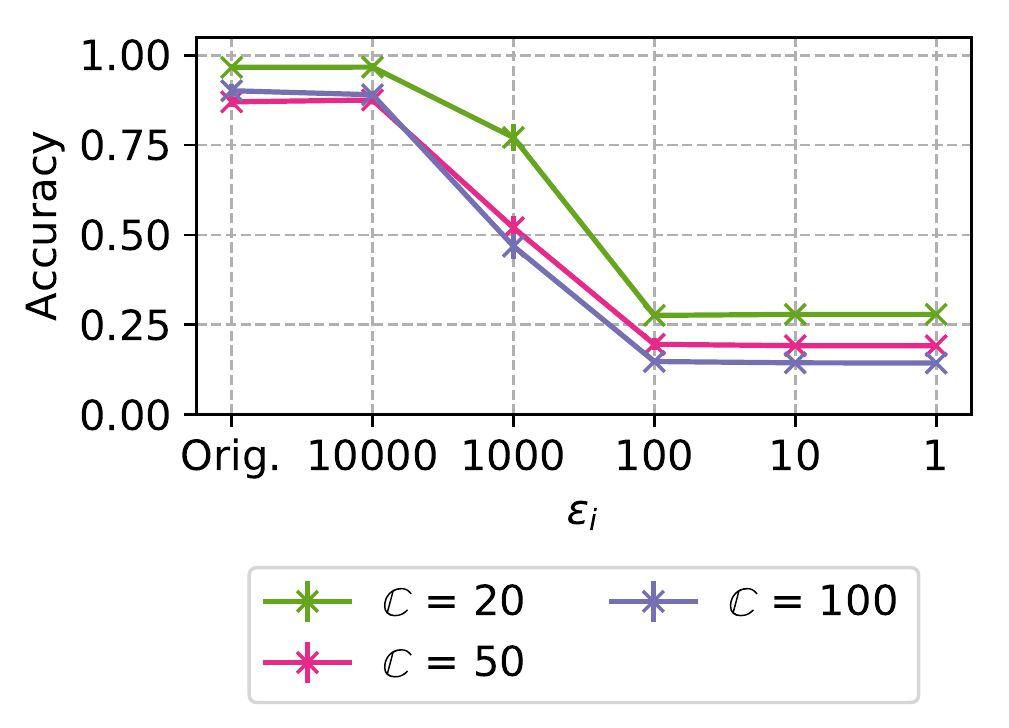}
		\caption{Target model accuracy (LDP)}
		\label{fig:eval:lfw_ldp_acc}
	\end{subfigure}%
	\begin{subfigure}{0.25\linewidth}
		\includegraphics[width=1\linewidth]{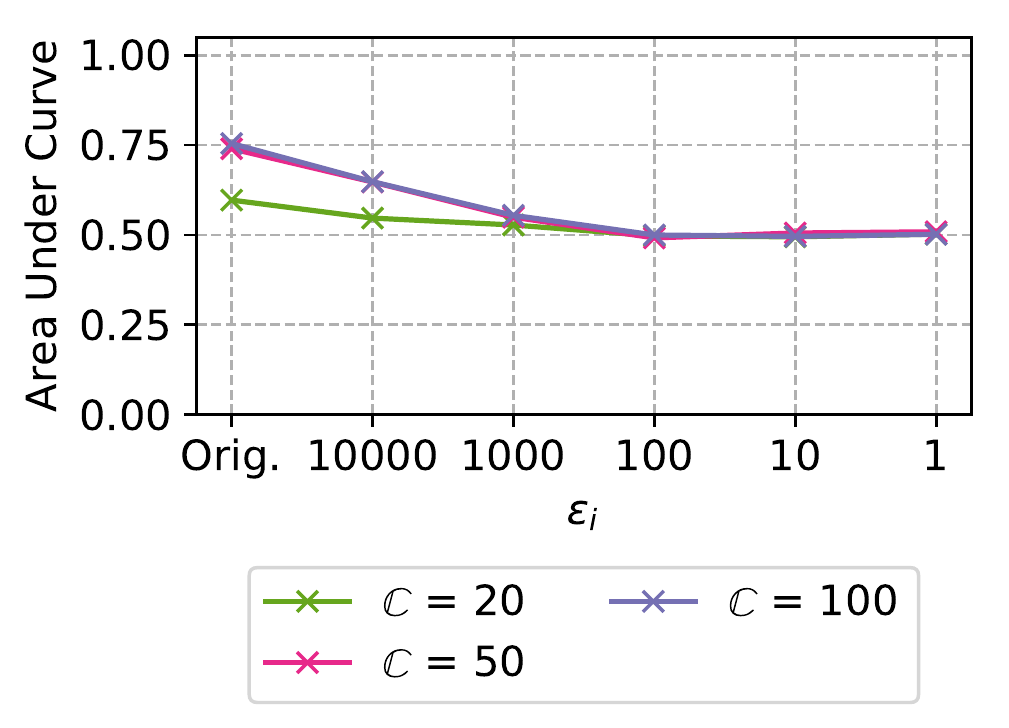}
		\caption{Black-box AUC (LDP)}
		\label{fig:eval:lfw_ldp_auc_bb}
	\end{subfigure}%
	\begin{subfigure}{0.25\linewidth}
		\includegraphics[width=1\linewidth]{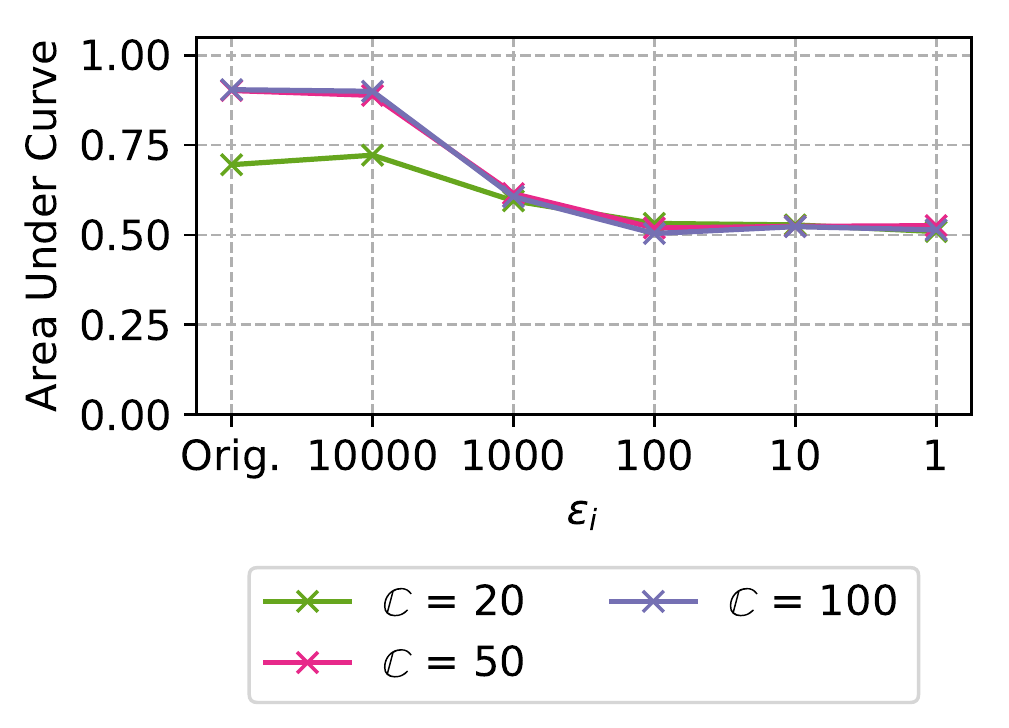}
		\caption{White-box AUC (LDP)}
		\label{fig:eval:lfw_ldp_auc_wb}
	\end{subfigure}%
	\begin{subfigure}{0.25\linewidth}
		\includegraphics[width=1\linewidth]{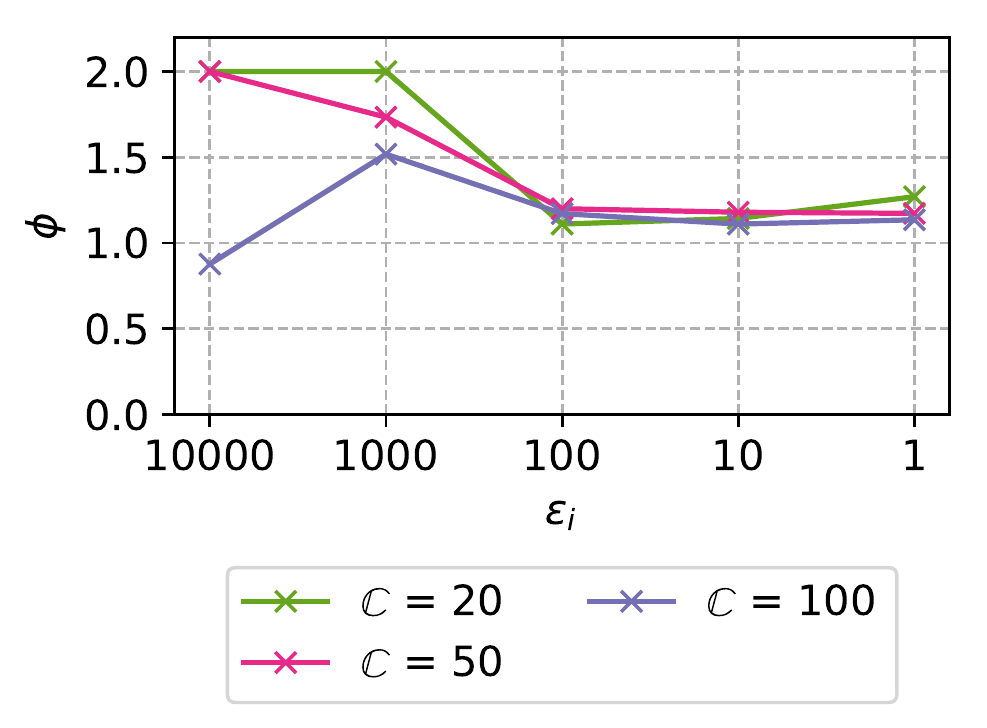}
		\caption{White-box $\varphi$ (LDP)}
		\label{fig:eval:lfw_ldp_phi_wb_new}
	\end{subfigure}\\
	\begin{subfigure}{0.45\linewidth}
		\includegraphics[width=1\linewidth]{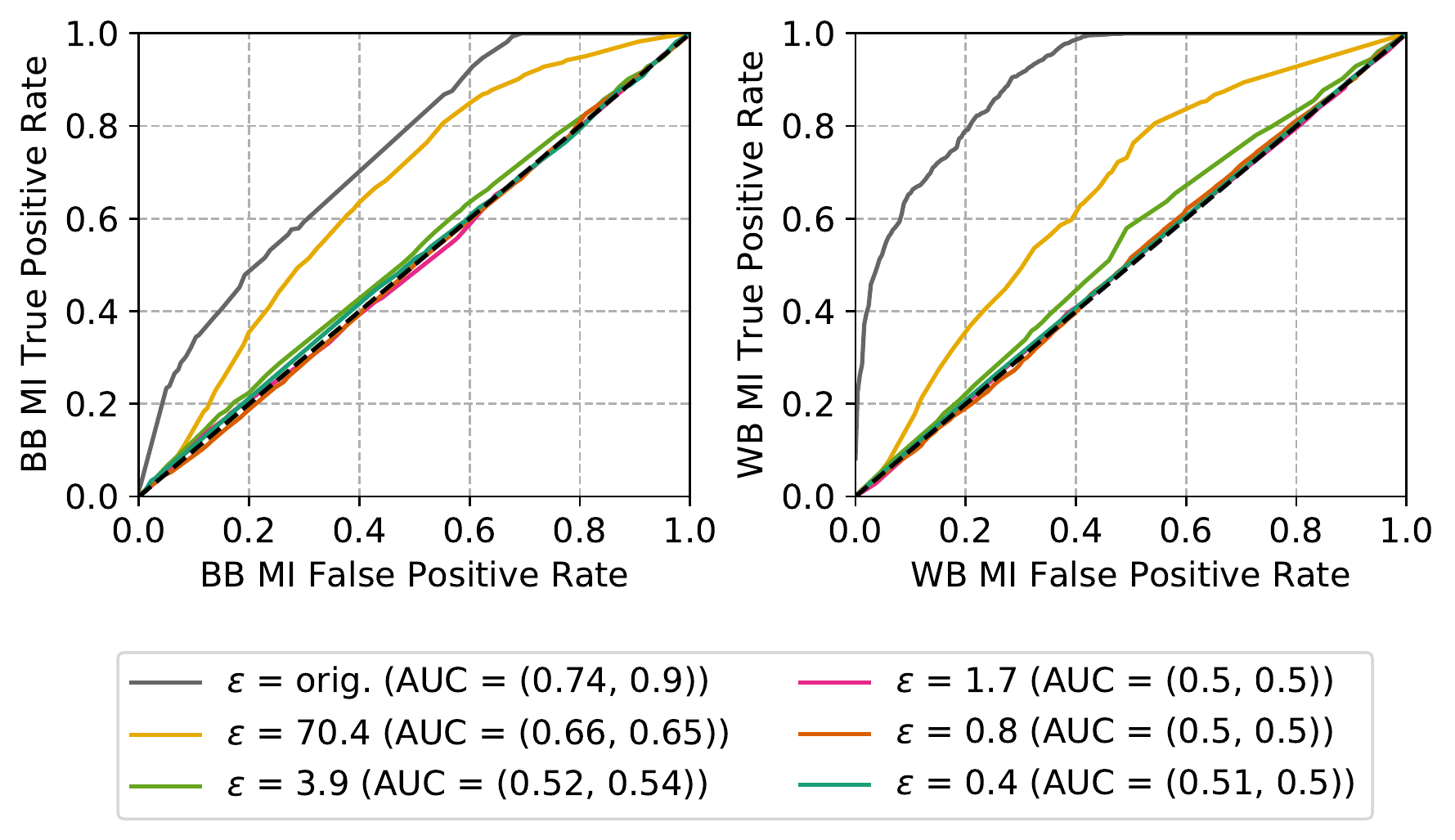}
		\caption{Receiver operating curve for $\mathbb{C}=50$ (CDP) }
		\label{fig:eval:lfw_unified_roc_cdp}	
	\end{subfigure}%
	\begin{subfigure}{0.45\linewidth}
		\includegraphics[width=1\linewidth]{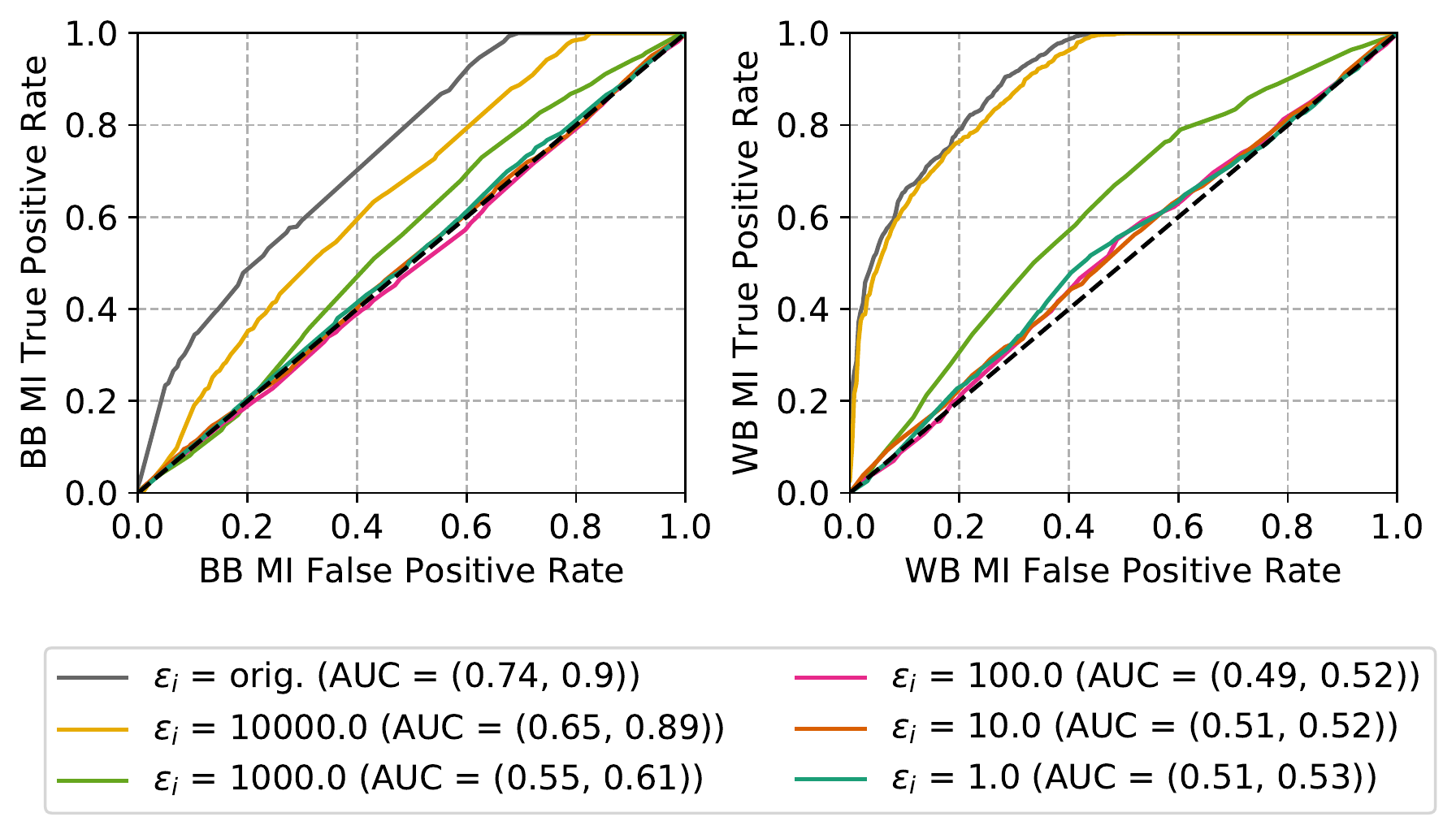}
		\caption{Receiver operating curve for $\mathbb{C}=50$ (LDP)}
		\label{fig:eval:lfw_unified_roc_ldp}	
	\end{subfigure}\\
	\begin{subfigure}{0.6\linewidth}
		\includegraphics[width=1\linewidth]{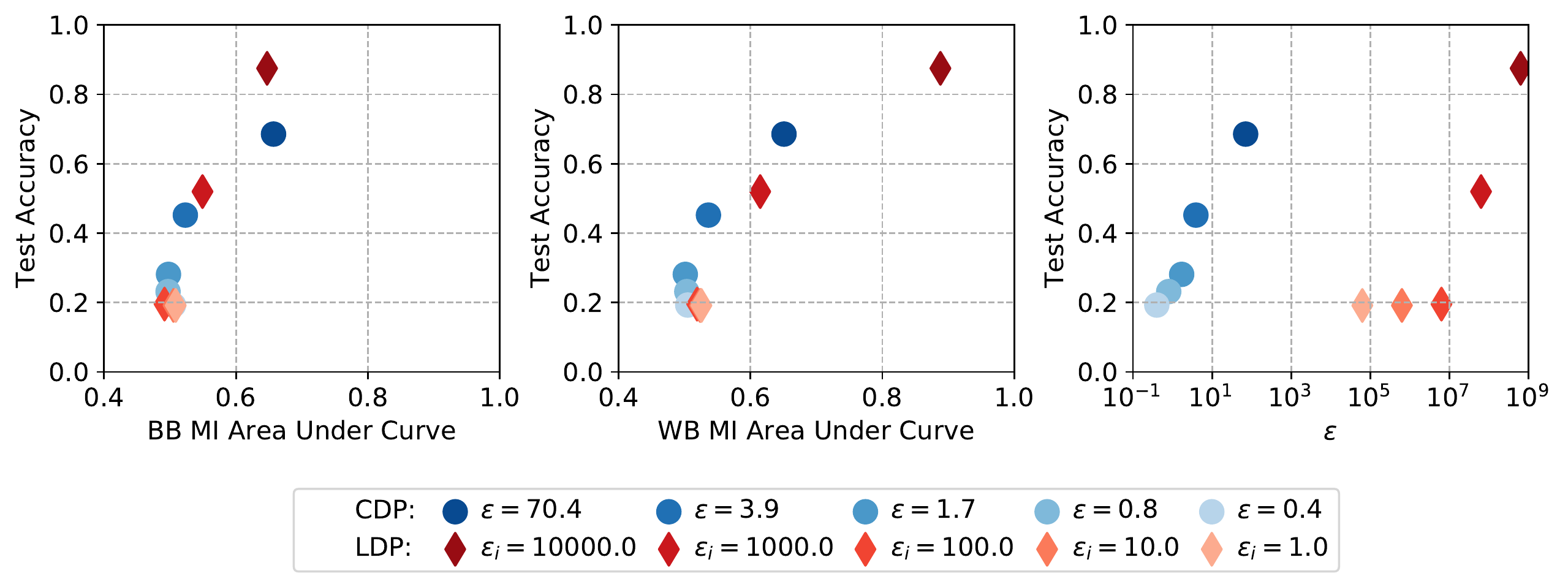}
		\caption{Privacy-accuracy trade-off comparison over black-box AUC, white-box AUC and $\eps$~for $\mathbb{C}=50$}
		\label{fig:eval:lfw_unified_scatter}	
	\end{subfigure}
	\caption{\cali{DO} accuracy and privacy analysis on LFW}
	\label{fig:eval:lfw:figures}
\end{figure*}

Again, the target model reference architecture converges for both CDP and LDP towards same test accuracy which is reflecting the majority class. However, the accuracy decay over increasing noise multipliers (decreasing epsilon) is much smoother for CDP when comparing Figures~\ref{fig:eval:lfw_cdp_acc} and \ref{fig:eval:lfw_ldp_acc}, and the structural changes, which LDP image pixelation causes, seem to lead to quicker losses in test accuracy. White-box MI AUCs are significantly higher for both LDP and CDP compared to BB AUC. However, under CDP the observable WB MI AUCs in Figure~\ref{fig:eval:lfw_cdp_auc_wb} quickly converge to the BB MI AUCs in Figure~\ref{fig:eval:lfw_cdp_auc_bb} under noise. Note that Figure~\ref{fig:eval:lfw_cdp_auc_bb} contains an outlier since the BB MI AUC is actually increasing for $\mathbb{C}=20$ when noise is added at $z=0.5$. The outlier is due to an increasing train-test-gap (cf.~train accuracy in Table~\ref{tab:comp:train_acc}) at this noise level which closes again at $z=2.0$. In contrast, the LDP WB MI AUCs are consistently higher than BB as stated in Figures~\ref{fig:eval:lfw_ldp_auc_bb} and \ref{fig:eval:lfw_ldp_auc_wb}. With respect to the relative privacy-accuracy trade-off $\varphi$ in Figures \ref{fig:eval:lfw_cdp_phi_wb_new} and~\ref{fig:eval:lfw_ldp_phi_wb_new} CDP outperforms LDP at higher noise levels. At MI AUC $=0.5$ CDP achieves $\varphi\approx1.5$ for all $\mathbb{C}$ while LDP yields $\varphi\approx1.1$ for all $\mathbb{C}$. The example ROCs for $\mathbb{C}=50$ in Figures~\ref{fig:eval:lfw_unified_roc_cdp} and \ref{fig:eval:lfw_unified_roc_ldp} illustrate that CDP already has a large effect on MI AUC at high $\eps$. In addition, we observe from the scatterplots in Figure~\ref{fig:eval:lfw_unified_scatter} that CDP realizes a strictly better privacy-accuracy trade-off under the WB MI for $\mathbb{C}=50$.

%% file: src/experiments-purch_skew.tex
\begin{figure*} 
	\centering
	\begin{subfigure}{0.25\linewidth}
		\includegraphics[width=1\linewidth]{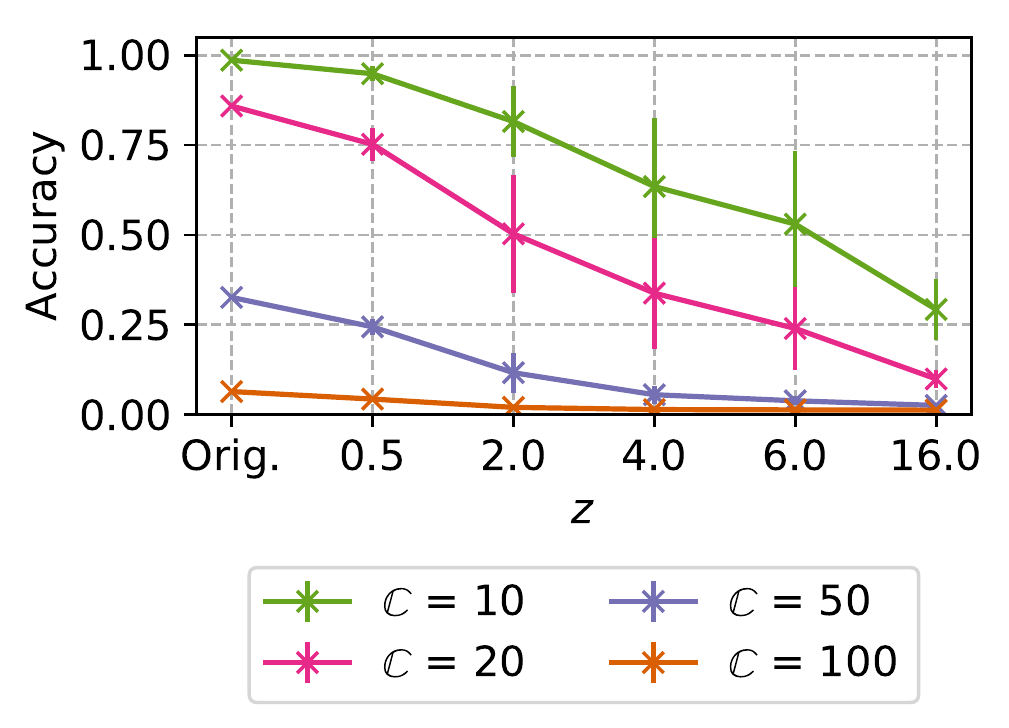}
		\caption{Target model accuracy (CDP)}
		\label{fig:eval:purch_skew_cdp_acc}
	\end{subfigure}%
	\begin{subfigure}{0.25\linewidth}
		\includegraphics[width=1\linewidth]{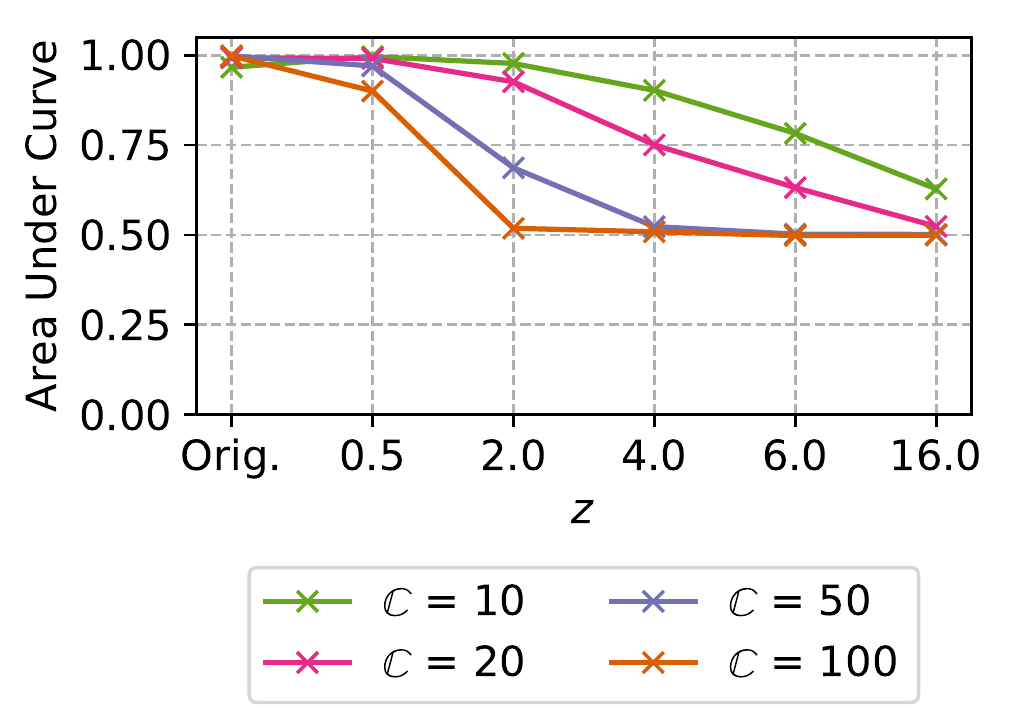}
		\caption{Black-box AUC (CDP)}
		\label{fig:eval:purch_skew_cdp_auc_bb}
	\end{subfigure}%
	\begin{subfigure}{0.25\linewidth}
		\includegraphics[width=1\linewidth]{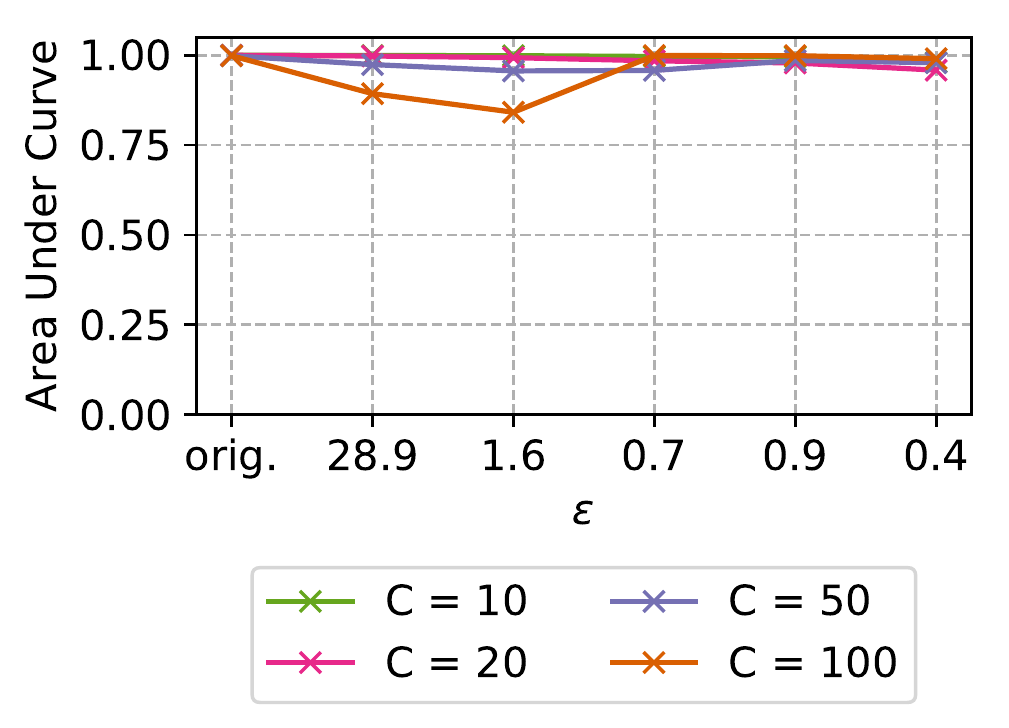}
		\caption{White-box AUC (CDP)}
		\label{fig:eval:purch_skew_cdp_auc_wb}
	\end{subfigure}%
	\begin{subfigure}{0.25\linewidth}
		\includegraphics[width=1\linewidth]{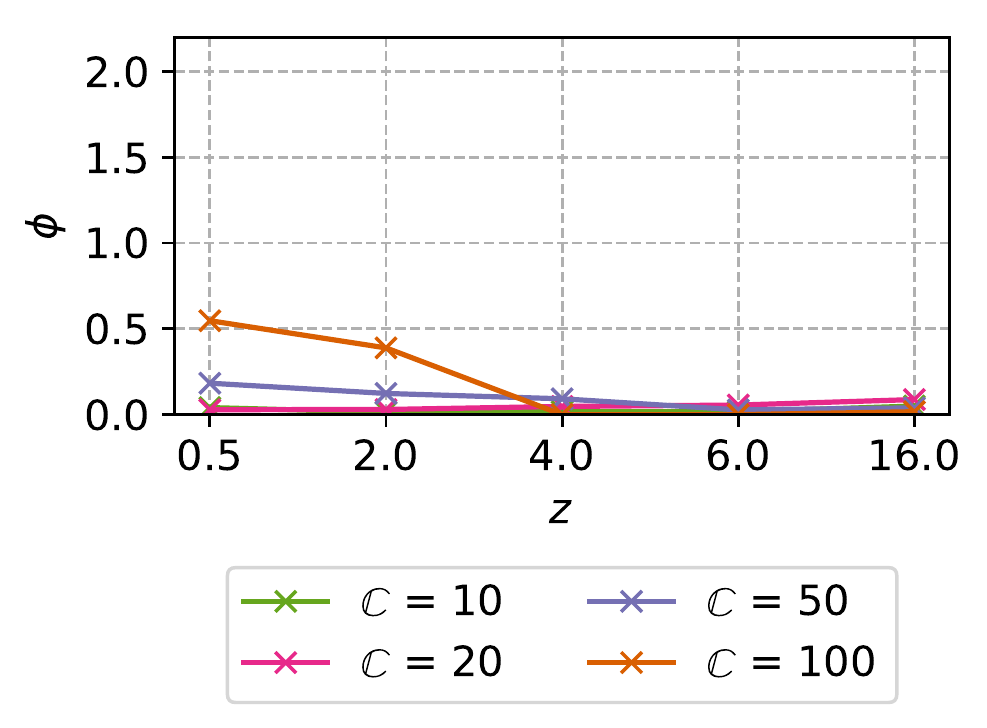}
		\caption{White-box $\varphi$ (CDP)}
		\label{fig:eval:purch_skew_cdp_phi_wb}
	\end{subfigure}\\
	\begin{subfigure}{0.25\linewidth}
		\includegraphics[width=1\linewidth]{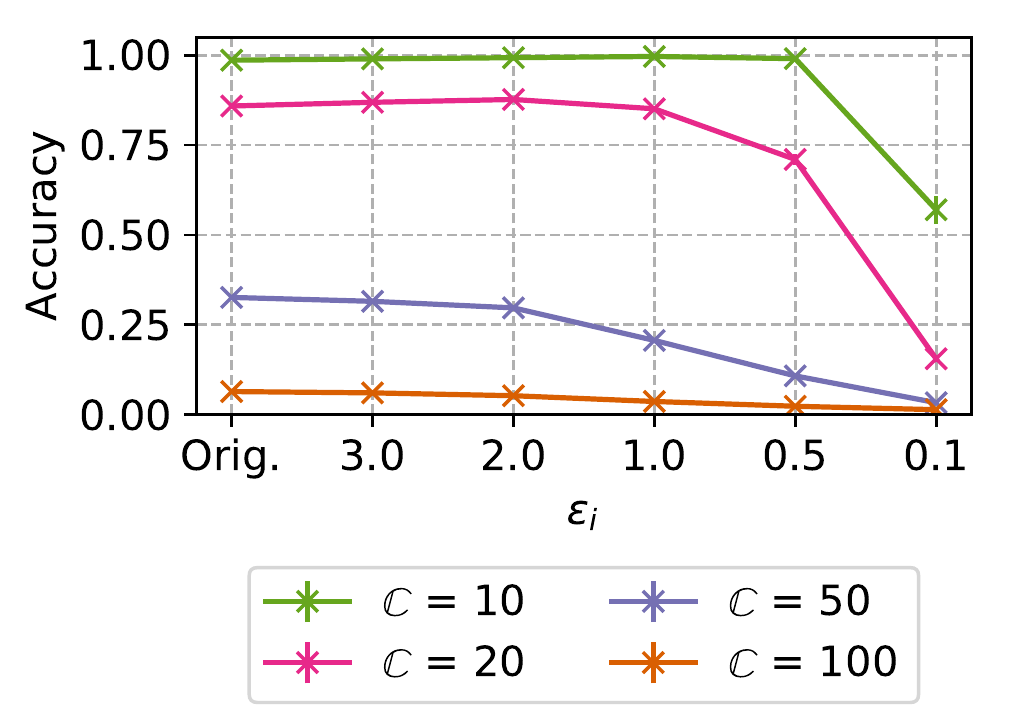}
		\caption{Target model accuracy (LDP)}
		\label{fig:eval:purch_skew_ldp_acc}
	\end{subfigure}%
	\begin{subfigure}{0.25\linewidth}
		\includegraphics[width=1\linewidth]{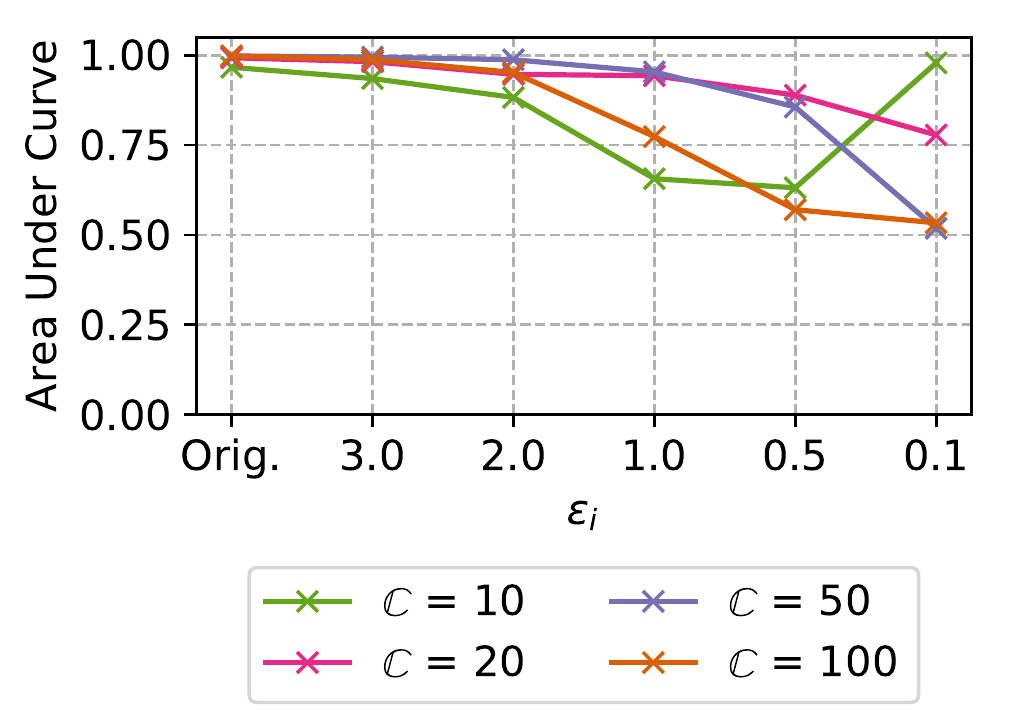}
		\caption{Black-box AUC (LDP)}
		\label{fig:eval:purch_skew_ldp_auc_bb}
	\end{subfigure}%
	\begin{subfigure}{0.25\linewidth}
		\includegraphics[width=1\linewidth]{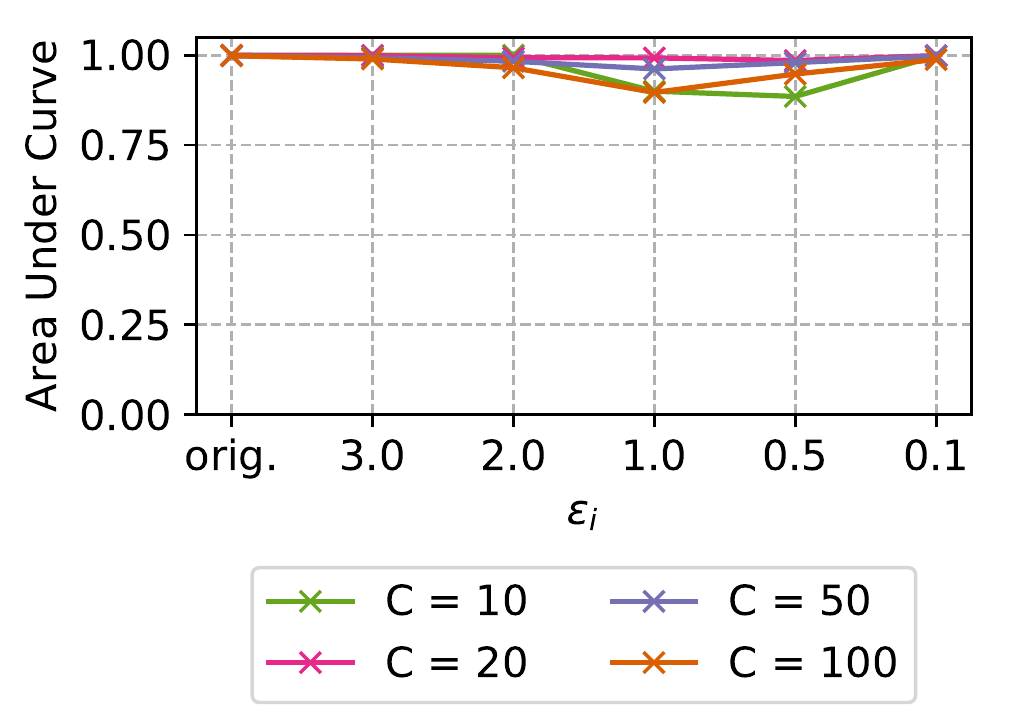}
		\caption{White-box AUC (LDP)}
		\label{fig:eval:purch_skew_ldp_auc_wb}
	\end{subfigure}%
	\begin{subfigure}{0.25\linewidth}
		\includegraphics[width=1\linewidth]{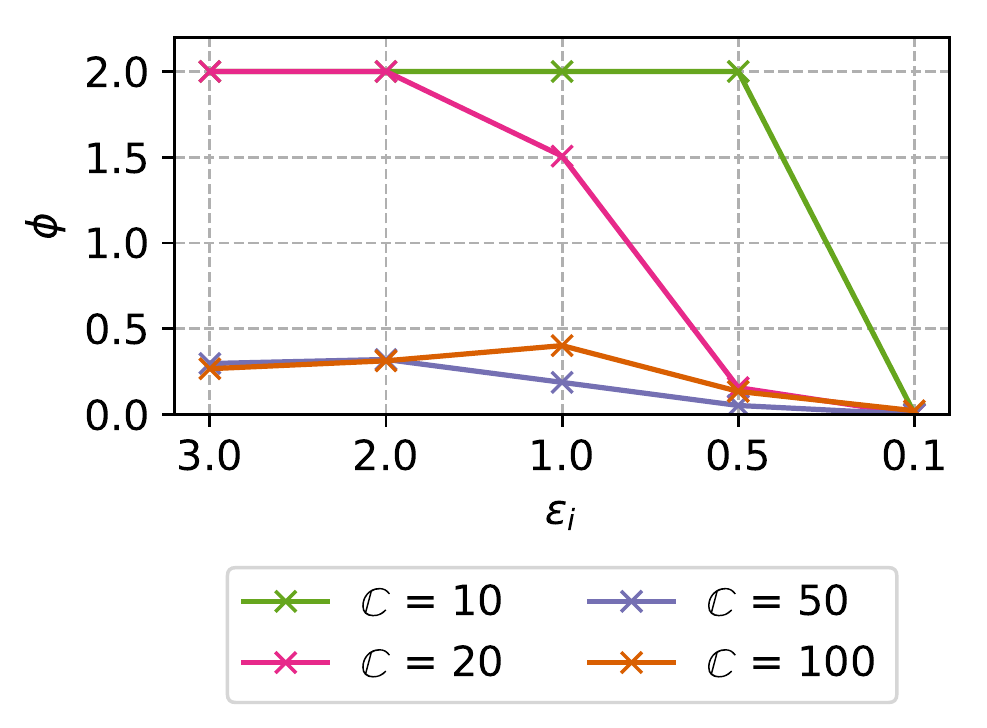}
		\caption{White-box $\varphi$ (LDP)}
		\label{fig:eval:purch_skew_ldp_phi_wb}
	\end{subfigure}\\
	\begin{subfigure}{0.45\linewidth}
		\includegraphics[width=1\linewidth]{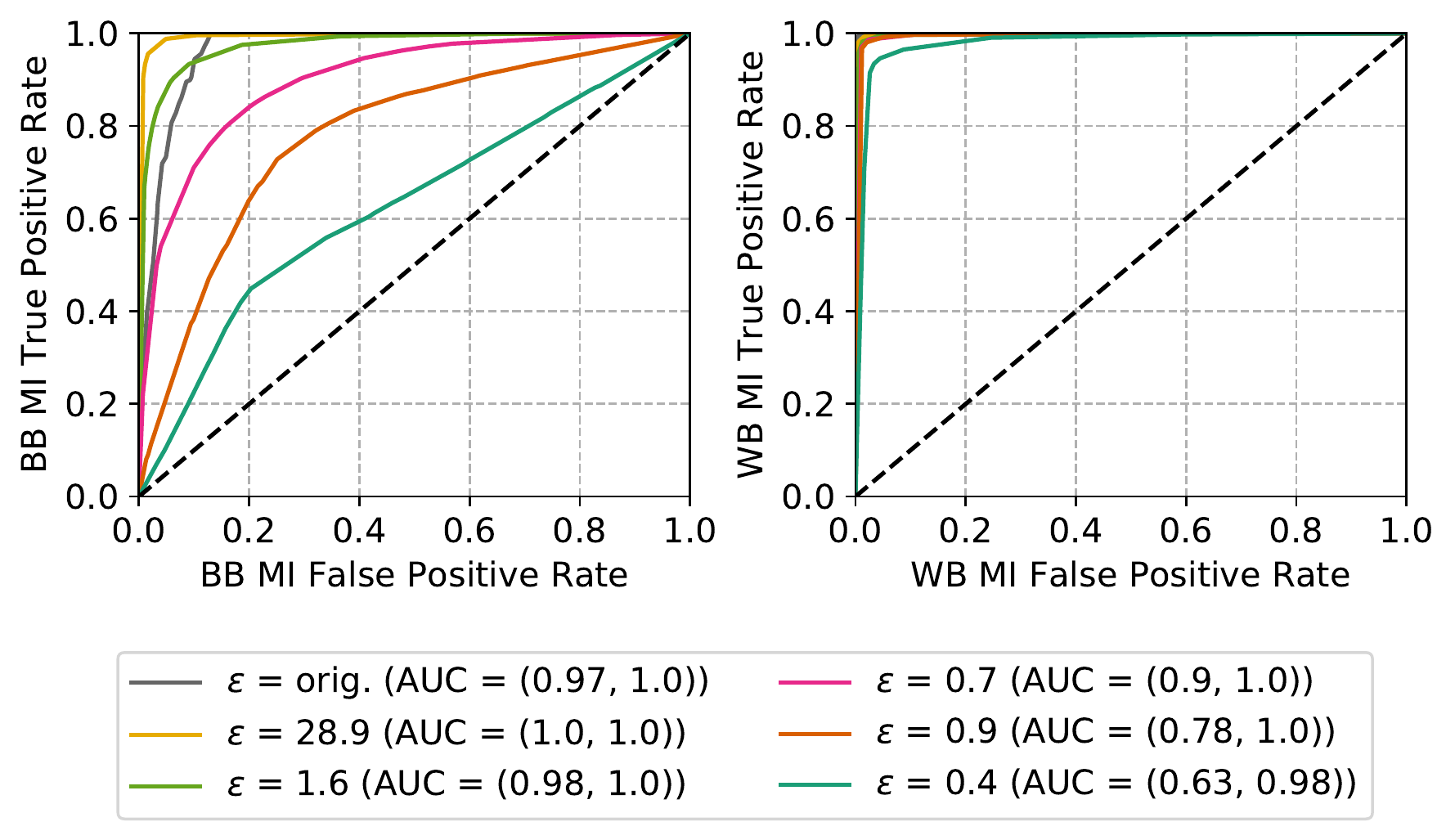}
		\caption{Receiver operating curve for $\mathbb{C}=10$ (CDP)}
		\label{fig:eval:purch_skew_unified_roc_cdp}	
	\end{subfigure}%
		\begin{subfigure}{0.45\linewidth}
		\includegraphics[width=1\linewidth]{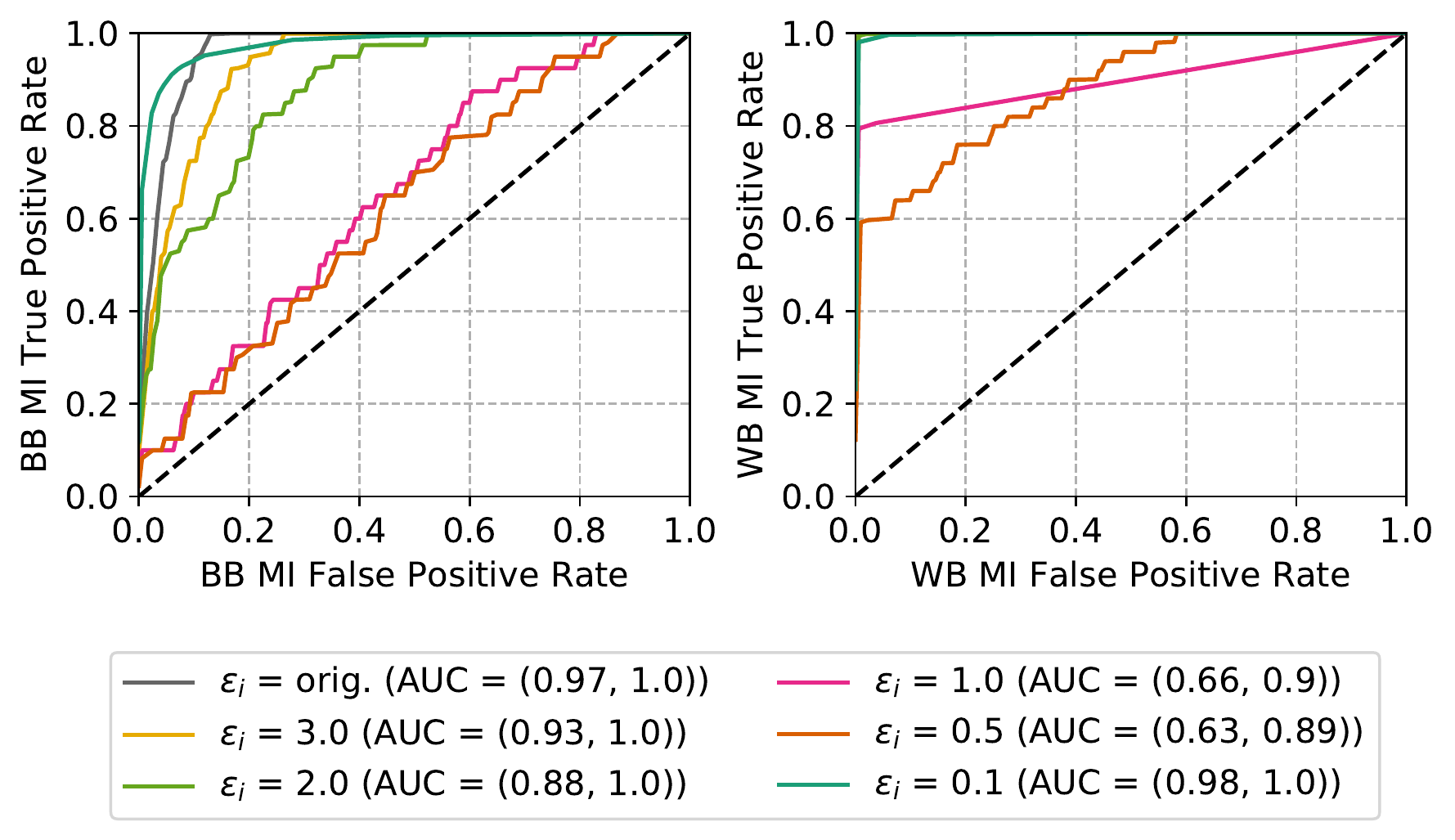}
		\caption{Receiver operating curve for $\mathbb{C}=10$ (LDP)}
		\label{fig:eval:purch_skew_unified_roc_ldp}	
	\end{subfigure}\\
	\begin{subfigure}{0.6\linewidth}
		\includegraphics[width=1\linewidth]{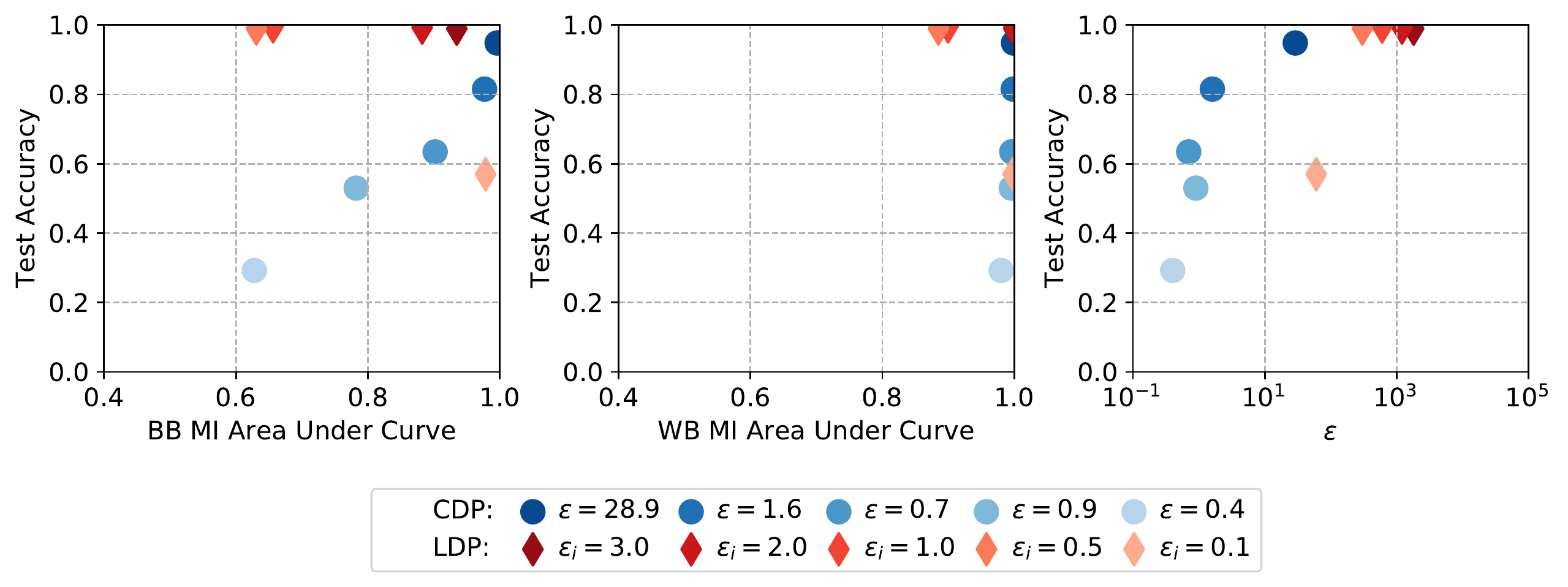}
		\caption{Privacy-accuracy trade-off comparison over black-box AUC, white-box AUC and $\eps$~for $\mathbb{C}=10$}
		\label{fig:eval:purch_skew_unified_scatter}	
	\end{subfigure}\\
    \begin{subfigure}{0.35\linewidth}
	    \includegraphics[width=1\linewidth]{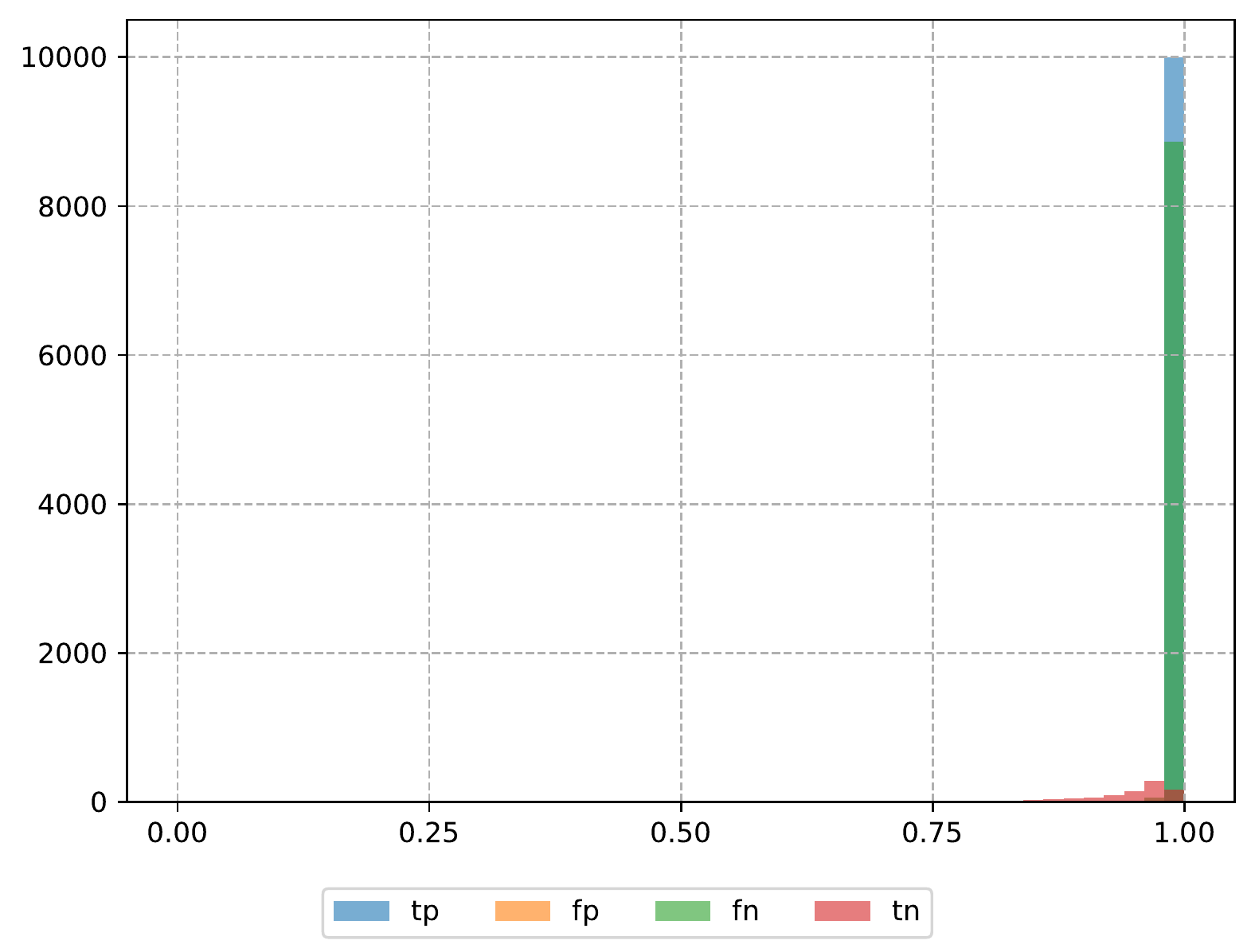}
	    \caption{Target model confidence according to attack model classification for $\mathbb{C}=10$, $\eps=1.0$}
	    \label{fig:eval:purch_skew_target_conf_1}	
    \end{subfigure}\hspace{6px}
    \begin{subfigure}{0.35\linewidth}
	    \includegraphics[width=1\linewidth]{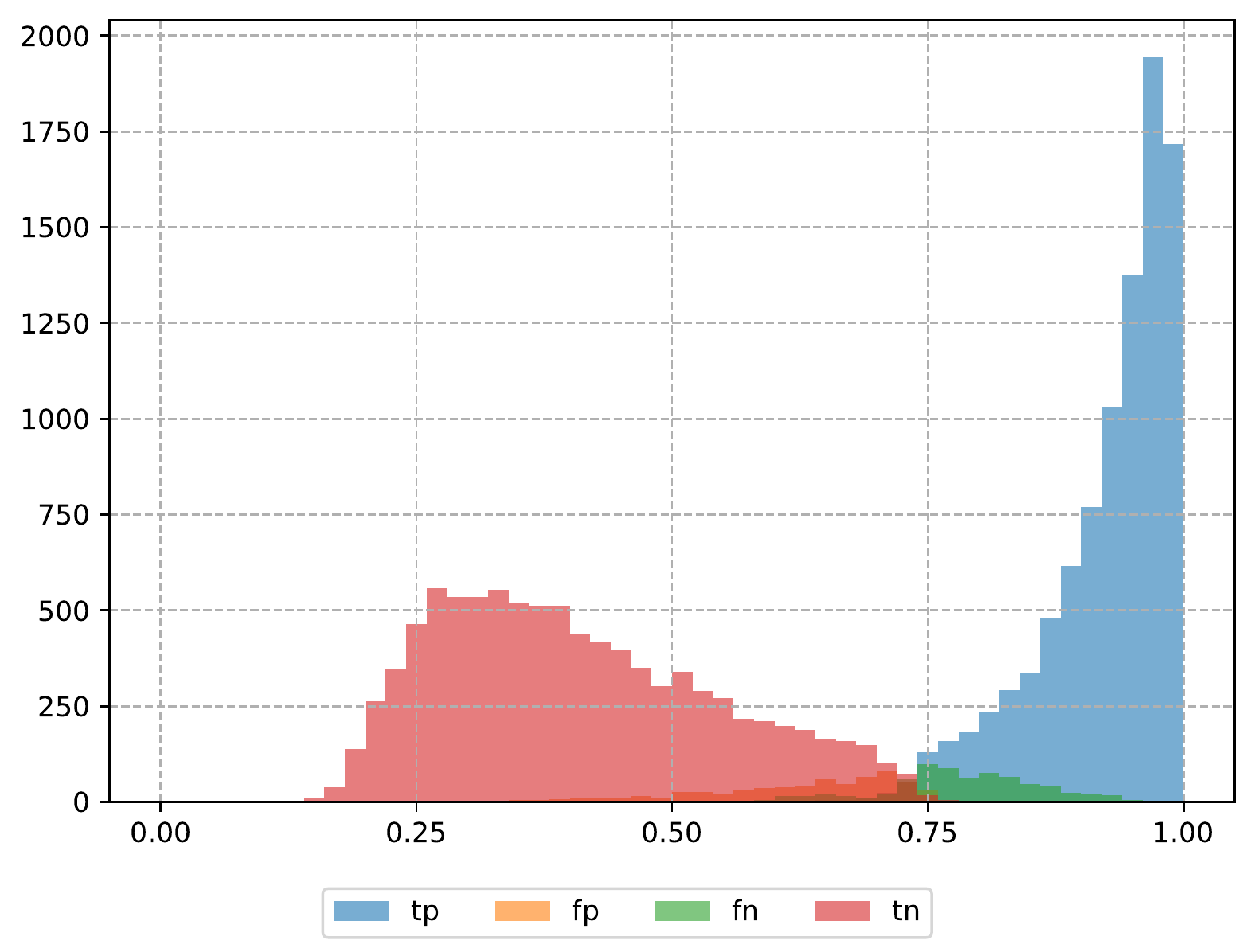}
	    \caption{Target model confidence according to attack model classificationfor $\mathbb{C}=10$, $\eps=0.1$}
	    \label{fig:eval:purch_skew_target_conf_01}	
    \end{subfigure}
    \caption{\cali{DO} accuracy and privacy analysis on Skewed Purchases (error bars lie within most points)}
    \label{fig:eval:purch_skew:figures}
\end{figure*}

The effects of dimensionality and imbalance of a dataset on MI have been addressed by related work~\cite{shokri2017,NSH18}. However, the effect of a domain gap between training and test data which is found in transfer learning when insufficient high-quality data for training is initially available and reference data that potentially follows a different distribution has not been addressed. For this task we consider the Skewed Purchases dataset. A general comment that we need to make for this dataset is that an outlier for \eps~in CDP and $\mathbb{C}=10$ is observable for this dataset in Table~\ref{tab:comp:train_acc}. Figures~\ref{fig:eval:purch_skew_cdp_acc} and~\ref{fig:eval:purch_skew_ldp_acc} show that the LDP test accuracy is in fact only decreasing at very small \eps~whereas CDP again gradually decreases over \eps. This leads to a consistently higher test accuracy in comparison to CDP. For the MI AUC we note that white-box MI outperforms the black-box MI for LDP and CDP on this transfer learning task when comparing Figures~\ref{fig:eval:purch_skew_cdp_auc_bb} and \ref{fig:eval:purch_skew_ldp_auc_bb} with Figures~\ref{fig:eval:purch_skew_cdp_auc_wb} and~\ref{fig:eval:purch_skew_ldp_auc_wb}. We again observe outliers where the MI AUC first decreases before recovering again. These outliers are also observable in the ROC curves presented in Figures~\ref{fig:eval:purch_skew_unified_roc_cdp} and \ref{fig:eval:purch_skew_unified_roc_ldp}. We can again reason about the cause of these outliers by visualizing the target model decisive softmax values in a histogram in Figures~\ref{fig:eval:purch_skew_target_conf_1} and \ref{fig:eval:purch_skew_target_conf_01}. LDP in any case generalizes the training data towards the test data. However, at LDP $\eps_i=1.0$ leads to nearly indistinguishable test and train distributions. Thus, the decisive softmax confidence of the target model increases in comparison to smaller and larger $\eps_i$. The same effect is observable at a lower level for white-box MI since the decisive softmax confidence values are only one feature in the white-box MI attack. W.r.t.~the relative privacy-accuracy trade-off $\varphi$ LDP outperforms CDP as depicted by $\varphi$ in Figures~\ref{fig:eval:purch_skew_cdp_phi_wb} and \ref{fig:eval:purch_skew_ldp_phi_wb}. Also, the absolute privacy-accuracy trade-off is favorable for LDP as depicted in Figure~\ref{fig:eval:purch_skew_unified_scatter}.

%% file: src/discussion.tex
\section{Discussion}
\label{sec:phi}
In the following we summarize our findings.\\

\textbf{Privacy parameter $\epsilon$ alone is unsuited to compare and select DP mechanisms.} 
We consistently observed that while \eps~in LDP is higher by a factor of hundreds or even thousands in comparison to CDP, the protection against black-box and white-box MI attacks is actually not considerably weaker. For Texas Hospital Stays LDP mitigates white-box MI at $\eps=6382$ (cf.~Table~\ref{tab:datasets}) whereas CDP lies between $\eps=0.9$ for $\mathbb{C}=100)$ and $\eps=0.3$ for $\mathbb{C}=300$. This observation at the baseline MI AUC also holds for Purchases Shopping Carts  where LDP $\eps=60$ and CDP is between $\eps=0.4$ for $\mathbb{C}=10$ and $\eps=0.3$ for $\mathbb{C}=100$), as well as COLLAB (LDP $\eps=2000$, CDP $\eps=1.7$) and LFW (LDP $\eps=62.5\times10^2$, CDP $\eps=2.1$ to $\eps=1.5$). Thus, we note that assessing privacy solely based on \eps~falls short. Given the results of the previous sections we would rather encourage data scientists to communicate privacy measured under an empirically evaluated attack, such as white- or black-box membership inference, in addition to privacy parameter \eps.

\textbf{LDP and CDP result in similar privacy-accuracy curves.} 
Our methodology allowed us to compare a wide area of privacy regimes in CDP and LDP under MI. We observed for most datasets that similar privacy-accuracy combinations are obtained for well generalizing models (i.e., use of early stopping against excessive overfitting) that were trained with LDP or CDP. The difference in black-box and white-box MI AUC were small, with white-box MI outperforming black-box MI on all datasets but COLLAB where both attacks achieve similar performance. However, the effect of the additional assumptions made by white-box MI (e.g., access to internal gradient and loss information) is small ($\approx3 - 5\%$). The privacy-accuracy scatterplots depict that LDP and CDP formulate very similar privacy-accuracy trade-offs for Purchases Shopping Carts, LFW and Texas Hospital Stays. At two occasions on the smaller classification tasks Purchases Shopping Carts ~$\mathbb{C}=\{10,20\}$ and Skewed Purchases ~$\mathbb{C}=\{10,20\}$ LDP realizes a strictly better privacy-accuracy trade-off. These observations lead us to conclude that LDP is an alternative to CDP for differentially private deep learning, since the privacy-accuracy trade-off is often similar despite the significantly larger \eps. Thus, data scientists should consider to use LDP especially when required to use optimizers without CDP or training ensembles (i.e., multiple models over one dataset), since the privacy loss would also accumulate over all ensemble target models when assuming that training data is reused between ensemble models. Here, we see one architectural benefit of LDP: flexibility. LDP training data can be used for all ensemble models withoutincreasing the privacy loss in contrast to CDP.

\textbf{The relative privacy-accuracy trade-off is favorable within a small interval.}
We observed that the privacy-accuracy trade-off stated in the previous scatterplots allows to identify whether CDP or LDP achieve better test accuracy at similar MI AUCs. However, the scatterplots do not reflect whether test accuracy is decreasing slower, similar or stronger than MI AUC over decreases in the privacy parameter \eps. For this purpose we introduced $\varphi$. We found that $\varphi$ allows to identify \eps~intervals in which the MI Accuracy loss is stronger than the test accuracy loss for all datasets except COLLAB. On the high dimensional datasets Texas Hospital Stays and LFW CDP consistently achieves higher $\varphi$ than LDP whereas LDP $\varphi$ are superior for CDP on Purchases and Skewed Purchases. 

%% file: src/related_work.tex
\section{Related Work}
\label{sec:rel}
Our work is related to DP and the interpretation of the privacy parameter \eps~ in neural networks, attacks against the confidentiality of training data and performance benchmarking of neural networks.

CDP is a common approach to realize differentially private neural networks by adding noise to the gradients during model training. Foundational approaches for perturbation with the differentially private gradient descent (DP-SGD) during model training were provided by Song et al.~\cite{SCS13}, Bassily et al.~\cite{BST14} and Shokri et al.~\cite{shokri2015}. 
Abadi et al.~\cite{abadi2016} introduced the implementation of the DP-SGD that was used in this work. Mironov~\cite{Mironov17} introduces Renyi DP for measuring the DP-SGD privacy loss over composition. Iyengar et al.~\cite{IST+19} suggest a hyperparameter free algorithm for differentially private convex optimization for standard optimizers.

Fredrikson et al.~\cite{fredrikson2015, fredrikson2014} formulate model inversion attacks that uses target model softmax confidence values to reconstruct training data per class. 
In contrast, MI attacks address the threat of identifying individual records in a dataset~\cite{sankararaman2009, backes2016}. 
Yeom et al.~\cite{YGF+18} have demonstrated that the upper bound on MI risk for CDP can be converted into an expected bound for MI. Jaymaran et al.~\cite{JE19} showed that in practice the theoretic CDP upper bound and the achievable MI lower bound are far apart.
We observe, that LDP can be an alternative to CDP as the upper and lower bounds are even farther apart from each other.
Hayes et al.~\cite{HMD+19} contribute MI attacks against generative adversarial models. Similar to our work they evaluate DP as mitigation. While we use LDP and CDP for feed-forward neural networks their work addresses generative models and suggests to apply CDP at the discriminator. 
Shokri et al.~\cite{Shokri2018} formulate an optimal mitigation against their MI attack~\cite{shokri2017} by using adversarial regularization. 
By applying the MI attack gain as a regularization term to the objective function of the target model, a non-leaking behavior is enforced w.r.t.~MI. While their approach protects against their MI adversary, DP mitigates any adversary with arbitrary background information. Carlini et al.~\cite{CLK+18} suggest \textit{exposure} as a metric to measure the extent to which neural networks memorize sensitive information. Similar to our work, they apply DP for mitigation. 
We agree that the desire to measure memorization of secrets is promising, however, in this work we focus on attacks against machine learning models targeting identification of members of the training dataset. In an empirical study, Rahman et al.~\cite{Rahman2018} applied the black-box MI attack against DP-SGD models on CIFAR-10 and MNIST. We additionally consider LDP and a wider rider range of datasets. 
They evaluate the severity of MI attack by the F1-score and thus may not reflect the optimal MI attack model in comparison to the ROC AUC. 
Abowd and Schmutte~\cite{abowd2019} describe a social choice framework to choose privacy parameter \eps~due to the production possibility frontier of the model and the social willingness to accept privacy and accuracy loss. We propose $\varphi$ to allow data scientists to assess whether losses in accuracy are outweighing gains in privacy and to compare LDP and CDP mechanisms aside from \eps.

MLPerf~\cite{MLPERF} is a benchmark suite for performance measurements related to machine learning.
Hay et al.~\cite{HAM+16} propose DPBench as an evaluation framework for privacy algorithms. We focus on comparing the privacy-utility trade-off and apply the core principles. 

%% file: src/conclusion.tex
\section{Conclusion}
\label{sec:conc}
This work compared LDP and CDP mechanisms for differentially private deep learning under MI attacks. The privacy-accuracy comparison comprises MI AUC and test accuracy to support data scientists in choosing among available DP mechanisms and selecting privacy parameter \eps. Our experiments on diverse learning tasks show that neither LDP nor CDP yields a consistently better privacy-accuracy trade-off. Instead, CDP and LDP are rather similar and when having the choice data scientists may compare the privacy-accuracy trade-off for LDP and CDP per dataset. Especially, since LDP does not require privacy accounting when training multiple models (i.e., ensembles) and offers flexibility with respect to usable optimizers. We furthermore suggest to consider the relative privacy-accuracy trade-off for LDP and CDP as the ratio of losses in accuracy and privacy over privacy parameters \eps, and show that it is only favorable within a small interval.

%% file: src/acknowledgements.tex
\section{Acknowledgements}
\label{sec:ack}
We thank Steffen Schneider for his instrumental contribution to implementation and analysis of white box MI attacks. This work has received funding from the European Union’s Horizon 2020 research and innovation program under grant agreement No.~825333 (MOSAICROWN). 